\documentclass[12pt,english]{paper}
\usepackage[T1]{fontenc}
\usepackage[colorinlistoftodos]{todonotes} 
\usepackage{graphicx}
\usepackage{hyperref}
\usepackage{natbib}
\usepackage{subcaption}
\usepackage{soul}
\usepackage{amsmath}
\usepackage{booktabs}
\usepackage{float}

\usepackage{siunitx}                        
\usepackage{threeparttable}  
\usepackage{color}

\begin{document}
\title{Becoming Immutable: How Ethereum is Made\thanks{We are grateful to Agnostic Relay for providing the data, especially to Ombre for patiently explaining how relays work.  We are also grateful to  Agostino Capponi, Hanna Halaburda, Fahad Saleh, Danning Sui, the participants to the workshops ``(Back to) The Future(s) of Money II'', the 2025 ``CBER CtCe'' conference and the 2025 Columbia CryptoEconomics Workshop for their comments and suggestions. }}
%
%
\author{Andrea Canidio\thanks{Unaffiliated, acanidio@gmail.com} \and
Vabuk Pahari\thanks{Max-Planck-Institute for Software Systems, 
vpahari@mpi-sws.org} 
}
%
%
%

\maketitle              
\begin{abstract}
Blockchain's economic value lies in enabling financial and economic transactions without relying on trusted, centralized intermediaries. In practice, however, transactions pass through a fragmented chain of intermediaries before being included on-chain. Because standard blockchain data reveal only the winning block, this process is largely unobservable. We address this limitation by constructing a novel dataset of 15{,}097 non-winning Ethereum blocks, that is, blocks proposed but not selected for inclusion. We show that 21\% of user transactions are delayed: they appear in candidate blocks but not in the winning block, implying that fragmented routing materially affects inclusion time. We further show that execution quality varies substantially across candidate blocks: for the same swap, both execution probability and execution price differ across proposed blocks. To study these differences, we examine competition between two arbitrage bots trading between decentralized and centralized exchanges. We find that, conditional on inclusion in a block that also contains transactions from these bots, user swaps in the same (opposite) direction are less likely (more likely) to execute and receive worse (better) prices. These results show that routing and block composition are central determinants of execution quality and market quality in on-chain markets.
\end{abstract}

\section{Introduction}

Blockchain systems are often described as ``immutable ledgers'' because, once a transaction is included in the canonical chain, it triggers a deterministic sequence of irreversible updates to the ledger. This immutability underpins blockchain's economic value by enabling transactions without trusted, centralized intermediaries. However, this static view overlooks the competitive process that determines which transactions become immutable and when.

In Ethereum, transactions pass through a fragmented chain of intermediaries before being recorded on-chain. This chain starts with either users or \textit{searchers}---automated bots that create transactions in response to, for example, price movements in financial markets. Users and searchers submit these transactions to one or more \textit{builders} through several channels: the public mempool, a publicly observable repository of pending transactions; direct bilateral communication; or intermediaries known as \textit{private mempools} (or private RPCs). Builders aggregate and order transactions into candidate blocks and submit them to \textit{relays}. Each relay forwards one block to the \textit{proposer}, who chooses which block to append to the canonical chain.\footnote{See Section~\ref{sec: background} for a detailed description of this process, commonly referred to as Proposer--Builder Separation (PBS).} 

This process resembles an auction in which builders compete to supply the proposer with the most valuable block, as measured by the sum of the fees paid by the transactions included in the block. The auction, however, is highly noisy. The value of including a transaction depends on rapid price movements in financial markets, builders continuously update and resubmit blocks, and communication latency plays a central role in determining the outcome. As a result, for each block ultimately added to the canonical chain, up to 5{,}000 candidate blocks may be constructed, propagated, and discarded along with their transactions.

This fragmented supply chain may materially affect the quality of blockchain-based markets. If a transaction is routed to only a subset of builders, it may appear in some candidate blocks but not in the winning one, therefore delaying its on-chain settlement. More subtly, there may be unobserved risk in transaction outcomes, especially for swaps: the same swap may be preceded by different transactions and therefore execute at different prices across different candidate blocks in which it appears.  At the same time, it is also possible that block construction and selection matter little for ordinary users if candidate blocks are sufficiently similar from their perspective. Distinguishing between these possibilities requires observing not only the winning block, but also the non-winning ones. Yet empirical analysis has until recently been infeasible because these non-winning blocks are not publicly disclosed.

In this paper, we overcome this limitation by assembling a novel dataset of \textit{non-winning} blocks submitted over 39 Ethereum blocks between 2:37:35 PM and 2:45:25 PM UTC on December 3, 2024 (blocks 21{,}322{,}622 through 21{,}322{,}660). The dataset is approximately 20~GB in size and contains 15{,}097 blocks, 10{,}793 unique transactions, and 2{,}380{,}014 transaction--block pairs.\footnote{The dataset includes all blocks submitted during the study period through the relay \textit{Agnostic}. See Section~\ref{sec: data description} for a detailed description of the data.} We simulate each candidate block to recover counterfactual transaction-level execution outcomes, that is, the outcomes that would have occurred had a non-winning block been selected for inclusion on-chain. We combine these data with several publicly available sources, including second-by-second price data from Binance. This enables us to compare the same transaction across alternative candidate blocks and to study how routing and block composition shape realized outcomes.

Our first finding is that fragmented routing generates substantial inclusion risk. Approximately 21\% of \textit{user} transactions appear in some proposed blocks but not in the corresponding winning block, postponing on-chain inclusion by one or more blocks.\footnote{We distinguish between users and searchers (see Section~\ref{sec: transactions} for the heuristic used to make this distinction). Searcher transactions are almost never delayed: they are either included immediately or discarded.} Strikingly, about 30\% of these delayed user transactions appear exclusively in blocks submitted by a single non-winning builder. This is hard to rationalize as users do not benefit from sharing transactions privately with only one builder rather than privately with multiple builders. We also identify 12 user transactions that remain exclusive to a non-winning builder for two or three consecutive auctions before later appearing in the public mempool. These transactions are not only delayed, but also subsequently exposed to potential attacks; one is in fact victim of a type of front-running called ``sandwich attack''. In four cases, the same builder that initially had exclusive access to the transaction ultimately included it on-chain after it appeared in the public mempool. This pattern is consistent with a private mempool operator initially sharing transactions exclusively with a single builder and reverting to the public mempool if that builder fails to win successive auctions.

Our second finding is that the process of block construction and selection introduces substantial variation in execution outcomes: for a nontrivial subset of user swaps, both execution success and execution prices vary substantially across candidate blocks. For each swap in our sample, we simulate every candidate block in which it appears and recover its execution outcome had that block been included on-chain. We then compute the fraction of blocks in which the swap executes successfully and how dispersed its execution prices are across blocks, conditional on successful execution. We find that the vast majority of user swaps (96.7\%) execute successfully in all blocks in which they appear. However, a small fraction fail very frequently: 1.7\% fail in at least half of the blocks in which they appear, and 1.1\% fail in more than 80\% of such blocks. Execution prices are similarly skewed. For half of the swaps in our sample, execution prices are tightly concentrated: at least two-thirds of realizations deviate by less than 4.3 basis points from the swap's average execution price. By contrast, for 7\% of swaps, execution prices are highly dispersed, with at least one-third of realizations deviating by more than 100 basis points from the swap's average.

Our third finding is that variation in the execution quality of user swaps is linked to competition from the two main arbitrage bots in our data (see the next paragraph). Restricting attention to swaps that sometimes fail and exploiting within-transaction variation across candidate blocks, we find that swaps trading the same asset pair and in the same direction as these bots are approximately 18\% less likely to execute when included in a block that also contains these bots' transactions. By contrast, swaps trading the same asset pair but in the opposite direction are about 1\% more likely to execute in such blocks. A similar, though quantitatively smaller, pattern emerges for execution prices: swaps in the opposite direction of the bots receive more favorable prices when included in blocks that also contain bot transactions, whereas swaps in the same direction receive worse prices.

We conclude by examining competition between the two main arbitrage bots in our data, which account for roughly one-third of the transactions in our dataset. These bots frequently compete for the same arbitrage opportunities between decentralized exchanges (DEX) and centralized exchanges (CEX) and appear to be associated with different major builders. Their transactions are also highly valuable, accounting on average for about 60\% of total transaction fees in the blocks in which they appear. Their competition therefore plays a central role in determining auction outcomes. Our data allow us to study how this competition evolves over the course of the auction and to compare bot behavior during the auction with price movements in traditional financial markets.

Specifically, we simulate each block that contains a swap transaction from either bot and extract the associated swap volume, on-chain execution price, and transaction fee---that is, the amount the bot is willing to pay to have its transaction included on-chain. Because this fee cannot exceed the bot's expected profit from inclusion, we use it to construct a bound on the implied CEX price, that is, the price of the off-chain leg of the arbitrage trade, at the time the block is constructed. We find that when the bots compete to rebalance either a USDT/WETH or a USDC/WETH DEX pool---that is, an on-chain liquidity pool holding the two tokens against which users trade---their implied CEX prices are nearly identical, indicating intense competition. Moreover, for both bots, these implied prices are at between 1 and 1.6 basis points \textit{better} than contemporaneous Binance prices.

\subsection{Background: the journey of an Ethereum transaction.}\label{sec: background}

In Ethereum, the right to create and append the next block to the blockchain is assigned to a proposer, selected from validators that have staked (i.e., locked) at least 32 ETH. In practice, however, proposers rarely construct blocks themselves. Instead, in more than 95\% of cases, they rely on specialized intermediaries known as builders, who assemble transactions and determine which ones to include in a block and in what order. Multiple builders then compete to have their block selected by the proposer for inclusion on-chain. This arrangement is known as Proposer--Builder Separation (PBS).

The interaction between builders and proposers is facilitated by relays. After constructing a block, a builder submits it to a relay together with a proposed payment to the proposer. The relay then forwards the highest-paying block and the associated payment to the proposer upon request. This process is similar to an ascending-price auction because the relay continuously broadcasts the value of the highest bid, although this information is observed with a delay due to network latency. Each auction cycle lasts approximately 12 seconds, after which the winning block is selected and appended to the canonical chain and a new auction begins. Latency is crucial because proposers are geographically dispersed and arbitrage transactions often depend on rapid price movements in traditional off-chain financial markets. To account for this geographic dispersion, multiple relays operate around the world, and builders typically submit blocks to several of them simultaneously. Our dataset, for example, is from a relay called Agnostic Relay.

PBS creates a layered transaction supply chain. At the top of the chain are users and searchers, who can submit transactions to builders through several channels. The most transparent is the public mempool, a publicly observable repository of pending transactions maintained by Ethereum nodes. Transactions may also be transmitted privately, either through direct bilateral channels or through intermediaries commonly called \textit{private mempools} or \textit{private RPCs}, which route transactions to one or more builders without broadcasting them publicly. During our study period, five major private mempool operators were active. Four of them---MEV Blocker, Flashbots Protect, Merkel, and Blink---state in their documentation that they privately share transactions with multiple builders (see \citealp{janicot2025privatemevprotectionrpcs} for a detailed comparison). By contrast, the fifth operator, MetaMask Smart Transactions, does not disclose its transaction-sharing policy.

This fragmentation matters because different submission channels expose transactions to different risks and opportunities. Public submission maximizes visibility, but it also exposes users to predatory trading. For example, when a user submits a swap to the public mempool, other traders may observe it and attempt to profit by front-running it or sandwiching it. Private routing is intended to reduce this exposure by hiding the transaction from the public mempool. However, it also implies that the transaction may be visible to only a subset of builders, depending on how the private mempool operator shares its order flow.

For users, the choice between submitting a transaction via the public mempool or through a private mempool is typically determined by their \textit{wallet}, a software application that automates transaction creation and broadcasting. Wallets generally provide a default submission option that users can modify, although this is not always the case, especially for mobile wallets. For example, users of the popular MetaMask wallet submit Ethereum transactions via MetaMask Smart Transactions by default.

\subsection{Relevant literature}

Our paper relates to three strands of literature.

First, it contributes to the empirical finance literature on order flow, market fragmentation, and execution outcomes. In traditional financial markets, fragmentation can affect spreads, execution speed, volatility, and price efficiency, depending on whether it intensifies competition across venues or instead fragments liquidity and access (see, for example, \citealp{BENNETT200649}, \citealp{OHARA2011459}, and \citealp{10.1093/rof/rfu027}). We study these issues in a blockchain setting in which routing matters for execution outcomes but is hidden in standard data. Conventional blockchain datasets reveal only the final outcome---the block ultimately appended on-chain---and not the routing process through which transactions are distributed across builders or the competing candidate blocks in which they may have appeared. In this respect, our paper is methodologically related to \cite{aquilina2022quantifying}, who use all messages received by the London Stock Exchange over a 9-week period---including those that fail to execute---to identify high-frequency trading races in traditional financial markets and estimate their welfare cost.

Second, our paper contributes to the emerging literature on proposer-builder separation (PBS), builder-searcher relationships, and order flow routing. \cite{JOHN2025102718} provide a broad overview of the economics of Ethereum and PBS. Related work studies competition among searchers and the relationship between builders and searchers, particularly when transactions are routed privately and some order flow is exclusive (see \citet{pai2024structural} and \citet{wu2025measuring}). \citet{lehar2023battle} and \citet{CAPPONI2025104132} analyze the trade-off between submitting transactions via the public mempool and via private channels, and characterize the resulting equilibrium sorting of transactions across submission routes; \citet{CAPPONI2025104132} also provides empirical support for its model using data from Flashbots Protect. \citet{halaburda2025impossibility} develop a theoretical model of transaction inclusion in blockchains that explicitly incorporates competition among builders and intermediaries that route user transactions. Relative to this literature, our contribution is to quantify how  hidden variation in routing and block composition affects user inclusion and execution outcomes.

Third, our paper relates to the growing literature on the microstructure of blockchain-based trading and decentralized exchanges. This literature studies how decentralized trading mechanisms shape liquidity, arbitrage, and execution outcomes. The most closely related paper is \cite{capponi2024price}, which also studies competition between arbitrage bots for the same CEX--DEX arbitrage opportunities. However, that paper examines an earlier period (i.e., pre-PBS) in which searchers competed by submitting transactions to the public mempool and engaging in \textit{priority fee auctions}. The current market structure, and the nature of competition between searchers, are very different: searchers now submit transactions directly to builders, builders decide which searcher transactions to include, and searchers and builders are often vertically integrated.\footnote{See also \cite{capponi2024price-private}, who develop a theory of searcher competition when transactions are sent privately to builders.} Other related papers include \cite{lehar2025decentralized}, which analyzes the economics of Uniswap as an automated market maker; \cite{milionis2022automated}, which shows how stale prices generate arbitrage opportunities and losses for liquidity providers; and \cite{capponi2025liquidity}, which studies decentralized exchanges from a market-microstructure perspective. By observing non-winning blocks, our paper identifies two outcomes that standard blockchain data cannot reveal directly: inclusion risk and cross-block variation in execution quality.

In our companion paper \cite{pahari2025exclusiveethereumtransactionsevidence}, we use the same data to measure the share of transactions that are \textit{exclusive} to a builder, meaning that they appear only in blocks proposed by the winning builder during a given auction cycle. This question is particularly important given the high concentration of the builder market: currently, the top builder (Titan) produces about 50\% of all blocks, while the top three account for more than 90\%. We find that between 77\% and 84\% of the total value of transactions included on-chain is attributable to exclusive transactions, confirming that exclusivity is the main driver of builder profits and PBS auction outcomes.

Finally, to the best of our knowledge, \cite{yang2025decentralization} is the only other paper that examines the content of non-winning blocks. They study builder competition by comparing bids with block revenues (i.e., the total fees paid by transactions included in a proposed block), but they do not analyze individual transactions in non-winning blocks. To our knowledge, \cite{mamageishvili2025searchercompetitionblockbuilding} is the only other paper that studies transactions proposed for inclusion on-chain but never actually included. However, it focuses on a much narrower setting: the backrunning auction (i.e., the auction for the right to place an arbitrage transaction immediately after a user's transaction) operated by Flashbots Protect, a private mempool operator.

The remainder of the paper is organized as follows. The next section describes our dataset and provides additional summary statistics, including an in-depth analysis of the auction cycle leading to the inclusion of Block 21322649 on the Ethereum blockchain. The following section presents the empirical analysis. The final section concludes.

\section{Dataset Description and summary statistics}\label{sec: data description}

As already mentioned, our primary dataset includes \textit{all} blocks (winning and non-winning) submitted via a relay called ``Agnostic relay'' during the study period.\footnote{Founded in 2022, Agnostic relay held 20\% of the relay market in 2023, declining to around 5\% during our study period.} 
We complement our primary dataset with several publicly available data sources:
\begin{itemize}
    \item Winning blocks for our study period, including those not supplied by Agnostic relay. We also track winning blocks beyond our study period to determine whether transactions in our primary dataset were eventually included on-chain.
    \item Hashes of blocks submitted to other relays (Flashbots, BloXroute, Manifold, Eden, Ultra Sound, SecureRpc, and Aestus), along with the name of the builders submitting them. Here, a block's hash serves as its unique identifier, as two blocks that differ in a single transaction have different hashes. We can, therefore, track whether a block in our primary dataset was also submitted to other relays.
    \item Data on transactions submitted via the public mempool, obtained from \url{https://mempool.guru/}.
    \item Data on transactions submitted via two private mempools, MEV Blocker and Flashbots Protect, obtained from \url{https://Dune.com}.
    \item Second-by-second price data from Binance (opening price).
\end{itemize}
We also simulate the non-winning blocks in our dataset to assess how the execution of a transaction --- particularly a swap on a blockchain-based financial market --- varies depending on the block in which it is included. We simulate the block using an Ethereum Archive Node running the Erigon client, forking at the appropriate block heights.

\subsection{Blocks}

Out of the 39 winning blocks in our study period, we observe no bids for 11 auction cycles.\footnote{Specifically, we observe no bids for the auction cycles that led to the inclusion of blocks 21,322,624; 21,322,625; 21,322,627; 21,322,629; 21,322,633; 21,322,634; 21,322,636; 21,322,644; 21,322,646; 21,322,647; 21,322,651.} This is because the proposer for those auction cycles was not registered with Agnostic Relay, so the corresponding auctions were not run through that relay. For the remaining 28 auction cycles, the number of blocks per auction cycle in our data ranges from 220 to 951, with a mean of 539.18 and a median of 509. These blocks account for 28.3\% of \textit{all} blocks submitted to the major relays during these 28 auction cycles. Finally, most blocks in our dataset (85\%) were also shared with at least one other relay.

We are able to match 14,043 of the 15,097 blocks in our data to 23 known builders. The most active builders in our dataset are Titan Builder (7,024 blocks, 46.5\%), Rsync Builder (2,359 blocks, 15\%), Flashbots (1,936 blocks, 12.8\%), and Beaverbuild (1,418 blocks, 9.4\%). However, the distribution of builders in our primary dataset differs from their distribution across \textit{all} relays. Across all relays, Titan Builder submits 13,679 blocks (25.6\%), Rsync Builder 7,936 blocks (14.9\%), Flashbots 2,626 blocks (4.9\%), and Beaverbuild 21,447 blocks (40.2\%). Thus, Beaverbuild is underrepresented in Agnostic Relay relative to its overall presence across relays. We acknowledge this imbalance as a limitation of our data.

To investigate this imbalance, we examine the timestamps of blocks that were shared between Agnostic and at least one other relay. We find that only 5\% of these blocks were received first by Agnostic. More specifically, among blocks received by both Agnostic and Ultrasound, only 1\% reached Agnostic first; among blocks received by both Agnostic and BloXroute, 15\% did. Agnostic therefore appears to have a latency disadvantage relative to the other major relays, which may help explain why some builders submit fewer blocks to it.

A methodological note on the timestamps used throughout the paper. Each block in our dataset has two timestamps: \textit{received at}, the time the relay received the block from the builder, and \textit{made available at}, the time the relay made the block available to the proposer. The gap between these timestamps arises because relays simulate blocks to verify their validity before making them available to the proposer.
This delay is non-negligible: the median gap is 0.76 seconds, and the 75th percentile is 1.5 seconds. In what follows, we focus on the \textit{received at} timestamp because it more accurately captures when builders submit blocks and when transactions first appear in them.

\subsection{Transactions}\label{sec: transactions}

Each transaction included in a block specifies either a fee or an explicit payment to the builder. The sum of these fees and payments constitutes the builder's revenue if the block wins the auction and is added to the chain. Similarly, builders pay proposers by appending a transfer to the proposer's address at the end of each block. In what follows, we analyze builder bidding behavior separately from user and searcher activity. To do so, we label these transfers from builders to proposers as ``bids,'' so that when we refer to ``transactions,'' we mean transactions that are \textit{not} bids. 

We identify transactions by their hash, a unique cryptographic identifier generated from the transaction's contents. This allows us to compare the set of proposed transactions with the subset that ultimately appears in winning blocks. The composition of these two sets differs markedly. Of the 5,576 unique transactions in our primary dataset that ultimately appear in winning blocks, the vast majority (87.7\%) are user transactions, while only a small share are searcher transactions.\footnote{We classify an address as a searcher if it trades on decentralized exchanges (DEXes) using a smart contract whose source code is not disclosed on \url{https://Etherscan.com}. In other words, part of the searcher's logic is embedded in a smart contract but hidden by making only its bytecode available (the bytecode is the compiled, machine-readable version of the smart contract). Trading on a DEX is necessary for common forms of ``searching,'' such as sandwich attacks, arbitrage, backrunning, and liquidations on lending protocols.} Also, only 14.4\% of transactions included in winning blocks are swaps. Instead, among all the 10,793 proposed transactions in our dataset, roughly half are swaps and roughly half originate from searchers.

A notable feature of our primary dataset is the concentration of activity in two addresses, which together account for 2,981 transactions, or 27\% of all unique transactions. The first address\footnote{\texttt{0x68d3A973E7272EB388022a5C6518d9b2a2e66fBf}.} is responsible for 1,873 transactions, of which only 31 are included in a winning block. Only 6 of these 1,873 transactions appear in blocks \textit{not} built by Titan, and all 6 are simple token transfers to Binance. The second address\footnote{\texttt{0x51C72848c68a965f66FA7a88855F9f7784502a7F}} is responsible for 1,108 transactions, of which only 97 are included in a winning block. Only 249 of these 1,108 transactions appear in blocks \textit{not} built by Rsync, and only 4 appear in blocks \textit{not} built by either Rsync or Titan.
Among those are 2 simple token transfers to addresses associated with Wintermute, an algorithmic trading firm that operates Rsync Builder. Based on these patterns, we refer to the two addresses as ``Titan-bot'' and ``Rsync-bot''.\footnote{Using a different methodology, \cite{heimbach2024non} also identify Rsync-bot as a non-atomic arbitrage bot, while Titan-bot did not yet exist during the period covered by their study. \cite{wu2025measuring} also identify Rsync-bot and Titan-bot as non-atomic arbitrage bots ``owned by'' or ``exclusive to'' their respective builders.}


To further validate our identification of ``Rsync-bot'' and ``Titan-bot,'' we examine whether these searchers tailor otherwise similar transactions to specific builders. Using transaction hashes, we find no case in which Rsync-bot submitted the same transaction to both Rsync and Titan. However, comparing transaction logs---that is, the sequence of operations executed by a transaction and their outputs---reveals 203 cases of economically identical transactions sent to both builders. The difference in hashes arises because Rsync-bot varies the fee offered across builders: on average, it pays 18\% more to Rsync than to Titan for executing the same transaction. Taken together, these patterns support our interpretation that Rsync-bot is integrated with Rsync and that its higher fee offers to Rsync reflect builder-specific routing incentives. We also identify a small number of Titan-bot transactions in Beaverbuild blocks; when comparable transactions appear in both Titan and Beaverbuild blocks, Titan-bot pays higher fees for inclusion with Titan, again consistent with builder-specific routing.

Almost all transactions from these bots are token swaps on decentralized exchanges (DEXs): 1,864 for Titan-bot and 1,095 for Rsync-bot. We later show that these bots frequently trade in the same DEX pools within the same auction cycle. Moreover, when they trade the same token pair in the same direction, their swaps are almost always mutually exclusive, in the sense that no proposed block contains transactions from both bots.\footnote{There is a single exception to this pattern: a Titan-bot transaction selling approximately USD~700 of MOG Coin for ETH appears in some blocks alongside an Rsync-bot transaction selling approximately USD~7{,}000 of MOG Coin for ETH.} Taken together, these patterns indicate that both bots primarily engage in arbitrage between on-chain and off-chain markets, and frequently compete for the same opportunities.




%
Swaps on decentralized exchanges (DEXes) are the most common type of transaction across all unique transactions in our primary dataset. We identify 5,282 swap transactions, of which 3,915 occur on Uniswap V3, 1,222 on Uniswap V2, 142 on Sushiswap, and 102 on Pancakeswap.\footnote{We observe 99 transactions that interact with multiple DEXes, which is why the total number of swaps on individual DEXes exceeds the number of unique swap transactions.} Despite their prevalence, only 808 swap transactions --- approximately 15\% of all swap transactions --- were ultimately included on-chain. A small number of addresses, all belonging to searchers, account for the majority of swaps: the top four addresses are responsible for 3,649 swap transactions (about 70\%), with Rsync-bot and Titan-bot alone responsible for 2,955 (56\%). Swaps are also concentrated among a few pools: the top 11 DEX pools account for 2,876 transactions (54.4\%).

\subsection{Auction Cycle for Block 21322649}

\begin{figure}[th!]
\centering
        \includegraphics[scale=0.4]{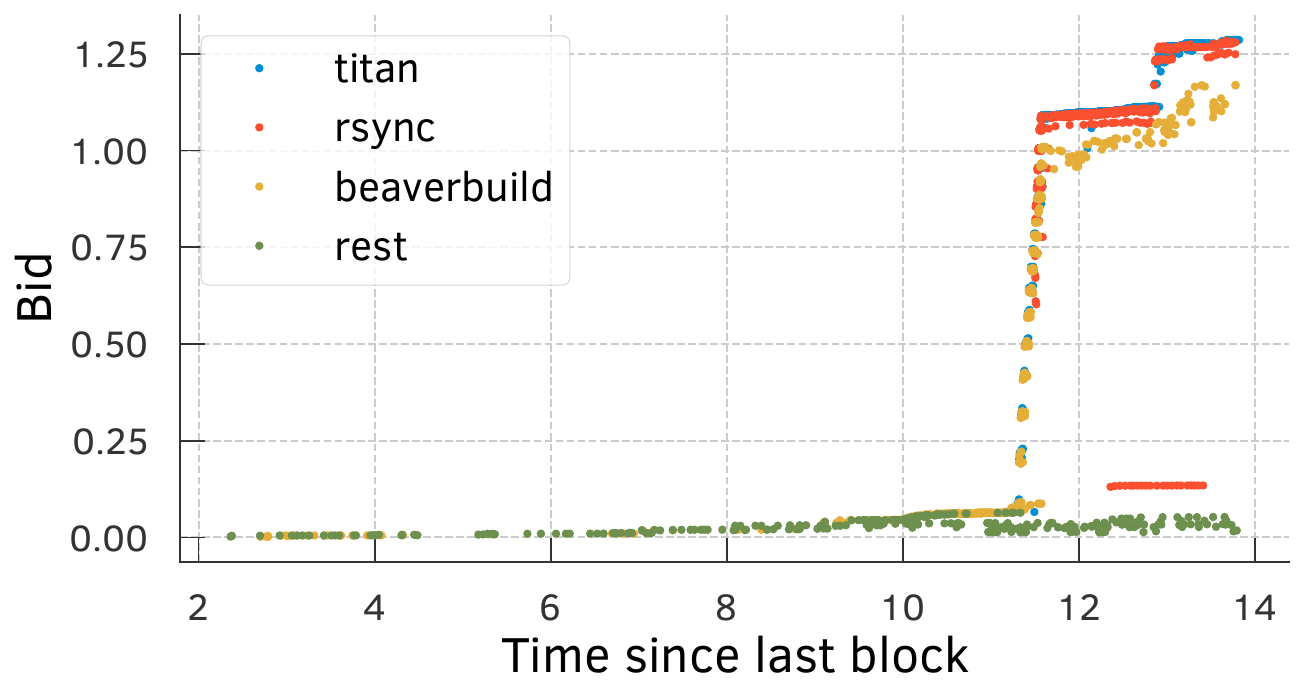}
    \caption{Builders' bids across all relays during the auction cycle for Block 21322649}
    \label{fig:builder-bid-across-relays}
\end{figure}

A novel feature of our dataset is that it provides a highly granular view of auction dynamics, that is, how the content of proposed blocks changes over the course of an auction cycle. To illustrate this feature, we examine in detail the auction cycle that resulted in the inclusion of block 21,322,649 on the Ethereum blockchain. This cycle is noteworthy for two reasons. First, bid data across all relays reveal intense competition between Titan Builder and Rsync Builder, with Rsync ultimately winning the auction (see Figures~\ref{fig:builder-bid-across-relays}). Second, Rsync-bot and Titan-bot compete actively in the same DEX pools and account for 94\%--97\% of the total revenue generated by blocks in which they appear. This auction cycle therefore provides a useful setting in which to study the interaction between the two bots' behavior and their builders' bidding strategies.

For this auction cycle, our dataset contains 669 blocks and 504 unique transactions, of which 299 (59\%) are swaps. Only 21 swaps---8 from users and 13 from searchers---were included in winning block 21,322,649, which was selected approximately 13 seconds after the previous winning block. A further 8 swaps were included in winning block 21,322,650, 3 in winning block 21,322,651, and 1 in winning block 21,322,652. All swaps included in these later blocks are from users. The auction cycle also contains 181 transactions from Titan-bot and 42 from Rsync-bot, all of which are DEX swaps. None of the Titan-bot transactions were included in a winning block, whereas 5 Rsync-bot transactions were included, all in winning block 21,322,649.



\begin{figure}[t]
    \centering
        \includegraphics[scale=0.4]{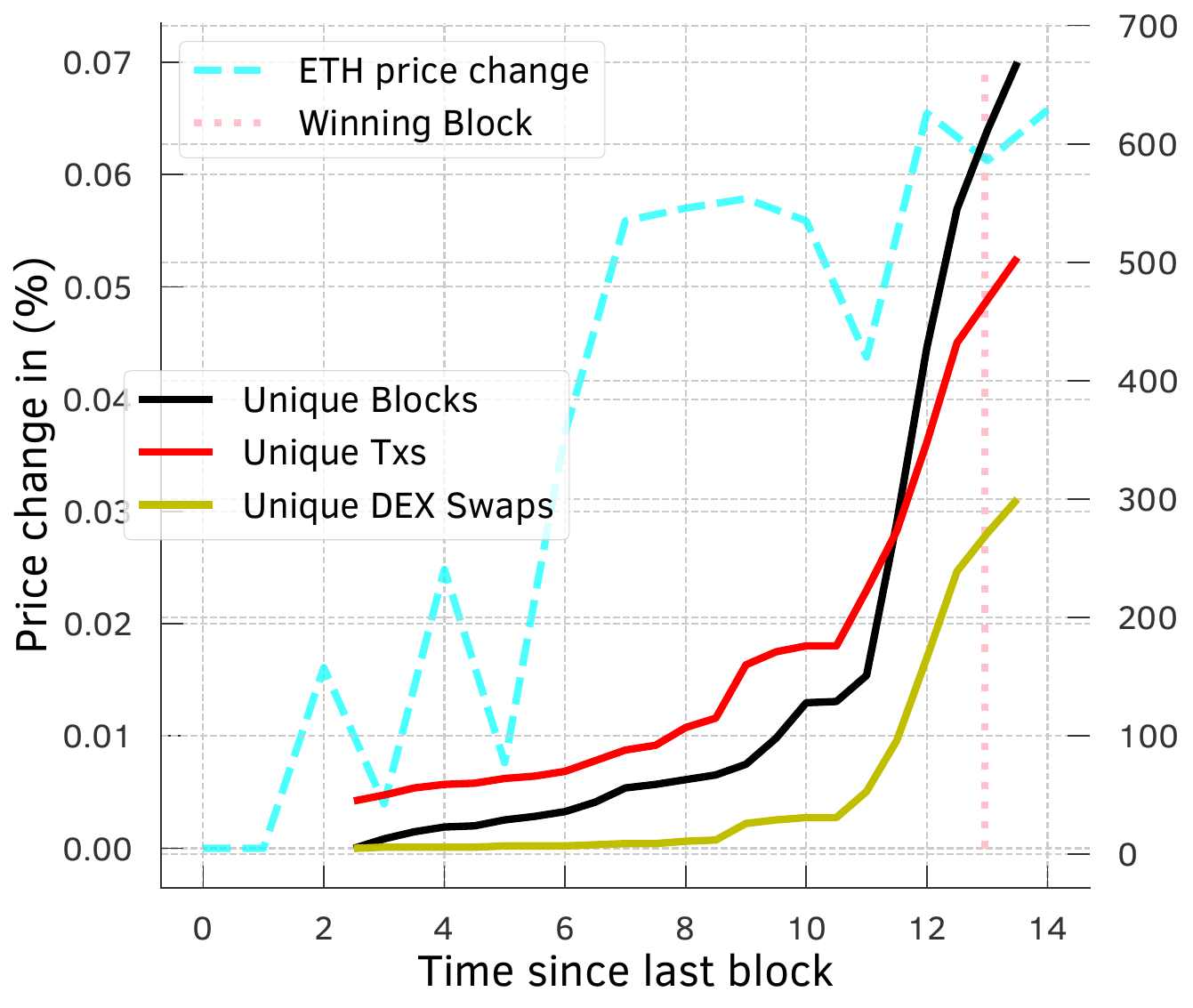}
        \caption{Number of Unique transactions, blocks and swaps vs ETH price change on Binance, during the auction cycle for Block 21322649}
        \label{fig:summary-of-block-21322649}
        \end{figure}

\begin{figure}[h!]
     \centering
\begin{subfigure}[t]{0.45\textwidth}
     \centering
        \includegraphics[scale=0.3]{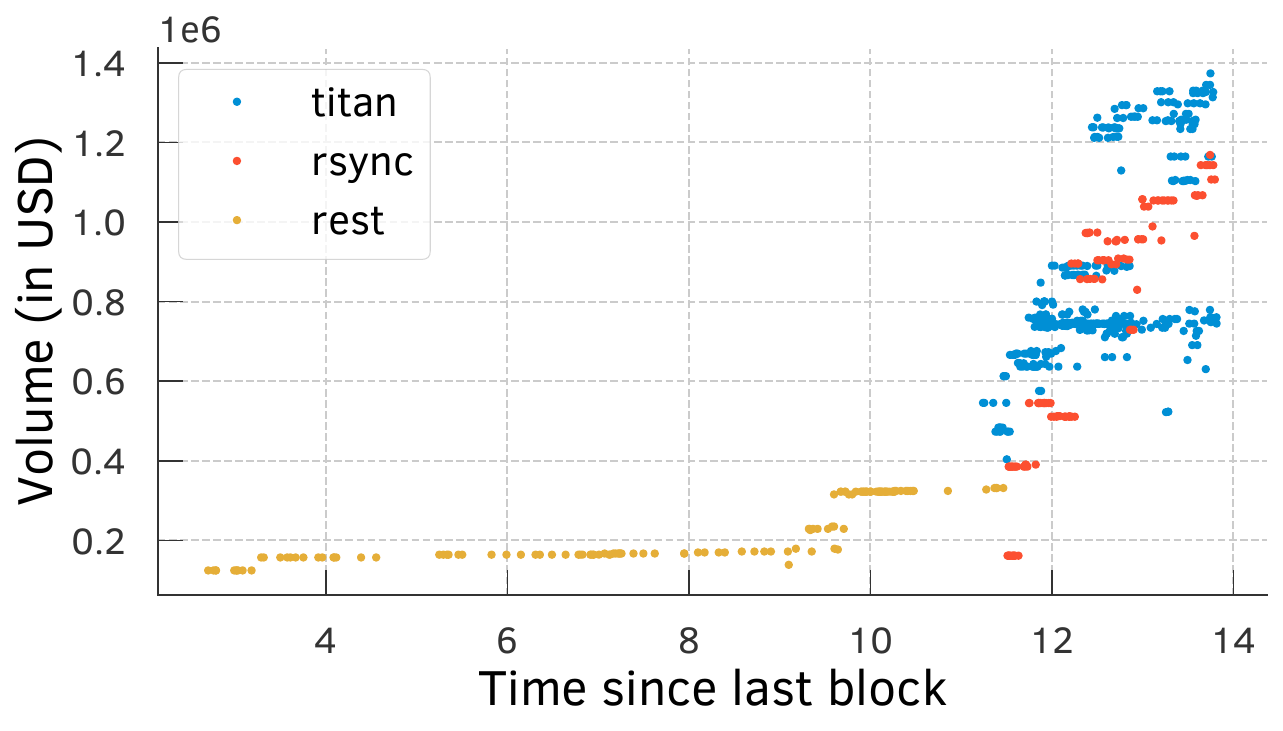}
        \caption{Swaps traded volume per block}
        \label{fig:volume_of_dex_swaps}
\end{subfigure}~~%
\begin{subfigure}[t]{0.45\textwidth}
        \centering
        \includegraphics[scale=0.3]{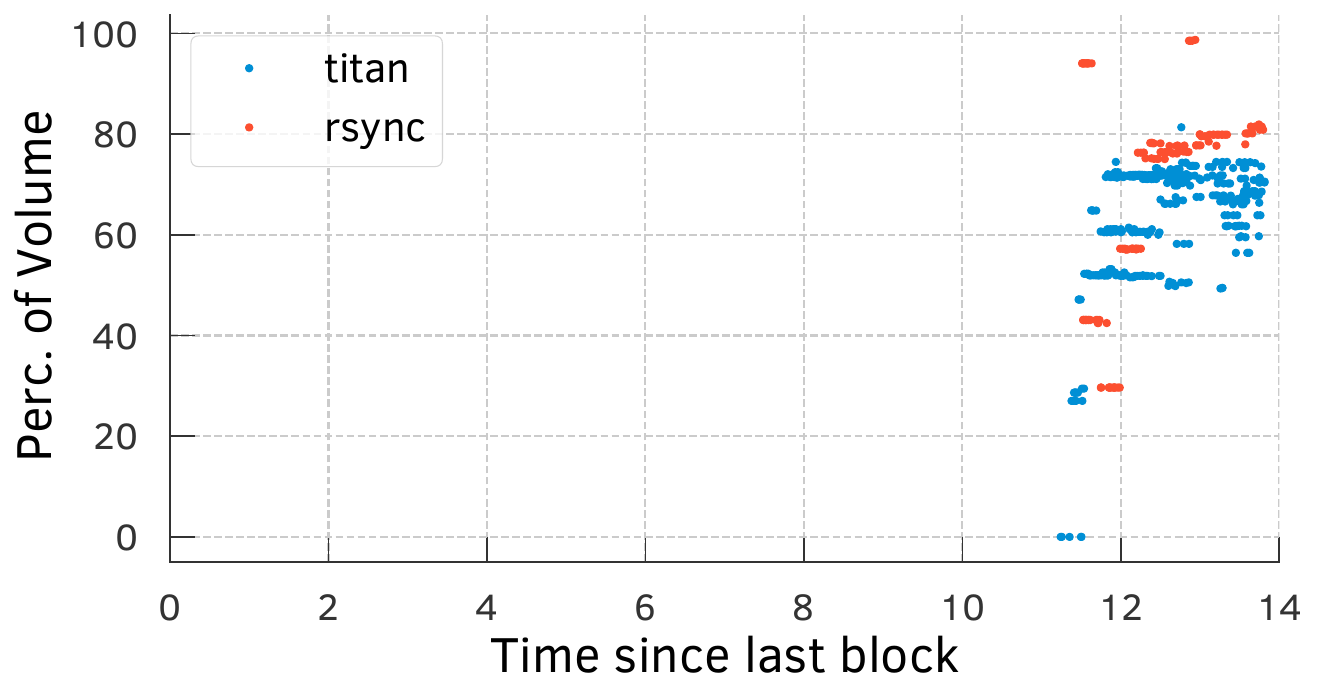}
        \caption{Percentage of the total DEX volume attributable to Rsync-bot and Titan-bot}
        \label{fig:perc-by-mev-bots}
    \end{subfigure}
    
    \begin{subfigure}[t]{0.45\textwidth}
    \centering
        \includegraphics[scale=0.3]{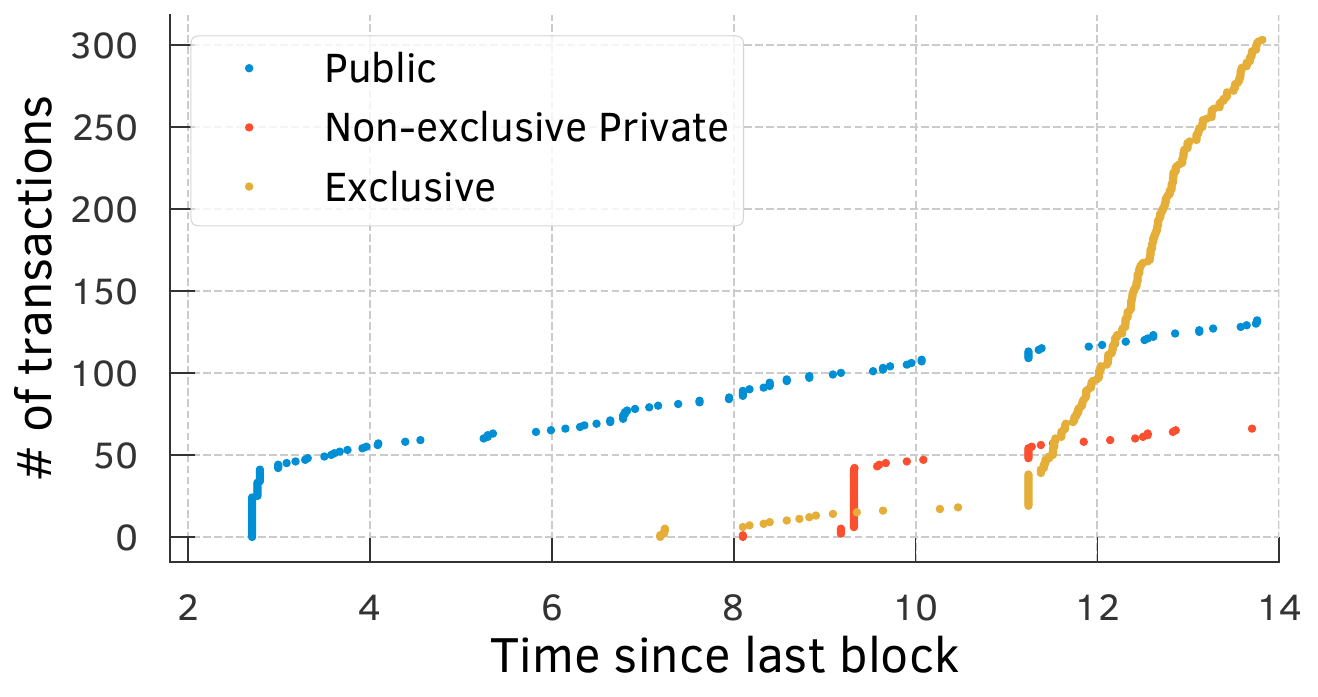}
        \caption{Number of Unique Public, Private, and Exclusive transactions}
        \label{fig:num-exclusive-txs-21322649}
\end{subfigure}~~%
\begin{subfigure}[t]{0.45\textwidth}
        \centering
        \includegraphics[scale=0.3]{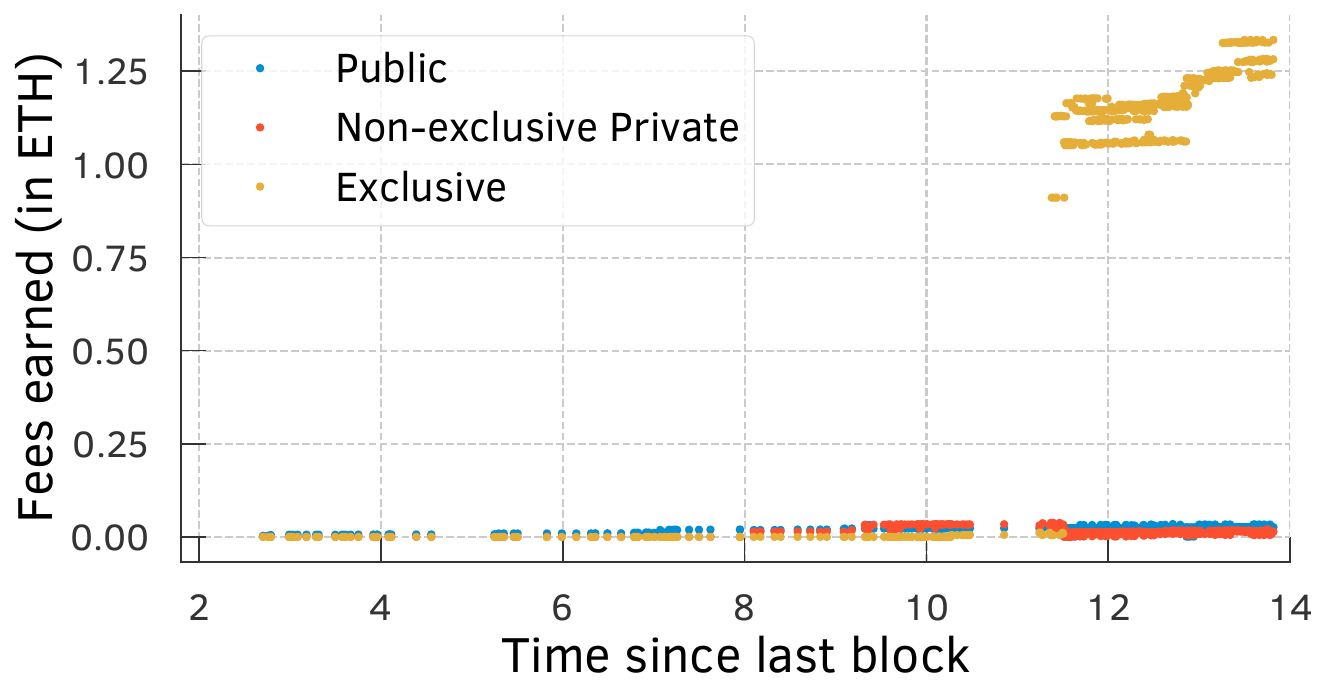}
        \caption{Fees earned from different types of transactions (in ETH)}
        \label{fig:fee-earned-exclusive-txs-per-block}
    \end{subfigure}
    \caption{Evolution of the auction cycle for block 21,322,649.}
    \end{figure}

As the auction cycle progresses, we observe a sharp increase in the number of swaps (see Figure~\ref{fig:summary-of-block-21322649}). This is accompanied by a rise in the volume swapped in each block (see Figure~\ref{fig:volume_of_dex_swaps}), as well as in the share of swap volume attributable to Titan-bot and Rsync-bot (see Figure~\ref{fig:perc-by-mev-bots}). The figure also shows the rate of arrival of different transaction types, distinguishing between public transactions (i.e., those that appear in the public mempool), private transactions (i.e., those that do not appear in the public mempool but appear in blocks built by multiple builders), and exclusive transactions (i.e., those that do not appear in the public mempool and appear in blocks built by a single builder).\footnote{Throughout the paper, for public transactions the relevant timestamp is the first time a transaction appears in a proposed block, not the time at which it appears in the public mempool.} In contrast to public transactions, which arrive at a relatively uniform rate throughout the auction cycle, private transactions---and especially exclusive transactions---tend to arrive toward the end of the cycle (see Figure~\ref{fig:num-exclusive-txs-21322649}). Despite arriving later, exclusive transactions are the most valuable, as reflected in the distribution of fees paid by different transaction types (see Figure~\ref{fig:fee-earned-exclusive-txs-per-block}).

\begin{figure*}[t!]
    \centering
    \begin{subfigure}[t]{0.5\textwidth}
        \centering
        \includegraphics[scale=0.3]{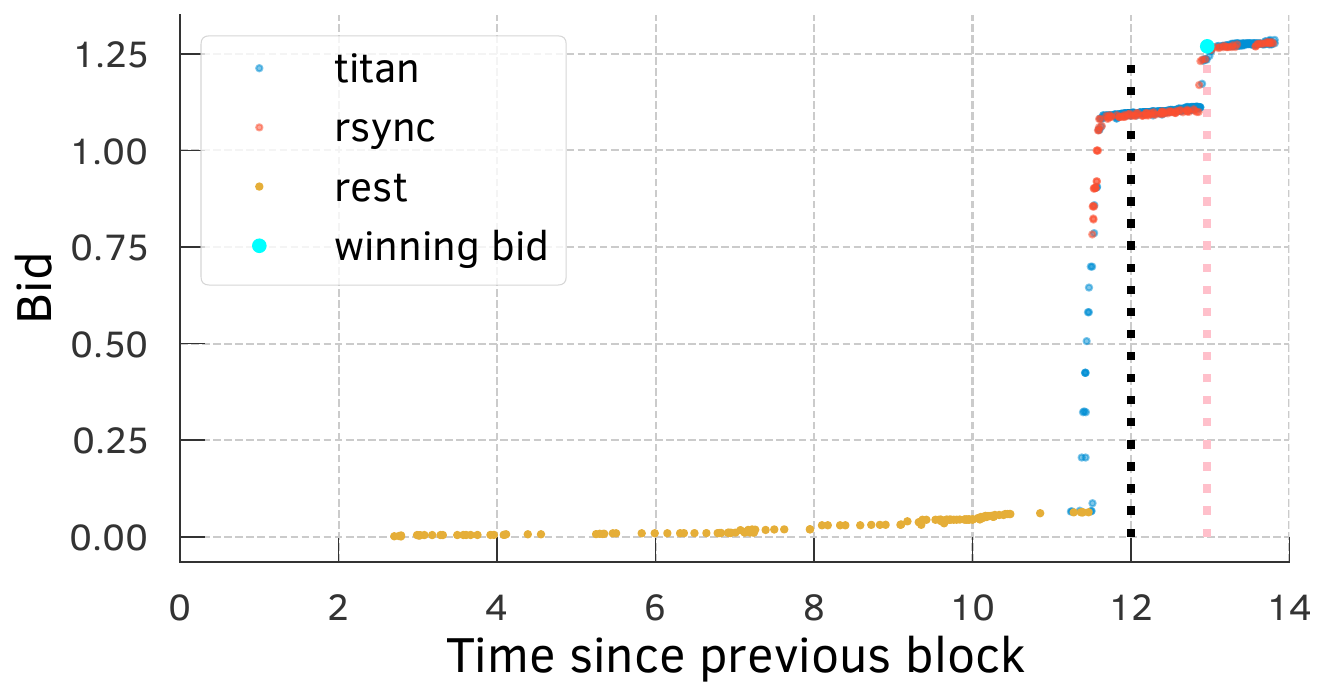}
        \caption{Bids by builders}
        \label{fig:bid-builders}
    \end{subfigure}%
    \begin{subfigure}[t]{0.45\textwidth}
        \centering
        \includegraphics[scale=0.3]{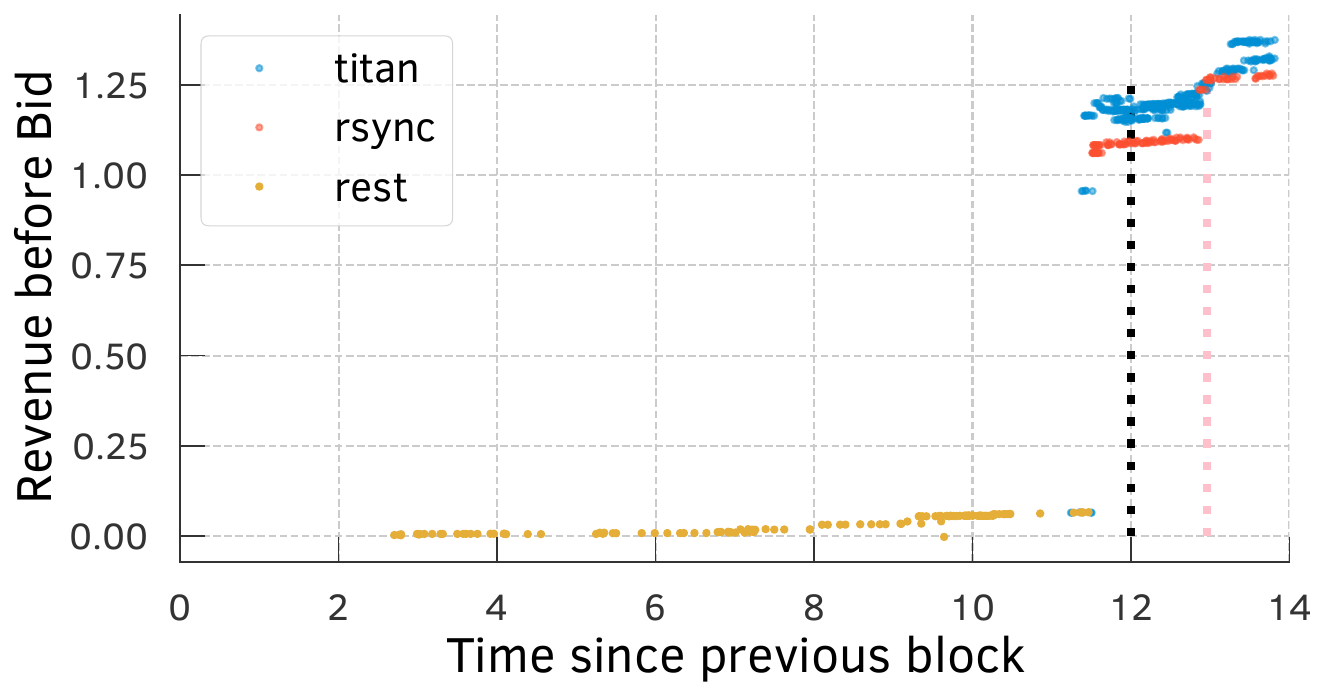}
        \caption{Block revenues by builders}
        \label{fig:revenue-builders}
    \end{subfigure}
\begin{subfigure}[t]{0.45\textwidth}
        \centering
        \includegraphics[scale=0.3]{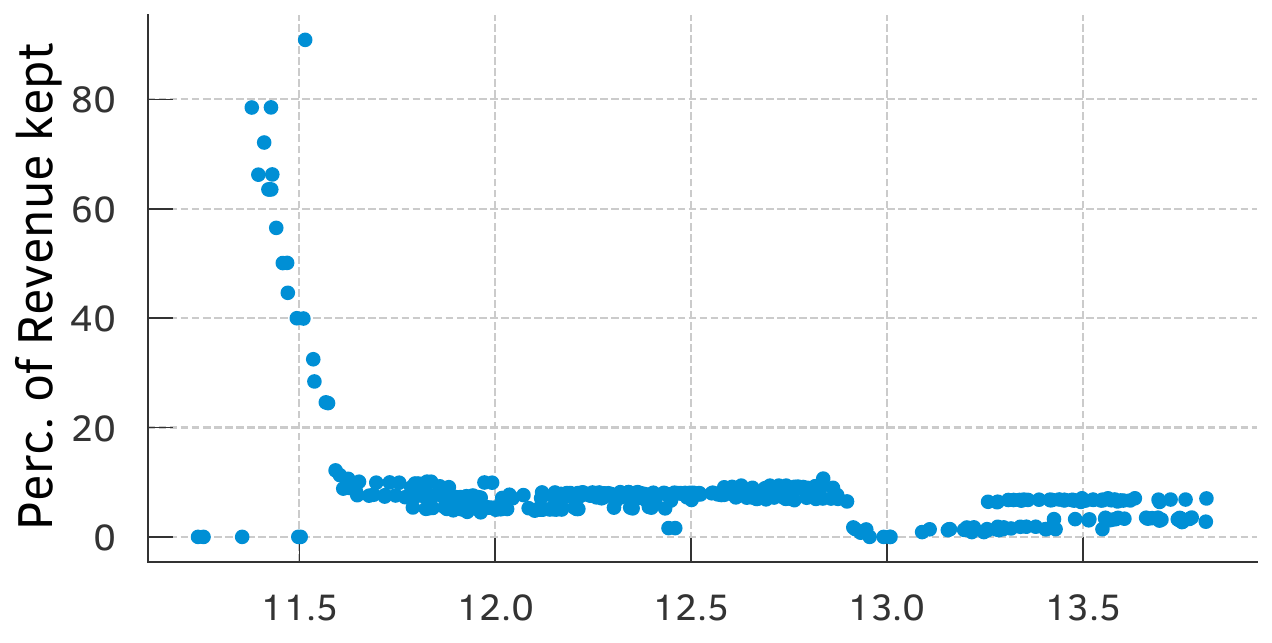}
        \caption{Perc. of block value kept by Titan }
        \label{fig:titan-perc-revenue}
    \end{subfigure}%
    \begin{subfigure}[t]{0.45\textwidth}
        \centering
        \includegraphics[scale=0.3]{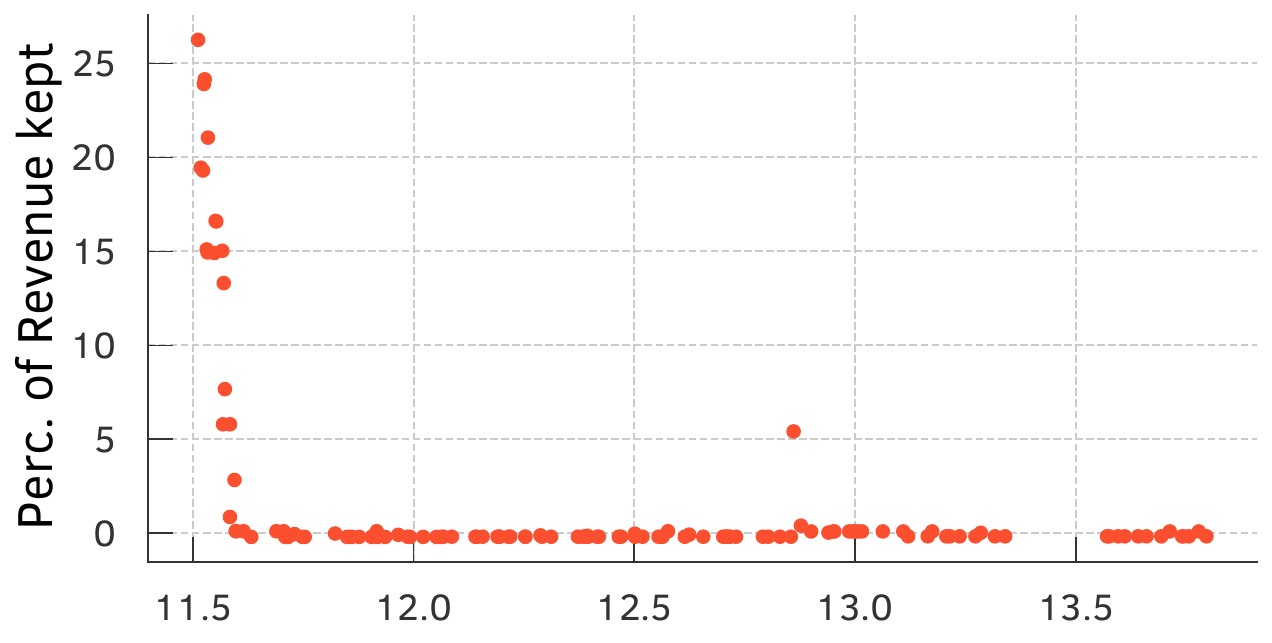}
        \caption{Perc. of block value kept by Rsync}
        \label{fig:rsync-perc-revenue}
    \end{subfigure}%
\caption{Bids and value of blocks for different builders (note the changes in the time scale)}
\label{fig:builder-bids}
\end{figure*}

Of the 669 proposed blocks in our primary dataset, 413 were submitted by Titan Builder and 123 by Rsync Builder. These two builders also account for all high-bid blocks, as shown in Panel (a) of Figure~\ref{fig:builder-bids}. Panel (b) shows that block revenues---measured as the sum of payments to the builder---closely track builders' bids over time. The remaining panels of Figure~\ref{fig:builder-bids} examine the share of block revenue retained by Titan and Rsync, rather than paid to the proposer as a bid. These panels show that, as the auction cycle progresses, both builders reduce the share of revenue they retain, eventually approaching zero. This convergence is faster and more pronounced for Rsync (Panel d) than for Titan (Panel c).

\subsubsection{Competition between Rsync-bot and Titan-bot during the auction cycle.}

As noted earlier, during this auction cycle Rsync-bot and Titan-bot account for 94\% to 97\% of the total revenues of submitted blocks, making their activity a central determinant of the auction outcome. Our data allow us to study competition between the two bots by comparing:
\begin{enumerate}
    \item each bot's trading volume and how it evolves over the course of the auction cycle;
    \item the fees each bot pays and how those fees change during the auction cycle;
    \item the price paid on the DEX, both in terms of the raw execution price (based on token in/out amounts) and the effective price net of the inclusion fee.
\end{enumerate}
Furthermore, because the two bots are integrated with their respective builders, we can interpret the fee attached to each transaction as the bot's expected profit if that transaction is included on-chain.\footnote{We discuss in Section \ref{sec: competition searchers} what happens when we relax this assumption, that is, when the fee attached to each transaction is \textit{at most} equal to the expected profit.} We can then infer a ``risk-adjusted implied price'' for the centralized exchange (CEX) side of the arbitrage path, a component that is typically unobservable. The method therefore provides new insight into the off-chain leg of arbitrage activity.

More precisely, suppose a bot buys a token for ETH on a DEX at price $p_{\text{DEX}}$ (quoted in ETH) and sells the same token for ETH on a centralized exchange (CEX) at price $p_{\text{CEX}}$. The bot's profit from this arbitrage trade, measured in ETH, is
\[
\pi = v \cdot (p_{\text{CEX}} - p_{\text{DEX}}),
\]
where $v$ denotes trade volume, measured in units of the non-ETH token. While we do not observe $p_{\text{CEX}}$ directly, we do observe the inclusion fee $f$, which under the integration assumption above we interpret as the bot's risk-adjusted expected profit. Substituting $f$ for $\pi$ yields the implied risk-adjusted CEX price,
\[
p_{\text{implied}} \equiv p_{\text{DEX}} + \frac{f}{v}.
\]
If the trade goes in the opposite direction---that is, the bot sells the token for ETH on the DEX and buys it on the CEX---then
\[
p_{\text{implied}} \equiv p_{\text{DEX}} - \frac{f}{v}.
\]
All prices are expressed in ETH. In some cases, such as trades involving stablecoins, it is more convenient to express prices in units of the other token; this simply requires inverting the expressions above. Finally, we assign to each observation of $p_{\text{implied}}$ the timestamp of the block containing the swap transaction that represents the DEX leg of the trade.


\begin{figure*}[htb!]
    \centering
    \begin{subfigure}[t]{0.5\textwidth}
        \centering
        \includegraphics[scale=0.3, trim={0 0 4.6cm 0}, clip]{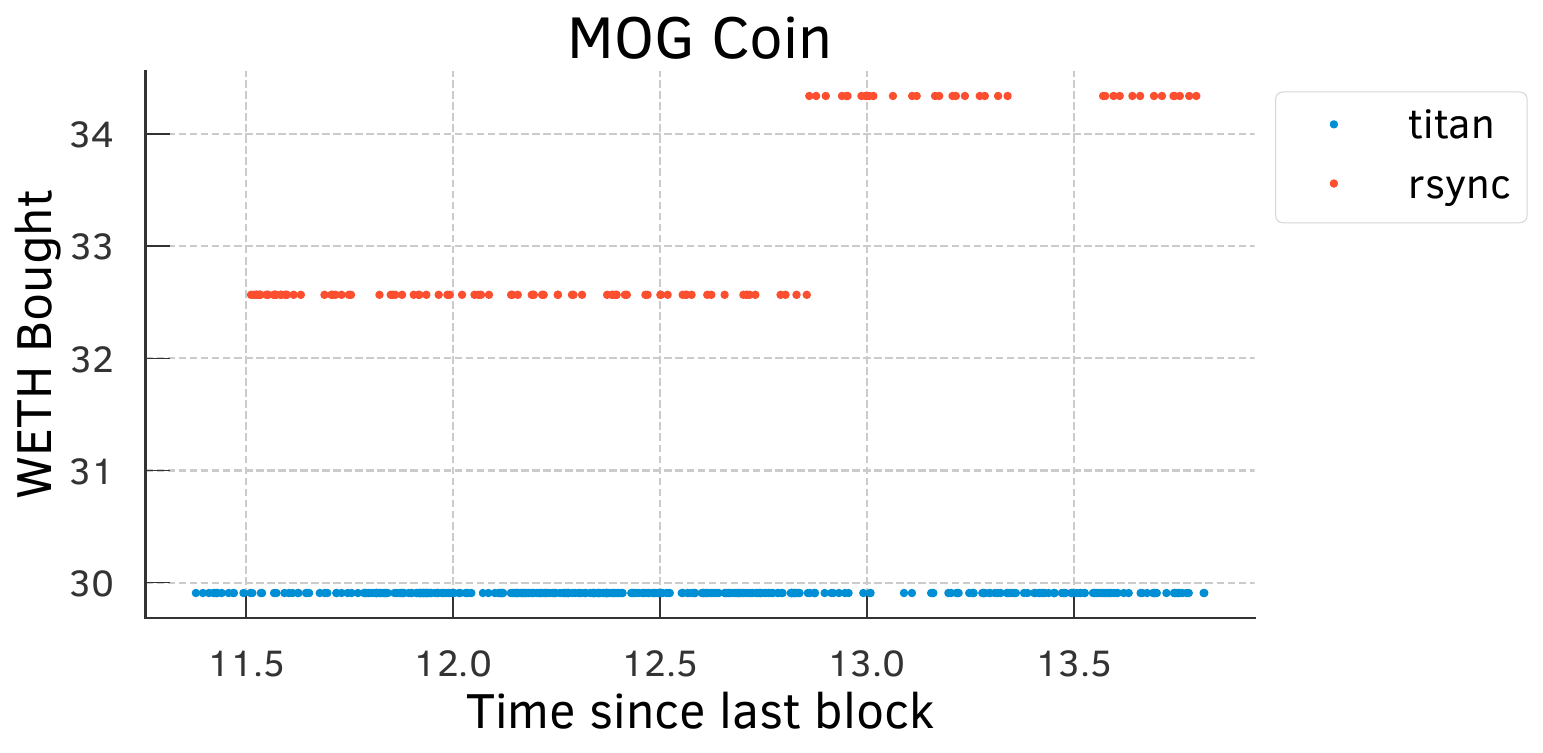}
        \caption{WETH bought on the DEX}
        \label{fig:mog-eth-amount-bought}
    \end{subfigure}~~%
    \begin{subfigure}[t]{0.5\textwidth}
        \centering
        \includegraphics[scale=0.3]{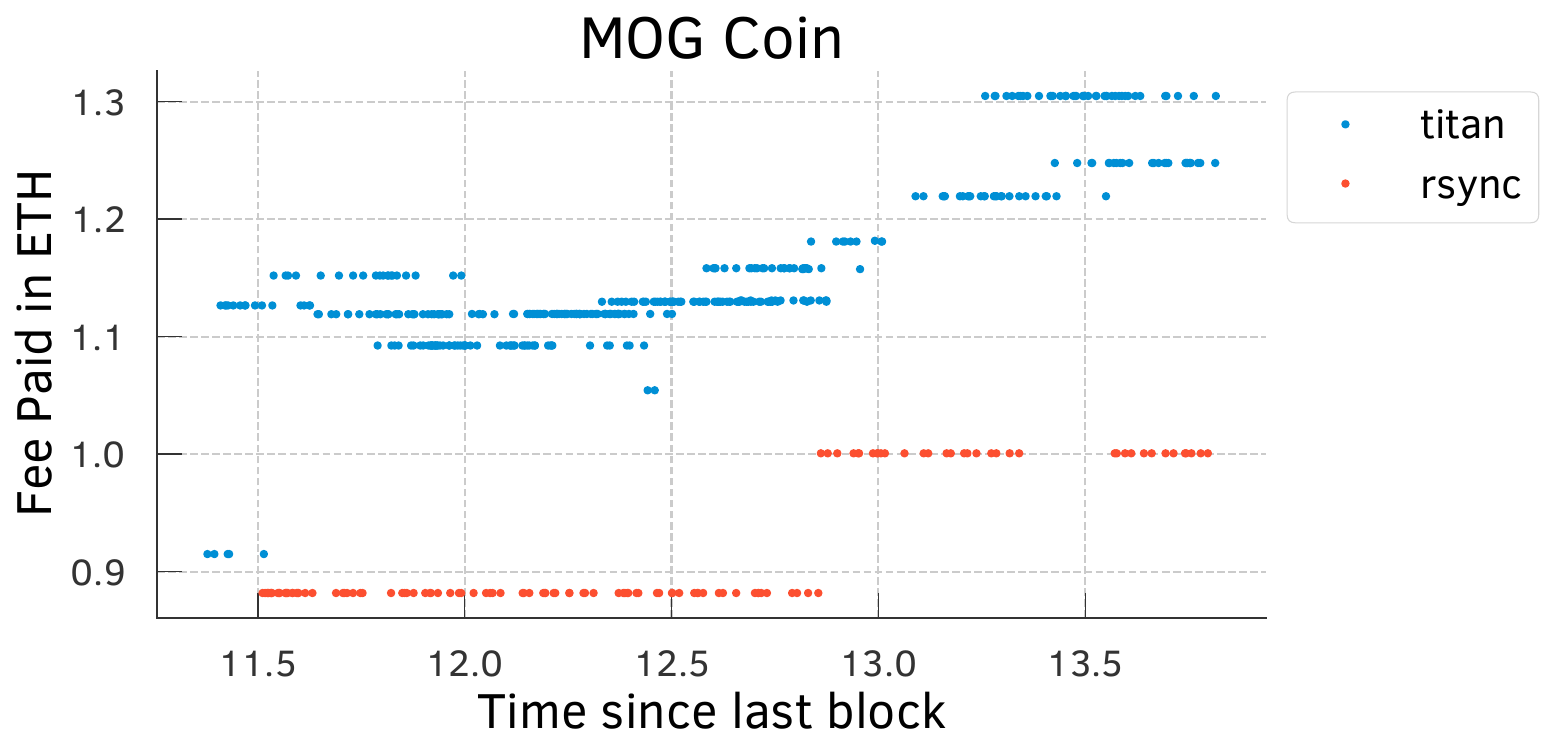}
        \caption{Fees paid for inclusion (in ETH)}
        \label{fig:mog-eth-fee-paid}
    \end{subfigure}
    \begin{subfigure}[t]{0.5\textwidth}
        \centering
        \includegraphics[scale=0.3, trim={0 0 4.6cm 0}, clip]{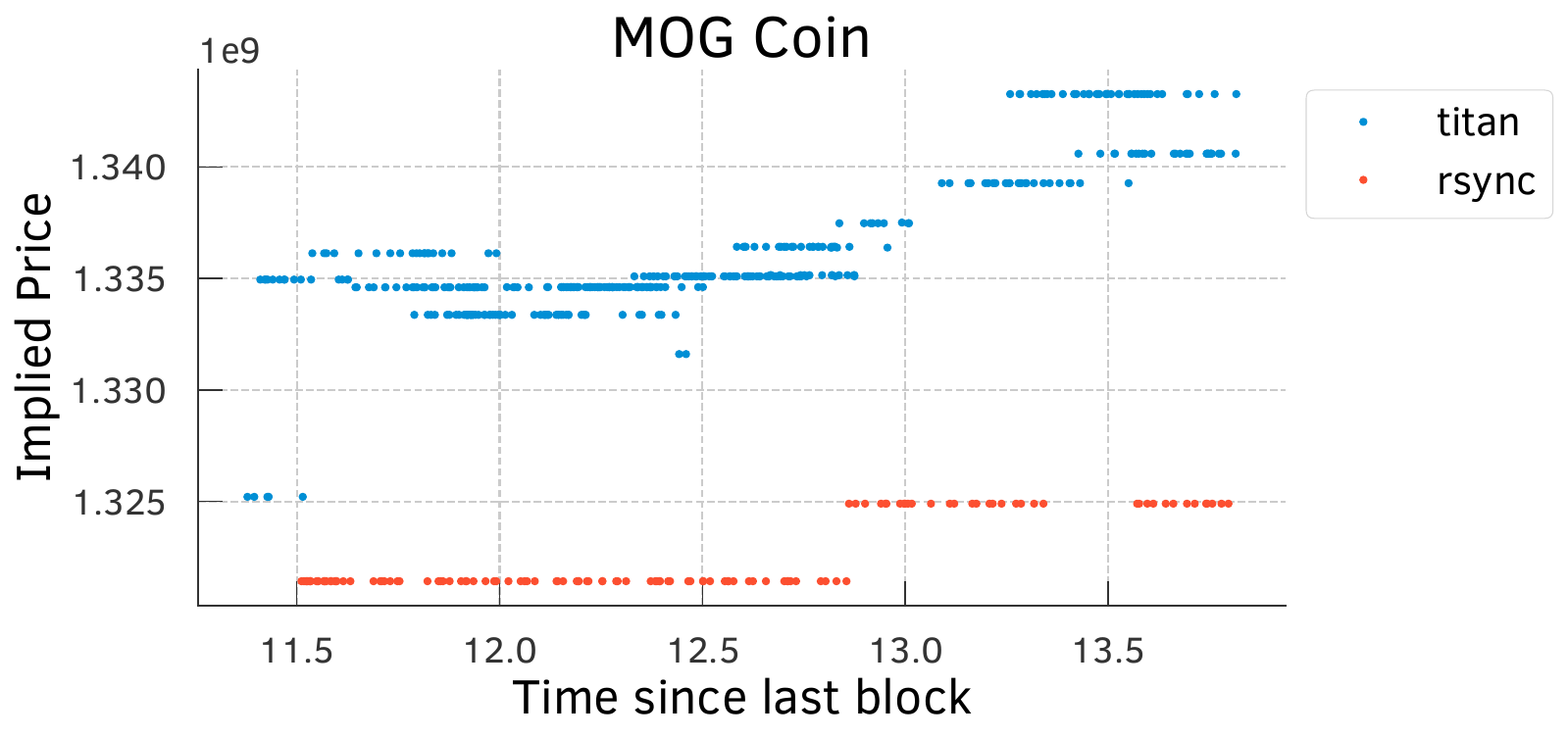}
        \caption{MogCoin / ETH implied DEX price}
        \label{fig:mog-eth-avg-price}
    \end{subfigure}
    \caption{Competition between Rsync-bot and Titan-bot on the WETH/MOG Uniswap v3 pool.}
    \label{fig:mog-eth-swaps}
\end{figure*}

\begin{figure*}[htb!]
    \centering
    \begin{subfigure}[t]{0.5\textwidth}
        \centering
        \includegraphics[scale=0.3, trim={0 0 4.6cm 0}, clip]{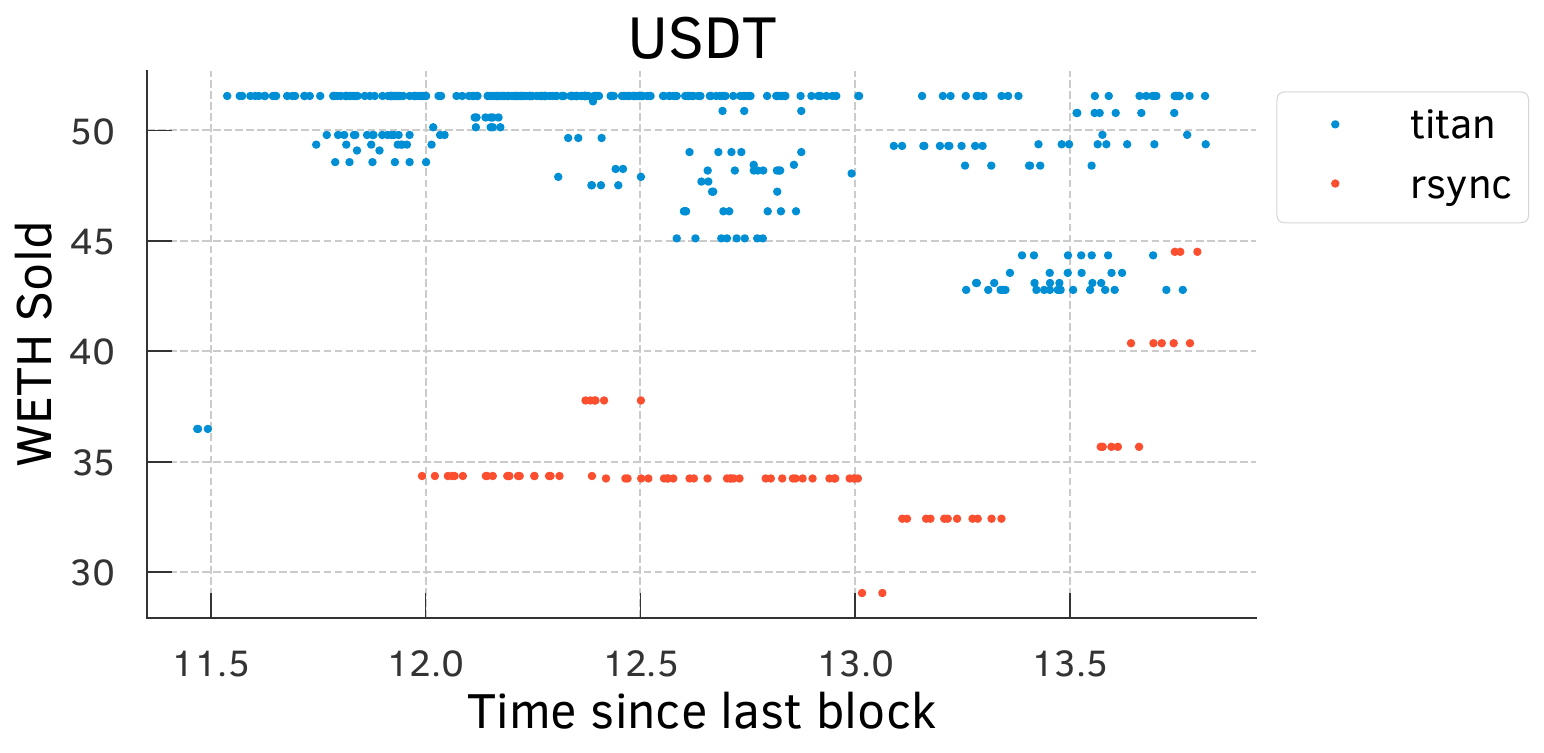}
        \caption{Amount of ETH sold on the DEX}
        \label{fig:usdt-eth-fee-paid}
    \end{subfigure}~~~%
\begin{subfigure}[t]{0.5\textwidth}
        \centering
        \includegraphics[scale=0.3, trim={0 0 4.6cm 0}, clip]{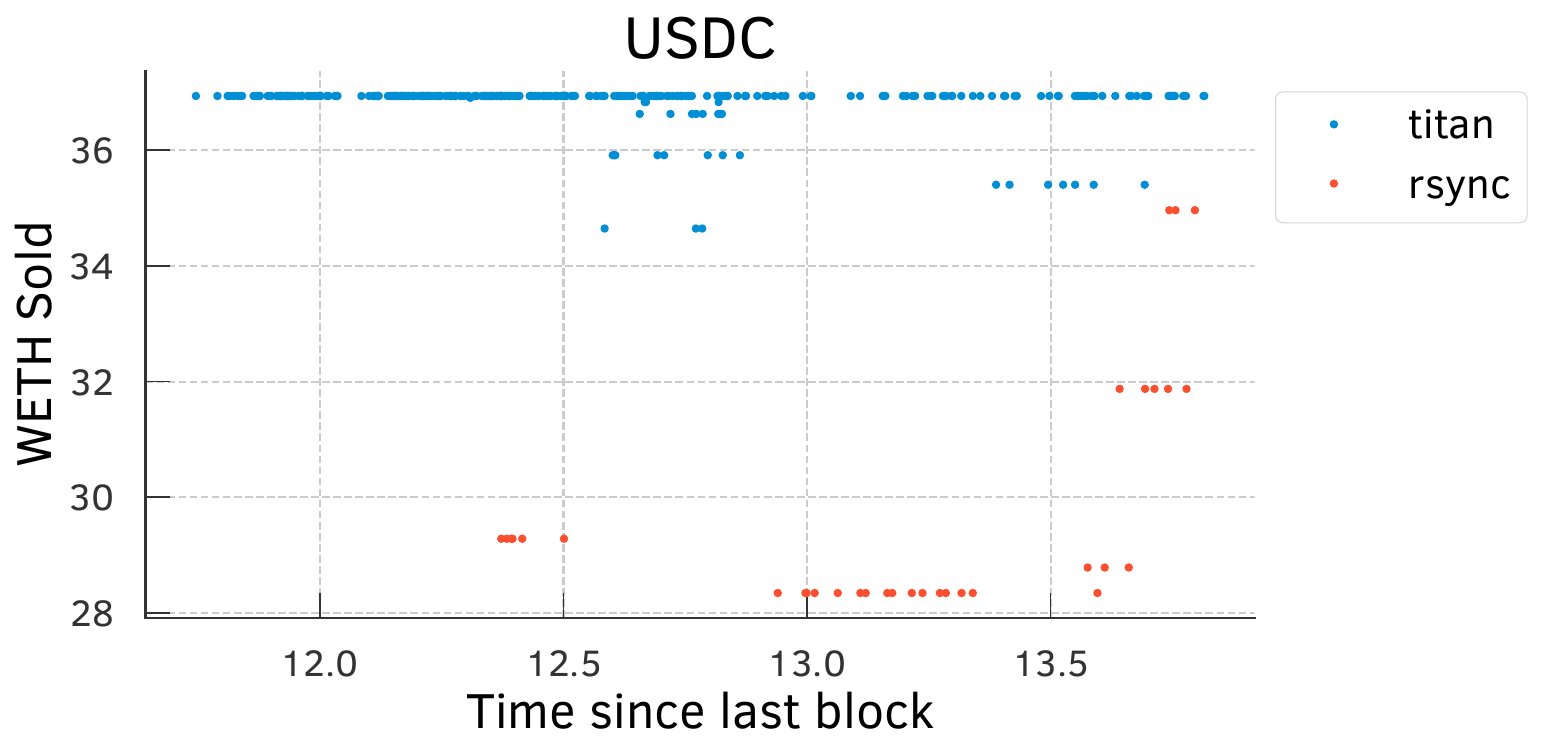}
        \caption{Amount of ETH sold on the DEX}
        \label{fig:usdc-eth-fee-paid}
    \end{subfigure}   
    \begin{subfigure}[t]{0.5\textwidth}
        \centering
        \includegraphics[scale=0.3]{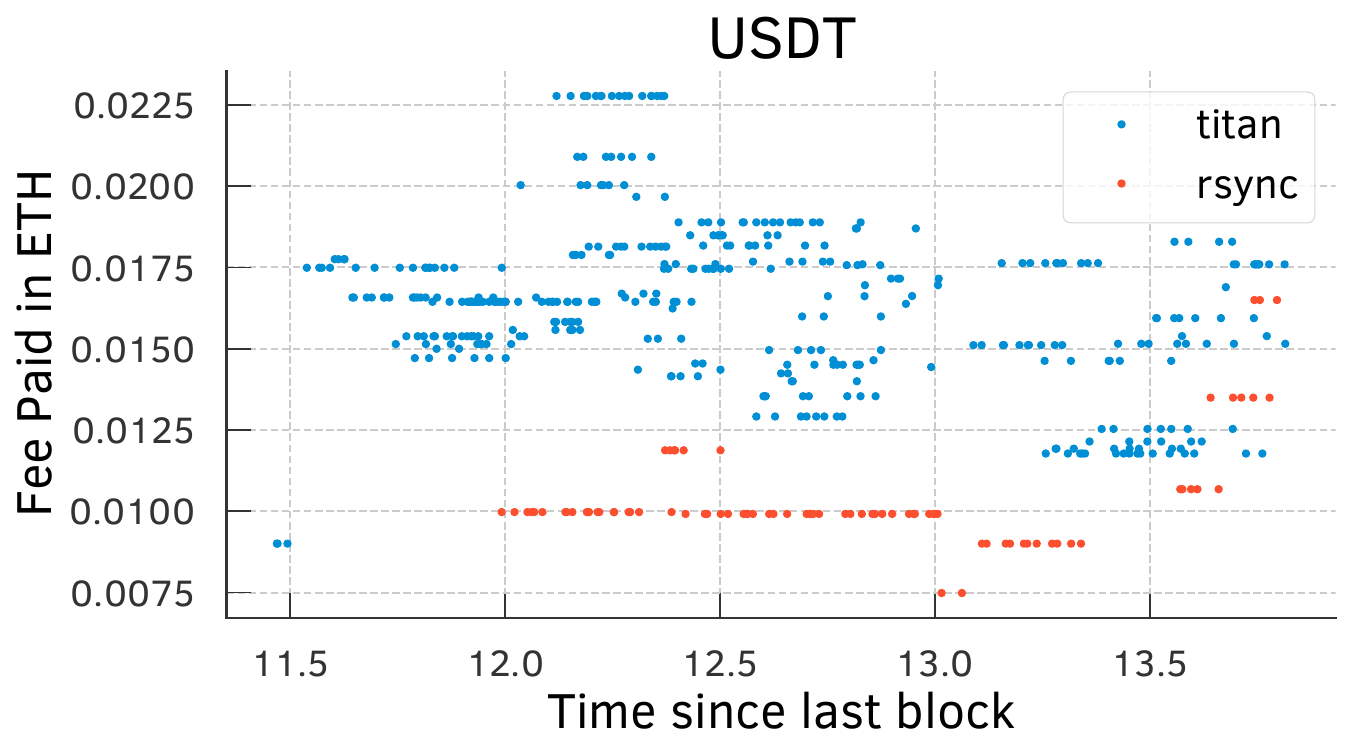}
        \caption{Fees paid for inclusion (in ETH)}
        \label{fig:usdt-eth-avg-price}
    \end{subfigure}~~~%
    \begin{subfigure}[t]{0.5\textwidth}
        \centering
        \includegraphics[scale=0.3]{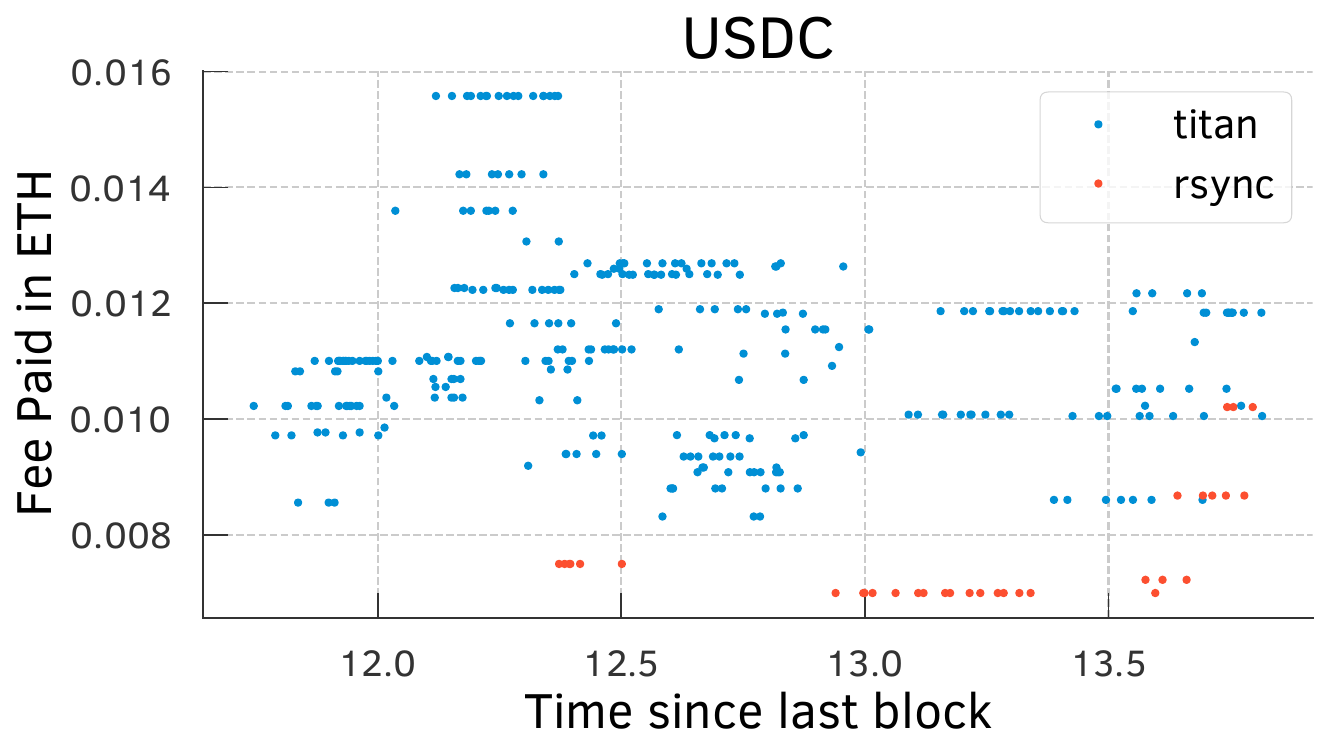}
        \caption{Fees paid for inclusion (in ETH)}
        \label{fig:usdc-eth-avg-price}
    \end{subfigure}  
    \begin{subfigure}[t]{0.5\textwidth}
        \centering
        \includegraphics[scale=0.3, trim={0 0 7.2cm 0}, clip]{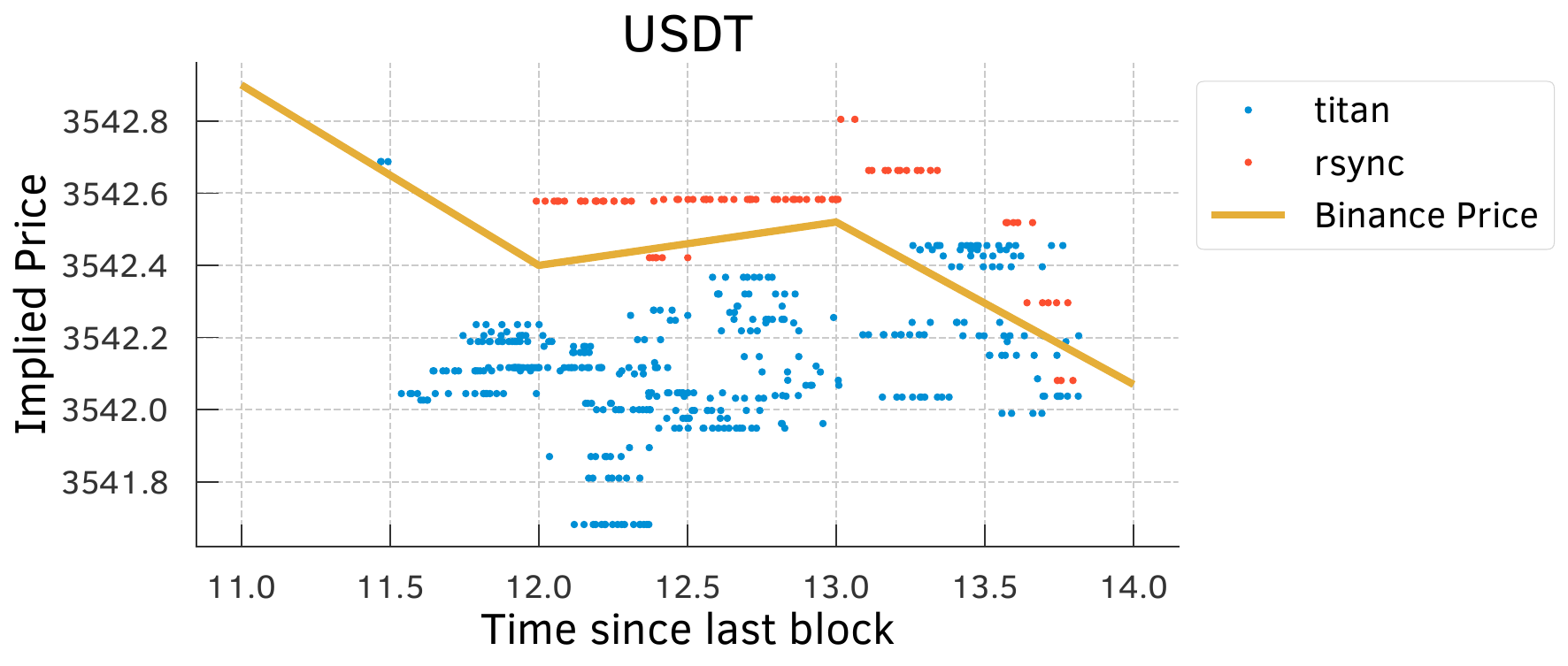}
        \caption{USDT / ETH implied DEX price with Binance price}
        \label{fig:usdt-eth-avg-price}
          \end{subfigure}~~~~%
          \begin{subfigure}[t]{0.5\textwidth}
        \centering
        \includegraphics[scale=0.3]{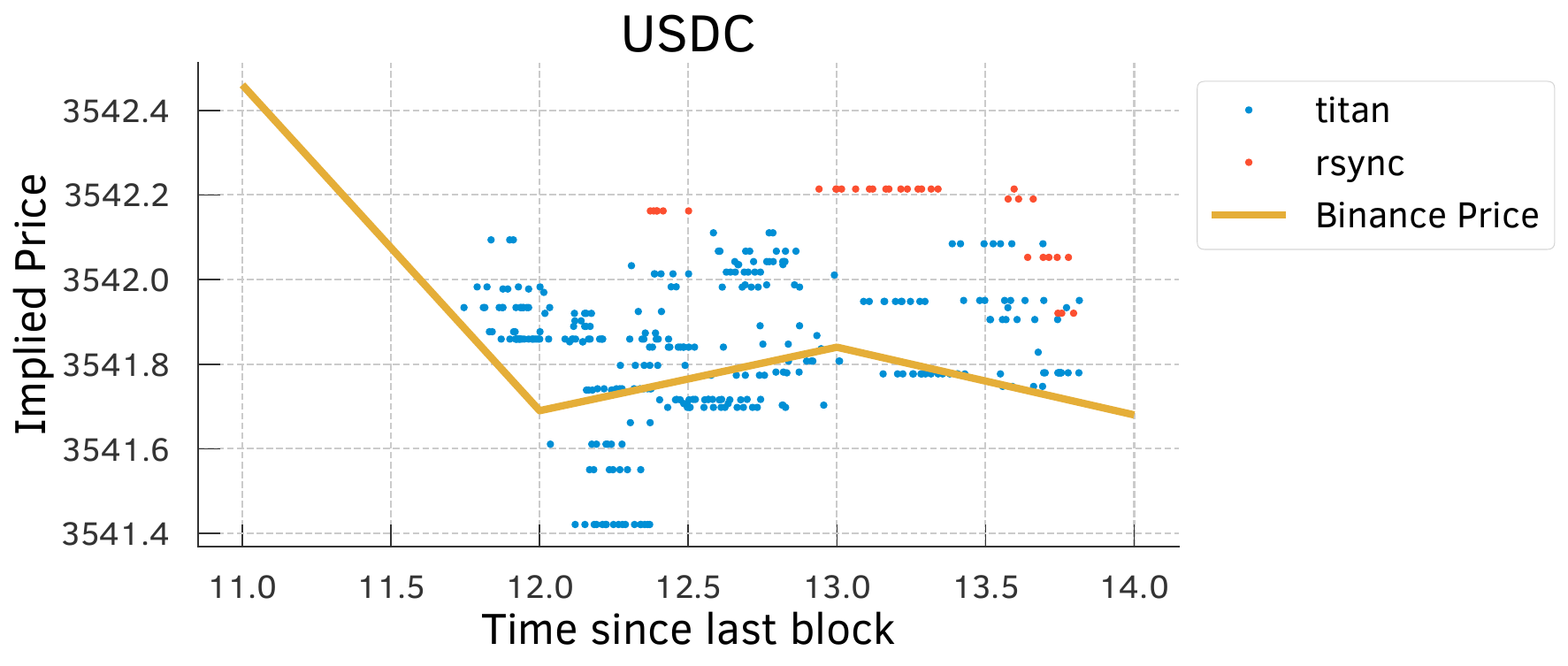}
        \caption{USDC / ETH implied DEX price with Binance price}
        \label{fig:usdc-eth-avg-price}
    \end{subfigure}
    \caption{Competition between Rsync-bot and Titan-bot on the WETH/USDT and WETH/USDC Uniswap v3 pools.}
    \label{fig:usdt-eth-swaps}
\end{figure*}

Figure~\ref{fig:mog-eth-swaps} presents the results of our analysis for MOG/WETH swaps on the Uniswap V3 pool, and Figure~\ref {fig:usdt-eth-swaps} presents the corresponding analysis for swaps on the USDT/WETH and USDC/WETH Uniswap V3 pools. A first observation is that, although traded volumes are similar across the three pools, the fees paid by the bots for swaps on the MOG/WETH pool are orders of magnitude larger than those paid on the other two pools. The MOG/WETH arbitrage opportunity is therefore far more valuable and appears to be the main source of value in this auction cycle. Focusing on MOG/WETH swaps, we find that Titan-bot keeps its trade volume roughly constant throughout the auction cycle while increasing its fees. This pattern is consistent with the risk associated with the arbitrage opportunity declining over the course of the cycle, thereby increasing expected profits. By contrast, Rsync-bot increases both traded volume on the DEX and fees as the auction cycle progresses.

Finally, in both figures we plot the implied centralized exchange (CEX) price, normalized so that higher values are more favorable to the bots. The main result is that, for the MOG/WETH arbitrage, Titan-bot's implied CEX price is 1.9\% better than Rsync-bot's (MOG Coin was not listed on Binance or Coinbase during our study period). For USDT/WETH and USDC/WETH, by contrast, Rsync-bot appears to have an advantage of approximately 0.17\% and 0.02\%, respectively. Note also that the implied CEX prices sometimes lie above and sometimes below Binance prices. This can arise for several reasons. The implied prices are risk-adjusted, so arbitrage risk lowers the implied CEX price. Also, Binance prices do not account for trading fees or price impact. Conversely, bots may trade across multiple centralized venues or use inventory, allowing them to achieve execution prices that are more favorable than those observed on Binance.

To summarize, the MOG/WETH arbitrage is substantially more valuable than the USDT/WETH and USDC/WETH opportunities, so Titan-bot appears to generate more value than its competitor in this auction cycle. Yet Rsync Builder wins the block. The reason is that, as the auction cycle progresses, Rsync raises its bid more aggressively and transfers more of the resulting block value to the proposer.

%
%

\section{Empirical Analysis}\label{sec: entire dataset}

We now turn to the full dataset and analyze all auction cycles in our sample. We first examine cross-block variation in execution quality and then study the behavior of Titan-bot and Rsync-bot.

\subsection{Time of inclusion in a winning block}

\begin{figure}[t]
    \centering
            \includegraphics[scale=0.4]{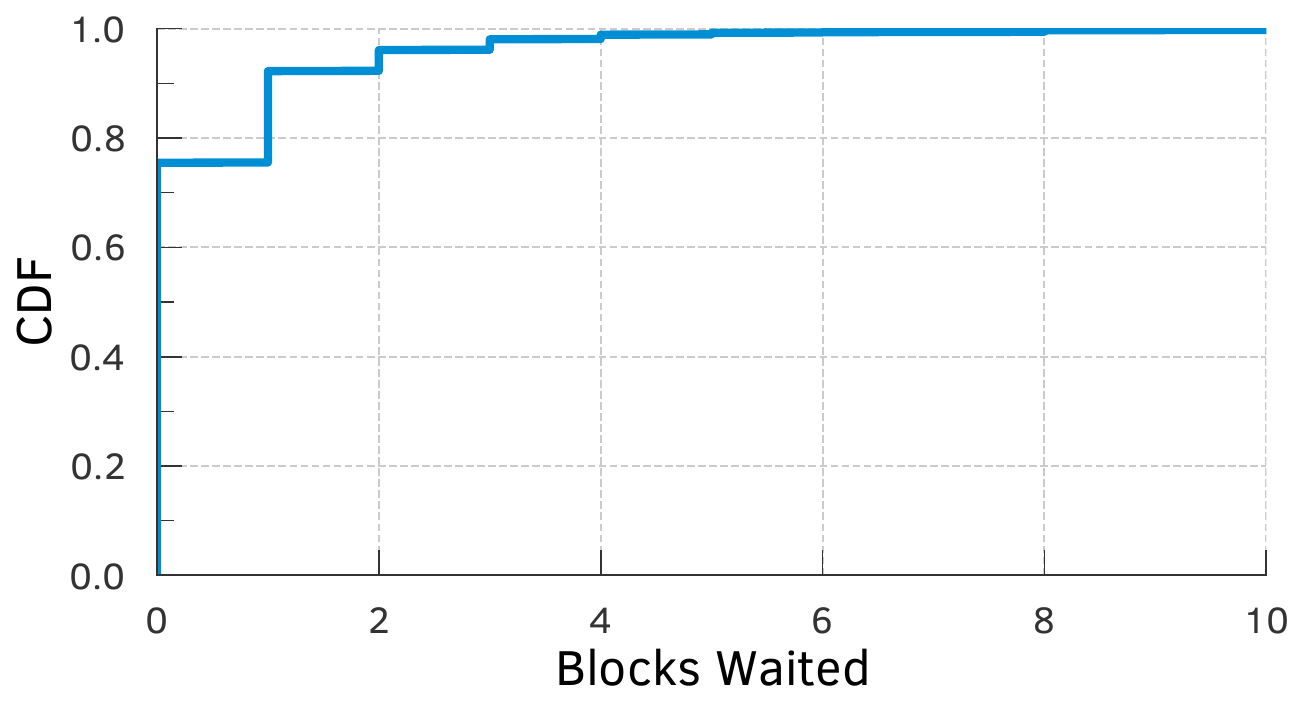}
    \caption{Empirical CDF of the waiting time to on-chain inclusion, measured in blocks, for users’ transactions.
}\label{fig: time of inclusion}
\end{figure}

We identify 1,288 transactions that appear in more than one auction cycle. These transactions appear before the winning block is selected in one cycle, are not included in that winning block, and then reappear in the next cycle. Of these, 871 appear for two cycles, 198 for three cycles, and 108 for four cycles before being included in a winning block (see Figure~\ref{fig: time of inclusion}). 
In addition, 28 transactions appear in multiple auction cycles but are never included in a winning block. Only 17 of the transactions that appear in multiple auction cycles are searcher transactions. Of the remainder, 701 first appear in our dataset as public (i.e., also present in the public mempool), 189 as private (i.e., present in blocks built by multiple builders but not in the public mempool), and 381 as exclusive to a single builder, of which 374 are exclusive to Titan Builder. None of the winning blocks during our study period is full, so these delays in on-chain inclusion cannot be explained by  capacity constraints.

Hence, the vast majority of searcher transactions appear in only a single auction cycle and are either included in the winning block or discarded. By contrast, transactions that persist across multiple auction cycles are overwhelmingly user transactions. Given that our dataset includes 5,968 user transactions, we estimate that approximately 21\% of user transactions are delayed. A key reason is that 30\% of these delayed transactions are, at least initially, exclusive to a single builder that does not win the auction.

In addition, we identify 12 transactions that were initially delayed because they were exclusive to a single builder and later appeared in the public mempool. Nine remained exclusive for two auction cycles, and three for three auction cycles. In four cases, after appearing in the public mempool, the transaction was ultimately included on-chain by the same builder that initially had exclusive access to it (Titan). These transactions appear to originate from ordinary users rather than searchers: three were submitted through the 0x router and four through the 1inch router, two widely used DEX-routing services popular among retail traders. The existence of these transactions is problematic. They were first delayed and then exposed in the public mempool, leaving users vulnerable to predatory trading (one of them appears to have been victim of a sandwich attacked).\footnote{These transactions are:
\begin{itemize}\vspace{-.2cm}
\item \texttt{0x1daea1584d684385fb25209bad3a49c54a1e2c500e27543334c7b3e695a12ffb},
\item \texttt{0x6605c1c36fcd3d696844519d6ff7199fe4ddd34fb4f3e0ed4da71aedfcf1c633},
\item \texttt{0x3dadbab8e85b6e48a2c33f9029ea4ebebdd3def8b9d907faff15ef81527de968},
\item \texttt{0x9c7364eddb94fe659bc8104585ff86a8a234480fe00a9f9aa3eb410e572ebdf2},
\item \texttt{0x9d476dbc524651cbe26bcafab99f6fe40a823a5588fa5cd5002e5b414a8a2f8c},
\item \texttt{0x417440b083bdcb529663c6615c32d89a92beb7863f1a5d6b12bca5b94f4a0f87} (victim of a sandwich attack according to \url{https://www.zeromev.org/block?num=21322643}),
\item \texttt{0xe56caee27e79dee8a4c7ed6f945ccbc2cc733d761e19bd68329dc364c61e73a7},
\item \texttt{0xae5b88bc0bffa04811eda0dfdcd7224599e2757339e18b2d57f4f40d61902a4c},
\item \texttt{0xd41139b67a193045b4d60266c084f6dadd0cb95dccd7b2a06ea68653d6a313ec},
\item \texttt{0x597fc0cb0610d1e3ea2baf0f6f90d0d741a6429bfdd9b375bfd2493b70d41dcd},
\item \texttt{0x44e8105d5c672195877dc1ed970511e0f0a15fc25cbcafe71d51d428d54c215d},
\item \texttt{0xd74718e6d8d58bc33334c0c2f6642d9dec49496329329d8fdcb1e698e81abc94}.
\end{itemize}
}

Unfortunately, our data do not allow us to determine with certainty which wallet originated a given transaction, which private mempool or other service relayed it to builders, or how the transaction subsequently propagated through the public mempool. This information is not recorded in Ethereum blocks. One plausible interpretation is that some private mempool operators initially share transactions exclusively with a single builder and revert to the public mempool when that builder fails to win two or three consecutive auction cycles.

\subsection{Successful execution}

A transaction's execution in a block depends on the transactions that precede it. Because the same transaction can be preceded by different sets of transactions across the blocks in which it appears, it may execute successfully in some proposed blocks but fail in others.

To study this possibility, we consider each transaction--block pair and classify it as a \emph{failure} if its simulated execution yields \texttt{status = 0} or if it produces no transaction log despite producing one when included in another block. A \emph{success} is any transaction--block pair not classified as a failure. Under this definition, failures include both outright execution failures and reverts, such as a swap reverting because its execution price would fall below the user's specified minimum acceptable price. By contrast, a success indicates that the transaction would have executed under the corresponding block composition had that block been added to the chain.

We identify 149 transactions that always fail and 49 that succeed in some blocks but fail in others. Of these 49 transactions, 44 are swaps: 10 generated by searchers and 34 by users. Because our dataset contains 10{,}793 transactions---including 4{,}565 searcher swaps and 717 user swaps---this implies that virtually all searcher swaps always execute successfully, whereas 4.7\% of user swaps fail in at least one of the blocks in which they appear. Cross-block variation in execution success is therefore concentrated almost entirely in user swaps; nearly all other transactions either always succeed or always fail.\footnote{A subtlety arises because transactions that always fail do not generate any log and therefore cannot be classified by type. As a result, identifying a transaction as a swap requires that it execute successfully in at least one block.}

\begin{figure*}[t]
        \centering
        \includegraphics[scale=0.4]{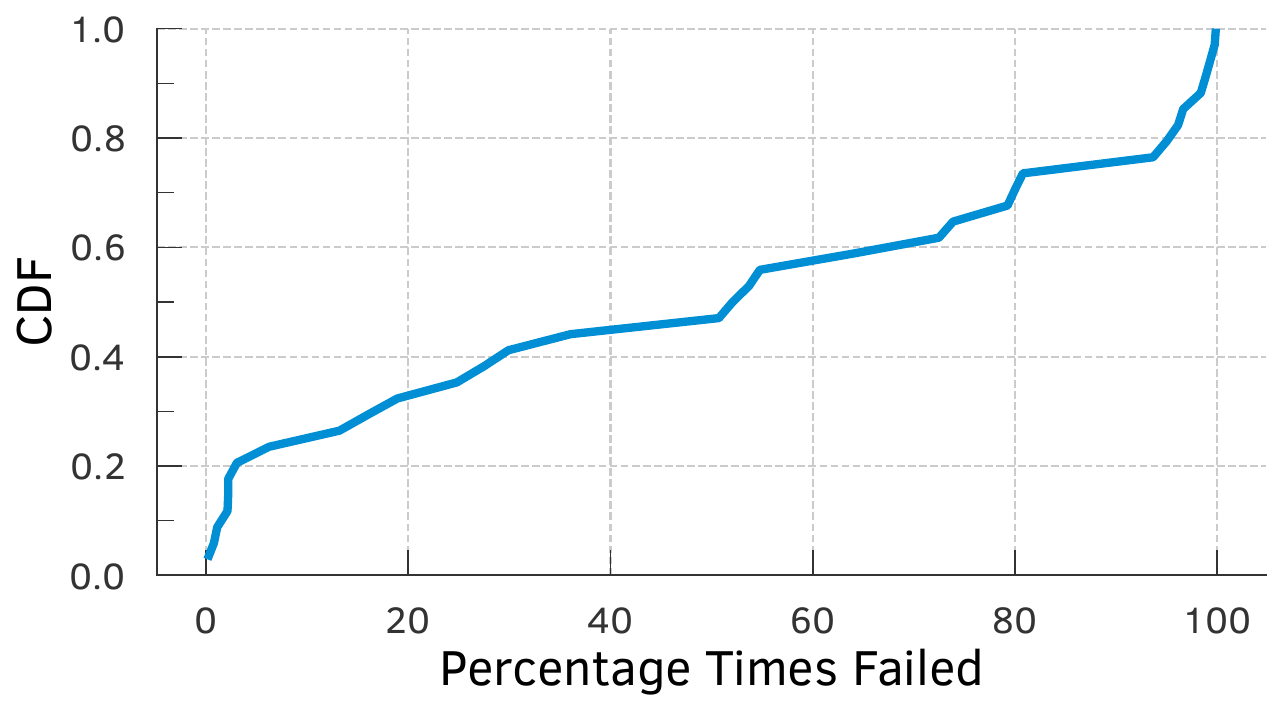}
    \caption{Empirical CDF of the fraction of candidate blocks in which a given user swap fails, conditional on the swap sometimes failing and sometimes succeeding.
}\label{fig: fail-success}
\end{figure*}

We therefore focus on user swaps that sometimes fail and plot the empirical CDF of their failure rates in Figure~\ref{fig: fail-success}. Among these swaps, approximately 50\% fail in at least half of the candidate blocks in which they appear, and roughly 30\% fail in more than 80\% of such blocks. When we consider the entire sample of user swap, therefore, we find that failure risk is highly skewed: 96.7\% of user swaps never fail, while a small fraction fail very frequently---1.7\% fail in at least half of the blocks in which they appear, and 1.1\% fail in more than 80\% of such blocks. Thus, execution risk is concentrated in a small subset of user swaps.

We then perform a regression analysis restricted to swaps that succeed in some blocks and fail in others, yielding 318{,}742 transaction--block observations. For each swap $i$ appearing in block $k$, we define a binary outcome equal to one if transaction $i$ executes successfully in block $k$, and zero otherwise. We regress this outcome on the following explanatory variables:
\begin{itemize}
    \item \textbf{Time since Last Block}: the time elapsed since the previous block, constructed from the block's ``received at'' timestamp.
    \item \textbf{Tx index}: the position of the transaction within the block, where lower values indicate earlier placement and therefore earlier execution. For example, \texttt{Tx index = 1} indicates the first transaction in the block, \texttt{Tx index = 2} the second, and so on.\footnote{We also tested a normalized specification in which transaction position is scaled from 1 to 100, with 1 indicating the first transaction and 100 the last. All results are robust to this alternative.}
    \item A dummy variable equal to one if the block was built by Titan.
    \item A dummy variable equal to one if the block was built by Rsync.
    \item A dummy variable equal to one if the block contains a transaction from Titan-bot.
    \item A dummy variable equal to one if the block contains a transaction from Rsync-bot.
\end{itemize}
We also include \textbf{transaction fixed effects}, that is, a dummy variable for each unique transaction hash. These fixed effects absorb transaction-specific characteristics and capture the average probability that a given transaction executes successfully across the blocks in which it appears. The coefficients on the remaining variables therefore measure how successful execution varies with block timing, builder identity, transaction ordering, and the presence of the two bots. We cluster the standard errors at the block level.\footnote{Our unit of observation is the transaction–block pair, and the main regressors vary at the block level. We therefore cluster standard errors at the block level, allowing for arbitrary dependence across transactions within the same candidate block. Because candidate blocks are nested within auction cycles and the number of auction cycles in our sample is limited, clustering also at the auction-cycle level produces near-singular covariance estimates and substantially reduces the effective number of independent clusters.}

Finally, we construct two subsamples: swaps in the same direction as Titan-bot and Rsync-bot, and swaps in the opposite direction. For each swap by either bot, we record the target pool, trade direction, and auction cycle (note that, in our data, when the bots swaps on the same pool within the same auction cycle they always swap in the same direction). We then identify all other swaps occurring in the same pool and auction cycle, and classify them according to whether they are in the same direction as the bots or in the opposite direction.

Table~\ref{tab:failures} reports three regressions, each estimated on a different sample. The pooled regression in Column 1 yields mixed results: inclusion in a block built by Titan or Rsync is associated with a lower probability of successful execution, whereas the presence of transactions from the corresponding bot is associated with a higher probability of success. The split-sample regressions in Columns 2 and 3 reveal a clearer pattern. Swaps in the \emph{same} direction as the bots are significantly less likely to execute successfully when included in blocks built by Rsync or Titan, especially when those blocks also contain transactions from the bots themselves. By contrast, swaps in the \emph{opposite} direction are more likely to execute successfully when included in blocks that also contain bot transactions. We also find that swaps appearing later in a block (i.e., with a higher transaction index) are slightly more likely to execute successfully. This result is difficult to interpret, however, because transaction ordering within a block is determined by the builder. Finally, transactions are less likely to execute successfully when included in blocks submitted later in the auction cycle.

Using the estimated coefficients, we perform simple back-of-the-envelope calculations to quantify the economic magnitudes. Relative to inclusion in a block built by neither Rsync nor Titan, a swap in the \emph{same} direction as Titan-bot or Rsync-bot is approximately 18\% less likely to execute successfully when included in a block that also contains a transaction from the corresponding bot. This estimate is economically relevant because almost all blocks containing transactions from these bots are built by the corresponding builder. By contrast, a swap in the \emph{opposite} direction is approximately 1\% more likely to execute successfully under the same conditions.

\begin{table}[ht!]
\hspace*{-2cm}
\centering
\sisetup{table-number-alignment = center,
         table-space-text-pre ={(},
         table-space-text-post={\textsuperscript{***}},
         input-open-uncertainty={[},
         input-close-uncertainty={]},
         table-align-text-pre=true,
         table-align-text-post=true}
\begin{threeparttable}
\caption{Probability of Success}
\label{tab:failures}
\setlength{\tabcolsep}{14pt}
\begin{tabular}{r
                S[table-format=-1.5]
                S[table-format=-1.5]
                S[table-format=-1.5]
                S[table-format=-1.5]}
\toprule
& {All Swaps} & {Same Dir. as Bots} & {Opp. Dir. as Bots} \\
\midrule
Time Since Last Block  & {-0.0038}\tnote{***} & {-0.0217}\tnote{***} & {-0.0012}\tnote{***} \\
                      & {(0.0002)}           & {(0.002)}            & {(0.0002)}                      \\
\addlinespace
Tx Index              & {0.0001}\tnote{***}  & {0.0006}\tnote{***}   & {5.74e-06}   \\
                      & {(0.0000)}           & {(0.0001)}        & {(6.26e-06)}                \\
\addlinespace
Is Titan Builder     & {-0.0078}\tnote{***} & {-0.1341}\tnote{***}  &{-0.0144}\tnote{***}   \\
                     & {(0.0009)}           & {(0.0098)}           & {(0.0016)}                \\
\addlinespace
Is Rsync Builder     & {-0.0148}\tnote{***} & {-0.1592}\tnote{***}  & {-0.0021}  \\
                      & {(0.0007)}           & {(0.0117)}          & {(0.0011)}                  \\
\addlinespace
Has Titan-bot tx    & {0.0024}\tnote{***}  & {-0.0423}\tnote{***}  & {0.0286}\tnote{***}   \\
                    & {(0.0006)}           & {(0.0061)}             & {(0.002)}                  \\
\addlinespace
Has Rsync-bot tx     & {0.0075}\tnote{***}  & {-0.0301}\tnote{***}  & {0.0095}\tnote{***}  \\
                     & {(0.0006)}           & {(0.0072)}          & {(0.0013)}                  \\
\midrule
Observations          & {318742}             & {17224}                & {26705}                       \\
R-squared             & {0.744}              & {0.556}               & {0.493}                       \\
Tx fixed effect       & {yes}                & {yes}                 & {yes}                            \\
\bottomrule
\end{tabular}
\smallskip
\footnotesize
\begin{tablenotes}[para,flushleft]
\item[*] $p < 0.1$, \item[**] $p < 0.05$, \item[***] $p < 0.01$
\end{tablenotes}\par
Note: Standard errors clustered at the block level in parentheses. Column titles refer to subsample type.
\end{threeparttable}
\end{table}

\subsection{Execution price}

Block composition may also affect a swap's execution price. To study this possibility, we consider all swaps in our dataset. This yields 272 unique transaction hashes, 112{,}747 transaction--block observations, and 342{,}528 swap--block observations, which constitute our unit of analysis (note that some transactions perform multiple swaps).

For each swap $i$, we compute its average execution price across all blocks, denoted by $p_{\text{avg},i}$. We then calculate, for each swap--block observation, its percentage deviation from this average:
\[
p_{dev,i,k} \equiv \frac{p_{i,k} - p_{\text{avg},i}}{p_{\text{avg},i}} \times 100,
\]
where $p_{i,k}$ is the execution price of swap $i$ in proposed block $k$. We normalize the base currency so that higher values of $p_{dev,i,k}$ always correspond to better execution from the user's perspective. The distribution of $p_{dev,i,k}$ across all $i$ and $k$ has a median of $0$, a minimum of $-10 \%$, and a maximum of $13\%$.

\begin{figure}[t]
    \centering
            \includegraphics[scale=0.4]{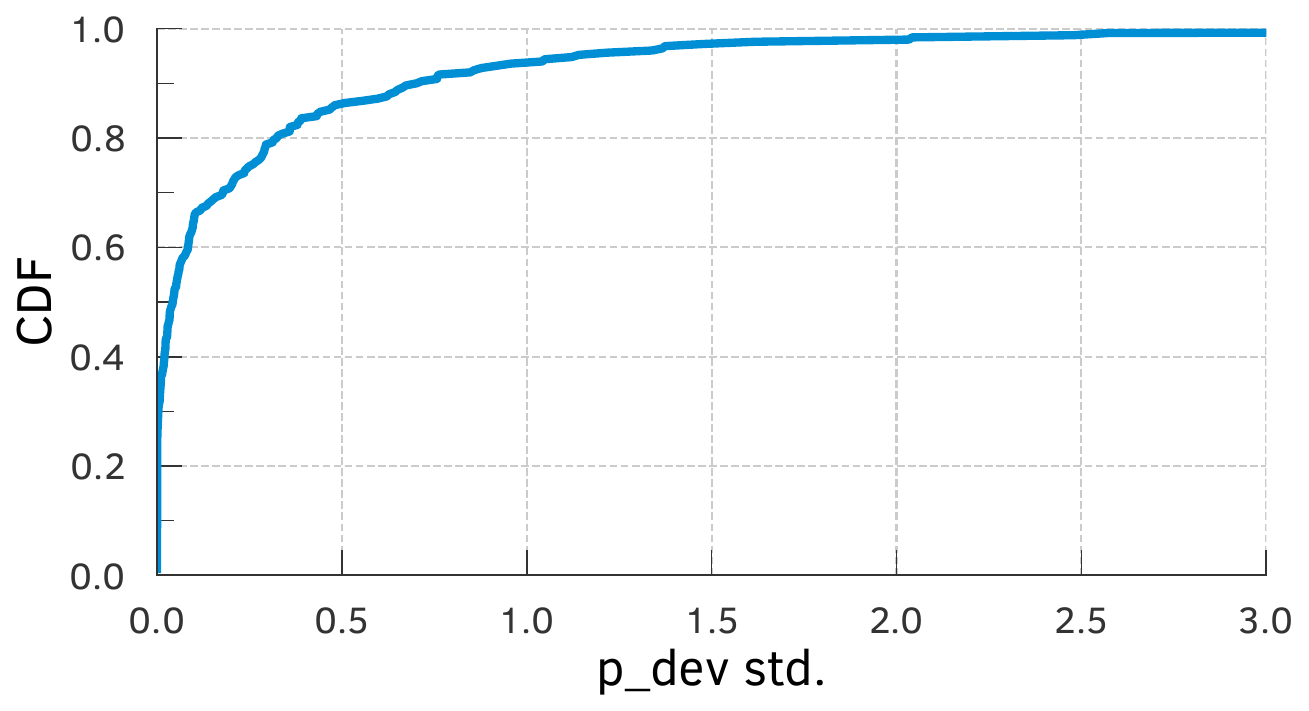}
    \caption{The CDF of the standard deviation of the $p_{dev, i, k}$.}\label{fig: std-price}
\end{figure}

For each swap $i$, we compute the standard deviation of $p_{\text{dev},i,k}$ across the candidate blocks in which it appears. This yields one observation per swap, capturing the dispersion of its execution price across candidate blocks. Figure~\ref{fig: std-price} plots the distribution of these standard deviations, showing that it is highly skewed: the median is 0.043, while the mean is 0.27. Moreover, for 15\% of swaps the standard deviation of $p_{\text{dev},i,k}$ is below $10^{-9}$, whereas for 7\% it exceeds 1. To interpret these magnitudes, suppose that, for each swap $i$, the distribution of $p_{\text{dev},i,k}$ is normal. Under this assumption, for half of the swaps in our sample the probability that the execution price deviates by more than 4.3 basis points from $p_{\text{avg},i}$ is less than one-third, whereas for 7\% of swaps the probability that the execution price deviates by at least 100 basis points from $p_{\text{avg},i}$ exceeds one-third.

We again use regression analysis to decompose this variation. Table~\ref{tab:pnorm} reports estimates from three regressions of $p_{dev,i,k}$ on the same explanatory variables and subsamples used in the success/failure analysis. As in the previous table, the regression on the full sample yields mixed results: inclusion in a block built by Titan is associated with slightly worse execution prices, while the presence of a Titan-bot transaction in the same block is associated with better execution; for Rsync, the pattern is reversed. These effects largely offset one another. By contrast, the regressions restricted to swaps in the same direction as the bots reveal a clearer pattern: execution prices are between 9 and 12 basis points worse when such swaps are included in blocks built by Rsync or Titan, while the presence of the two bots has a negligible additional effect. For swaps in the opposite direction, the results mirror those from the success/failure regressions: inclusion in a block built by Rsync or Titan improves execution prices by between 9 and 27 basis points. This effect remains positive, though smaller, when the block also contains a transaction from Rsync-bot or Titan-bot. In addition, later placement within the block (i.e., a higher transaction index) has a small but consistently positive effect on execution prices. Finally, although the regressions explain only a modest share of the variation in $p_{dev,i,k}$, as reflected in the low $R^2$ values, the estimated effects show that block composition has economically meaningful consequences for the execution prices of a subset of user swaps.

\begin{table}[ht!]
\hspace*{-2cm}
\centering
\sisetup{table-number-alignment = center,
         table-space-text-pre ={(},
         table-space-text-post={\textsuperscript{***}},
         input-open-uncertainty={[},
         input-close-uncertainty={]},
         table-align-text-pre=true,
         table-align-text-post=true}
\begin{threeparttable}
\caption{Price deviations $p_{dev, i,k}$}
\label{tab:pnorm}
\setlength{\tabcolsep}{14pt}
\begin{tabular}{r
                S[table-format=-1.4]
                S[table-format=-1.4]
                S[table-format=-1.4]
                S[table-format=-1.4]}
\toprule
& {All Swaps} & {Same Dir. as Bots} & {Opp. Dir. as Bots} \\
\midrule
Time since Last Block & {-0.0022}\tnote{***} & {-0.0430}\tnote{***} & {0.0163}\tnote{***} \\
                      & {(0.0007)}           & {(0.0031)}            & {(0.0022)}  \\
\addlinespace
Tx Index            & {0.0003}\tnote{***}  & {0.0001}\tnote{***}   & {0.0007}\tnote{***}  \\
              & {(0.0000)}           & {(3.2e-05)}            & {(0.0001)}  \\
\addlinespace
Is titan builder      & {-0.0053}\tnote{**}  & {-0.1206}\tnote{***}  & {0.0923}\tnote{***} \\
                      & {(0.0024)}           & {(0.0101)}            & {(0.0085)} \\
\addlinespace
Is rsync builder      & {0.0198}\tnote{***}  & {-0.0924}\tnote{***}  & {0.2758}\tnote{***}  \\
                      & {(0.0043)}           & {(0.0122)}            & {(0.0114)} \\
\addlinespace
Has titan-bot tx      & {0.0048}\tnote{**}   & {0.0068}              & {-0.0640}\tnote{***}  \\
                      & {(0.0021)}           & {(0.0076)}            & {(0.0096)}  \\
\addlinespace
Has rsync-bot tx      & {-0.0235}\tnote{***} & {-0.0038}              & {-0.0402}\tnote{***} \\
                      & {(0.0022)}           & {(0.0052)}            & {(0.0077)}  \\
\addlinespace
\midrule
Observations          & {342528}             & {15404}                & {25718}           \\
R-squared             & {0.002}              & {0.142}               & {0.073}           \\
Tx fixed effect       & {yes}                & {yes}                 & {yes}    \\
\bottomrule
\end{tabular}
\smallskip
\footnotesize
\begin{tablenotes}[para,flushleft]
\item[*] $p < 0.1$, \item[**] $p < 0.05$, \item[***] $p < 0.01$
\end{tablenotes}\par
Note: Standard errors clustered at the block level in parentheses. Column headers refer to different subsamples.
\end{threeparttable}
\end{table}

\subsection{Competition between Rsync-bot and Titan-bot}\label{sec: competition searchers}

Competition between Rsync-bot and Titan-bot i a defining feature of our data: every auction cycle for which we have data includes at least one transaction from these bots, and together they account for approximately 60\% of builder revenue in the blocks in which they appear. Their activity therefore plays a central role in determining which builder wins the auction.


We identify 57 instances in which Titan-bot and Rsync-bot compete for execution in the same DEX pool during the same auction cycle. In 52 cases, one of the swapped tokens is WETH, which allows us to apply the methodology introduced in the previous section---namely, to derive an implied risk-adjusted CEX price for each bot at different points during the auction cycle. To enable direct comparison with observed market data, we further restrict the sample to tokens traded on Binance. This yields 29 auction cycle--DEX pool combinations for which we can compare the bots' implied prices with contemporaneous Binance prices.

We plot the implied CEX price alongside the Binance second-by-second opening price for all 29 cases in Appendix~\ref{sec: appendix-searchers}. As expected, the implied CEX prices closely track Binance prices in most cases. This relationship is particularly strong for the USDT/WETH and USDC/WETH pairs, which account for the majority of the sample. A weaker but still visible relationship emerges for the UNI/WETH and AAVE/WETH pairs. By contrast, for the DAI/WETH pair, the implied CEX and Binance prices do not exhibit a consistent relationship, likely because Binance is not the primary centralized exchange for DAI trading. This interpretation is supported by the fact that total volume on the DAI/ETH market on Binance during our study period was only 2.5 ETH.\footnote{For comparison, total volume on Binance during the study period for the other markets was 1,487 ETH for USDC/ETH, 16,209 ETH for USDT/ETH, 1,374 ETH for WBTC/ETH, 6.9 ETH for UNI/ETH, and 36 ETH for AAVE/ETH.}

We complement the visual analysis with a regression approach that compares the bots' implied CEX prices with Binance prices. We start by aggregating the bots' implied prices into second-by-second averages: for each second \(s\), we compute the average implied CEX price and trade volume for Rsync-bot and Titan-bot using all trades executed by each bot between seconds \(s-0.5\) and \(s+0.5\). We denote the average implied CEX price in second \(s\) by \(p_{\text{implied},s}\). To ensure a sufficiently homogeneous sample, we restrict the analysis to the USDT/WETH and USDC/WETH trading pairs.\footnote{As discussed earlier, Binance prices for some token pairs appear unreliable. More importantly, the number of observations, values of \(p_{\text{implied},s}\), and trade volumes vary considerably across tokens. We have 104 observations for WETH/USDC and WETH/USDT combined, but only 24 for WETH/DAI, 18 for WETH/UNI, and 5 or fewer for WBTC/WETH and AAVE/WETH. Moreover, the average value of \(p_{\text{implied},s}\) is around 3{,}500 for stablecoin pairs (USDC, USDT, DAI), but drops to 261 for UNI/WETH, 15 for AAVE/WETH, and 0.037 for WBTC/WETH. Trade volumes also vary: average volume is 320.3 ETH for WETH/WBTC, compared to 43.1 ETH for other assets. As a result, separate regressions on the less common pairs are underpowered, while pooled regressions suffer from multicollinearity due to token-specific heterogeneity.} This yields 89 observations in total: 43 for Rsync-bot and 46 for Titan-bot. We regress \(p_{\text{implied},s}\) on the contemporaneous Binance opening price, its one- and two-second lags, and its one-second lead. The results, reported in Table~\ref{tab-implied-price}, show that only the contemporaneous and one-second-lagged Binance prices are significantly correlated with the bots' implied CEX prices, with the contemporaneous price having the greatest explanatory power. We therefore use the contemporaneous Binance price as the benchmark for evaluating \(p_{\text{implied},s}\).

\begin{table}[t!]
    \centering
\sisetup{
    table-number-alignment = center,
    table-space-text-pre ={(},
    table-space-text-post={\textsuperscript{***}},
    input-open-uncertainty={[},
    input-close-uncertainty={]},
    table-align-text-pre=false,
    table-align-text-post=false
}
\begin{threeparttable}
    \caption{Implied Price on Different Binance Prices}
    \label{tab-implied-price}
    \setlength{\tabcolsep}{10pt}
    \begin{tabular}{l
                    S[table-format=-2.3]
                    S[table-format=2.3]}
        \toprule
        & \multicolumn{1}{c}{Coefficient} & \multicolumn{1}{c}{Standard Error} \\
        \midrule
        Binance Price (0 sec.)   & 0.7945\tnote{***} & 0.108 \\
        Binance Price (+1 sec.)   & -0.1491            & 0.129 \\
        Binance Price (-1 sec.)   & 0.2513\tnote{**}  & 0.108 \\
        Binance Price (-2 sec.)   & 0.0972            & 0.063 \\
        Constant                  & 21.5424   & 23.070 \\
        \midrule
        Observations              & \multicolumn{2}{c}{89} \\
        R-squared                 & \multicolumn{2}{c}{0.998} \\
        \bottomrule
    \end{tabular}
    
    \smallskip
    \footnotesize
    \begin{tablenotes}[para, flushleft]
        \item[*] $p < 0.1$, \item[**] $p < 0.05$, \item[***] $p < 0.01$ \\
    \end{tablenotes}
\end{threeparttable}
\end{table}

We then compare each observation of \(p_{\text{implied},s}\) with the corresponding contemporaneous Binance price, \(p_{\text{binance},s}\). To interpret execution quality from the bot's perspective, we distinguish between two cases. If a bot \emph{sells} ETH on-chain, it must be \emph{buying} ETH on the CEX, so a lower CEX price is better. In this case, we define
\[
p_{\text{diff},s} \equiv p_{\text{binance},s} - p_{\text{implied},s}.
\]
Conversely, if the bot \emph{buys} ETH on-chain, we define
\[
p_{\text{diff},s} \equiv p_{\text{implied},s} - p_{\text{binance},s}.
\]
In both cases, a higher \(p_{\text{diff},s}\) indicates better execution for the bot. We then compute
\[
p_{\text{improvement},s} \equiv \frac{p_{\text{diff},s}}{p_{\text{binance},s}} \times 100
\]
as the percentage price improvement relative to the contemporaneous Binance price. The minimum, maximum, and mean values of \(p_{\text{improvement},s}\) are \(-1.4\), \(4.6\), and \(1.3\), respectively.

Table~\ref{tab:price_improvement_centered} reports estimates from a regression of 
\(p_{\text{Improvement},s}\) on a constant, a dummy equal to one for observations 
from Titan-bot, time since the last block, and swap volume measured in WETH. 
Time and volume are centered at their sample means, and standard errors are 
clustered at the auction-cycle level.\footnote{Clustering at the block level is 
not feasible in this specification because observations are constructed by 
aggregating across candidate blocks. We therefore cluster standard errors at the 
auction-cycle level, allowing for arbitrary correlation in residuals within the 
same auction cycle. Given the limited number of auction cycles in our sample, we 
also compute wild-cluster-bootstrap p-values as a robustness check; these yield 
similar results.} The constant therefore captures the average percentage price 
improvement for Rsync-bot at the sample-average time and volume, while the 
coefficient on the Titan-bot dummy captures the difference in execution quality 
between Titan-bot and Rsync-bot. The estimates indicate that both bots obtain 
better execution than the contemporaneous Binance price, with Rsync-bot obtaining 
a larger improvement. Quantitatively, the estimated coefficients imply that the 
bots' CEX execution price for ETH is approximately 1.0 to 1.6 basis points better 
than the contemporaneous Binance price.

Finally, recall that our derivations assume that the fee paid by the bots for inclusion equals their risk-adjusted expected profit conditional on inclusion on-chain. If this assumption is relaxed, the observed fee instead provides a lower bound on expected profits, since bots may offer builders less than their total expected profit in order to retain part of the surplus. Under this interpretation, our regression estimates should also be viewed as lower bounds: the two bots trade on the CEX leg of the arbitrage at prices \textit{at least} 1 to 1.6 basis points better than the contemporaneous Binance price.

\begin{table}[ht!]
    \centering
\sisetup{table-number-alignment = center,
         table-space-text-pre ={(},
         table-space-text-post={\textsuperscript{***}},
         input-open-uncertainty={[},
         input-close-uncertainty={]},
         table-align-text-pre=false,
         table-align-text-post=false}
\begin{threeparttable}
    \caption{Regression of percentage price improvement over Binance ($p_{\text{Improvement},s}$)}
    \label{tab:price_improvement_centered}
    \setlength{\tabcolsep}{12pt}
\begin{tabular}{r 
                S[table-format=-1.5]}
\cmidrule(lr){1-2}
                    &   {(1)}       \\
\midrule
Time since Last Block, centered & {-0.0010} \\
                           & {(0.001)} \\
\addlinespace
Volume, centered & {0.00007} \\
                           & {(0.00006)} \\
\addlinespace
Is Titan Bot & {-0.0070}\tnote{***} \\
                           & {(0.002)} \\
\addlinespace
Constant & {0.0165}\tnote{***} \\
                           & {(0.002)} \\
\addlinespace
\midrule
Observations              & {89}       \\
R-squared                 & {0.132}    \\
\bottomrule
\end{tabular}
    \smallskip
    \footnotesize
\begin{tablenotes}[para,flushleft]
    \item[*]    $p < 0.1$,
    \item[**]   $p < 0.05$,
    \item[***]  $p < 0.01$
\end{tablenotes}\par
Note: standard errors clustered at the auction-cycle level in parentheses. Time since last block and volume are centered at their sample means.
\end{threeparttable}
\end{table}

\section{Conclusions}

We study Ethereum by analyzing blocks that were proposed for inclusion in the canonical chain but were ultimately discarded. To the best of our knowledge, together with the companion paper \cite{pahari2025exclusiveethereumtransactionsevidence}, this is the first systematic analysis of non-winning Ethereum blocks. Our findings show that hidden fragmentation in transaction routing and block construction has economically meaningful consequences for on-chain market quality. Transactions are often delayed because they are routed only to builders that do not win the auction, and, for a subset of swaps, both execution probability and execution price vary substantially across candidate blocks. Competition between major searchers further amplifies these effects and helps determine which blocks win. Taken together, these results imply that standard blockchain data provide only a partial view of execution outcomes and may understate the importance of intermediation in the market for on-chain inclusion.

Our results should nevertheless be interpreted in light of several limitations. As noted, the data underrepresent some major builders, notably Beaverbuild, and span only eight minutes. However, to our knowledge, these are the best currently available data for studying how routing, builder competition, and block composition affect on-chain execution outcomes. Until more comprehensive data become available, they provide a uniquely informative view of the market for on-chain inclusion.

\bibliography{bib}
\bibliographystyle{chicago}

\newpage
\appendix

\section{Competition between Rsync-bot and Titan-bot}\label{sec: appendix-searchers}

\begin{figure*}[h]
    \centering
    \begin{subfigure}[t]{0.5\textwidth}
        \centering
        \includegraphics[scale=0.25]{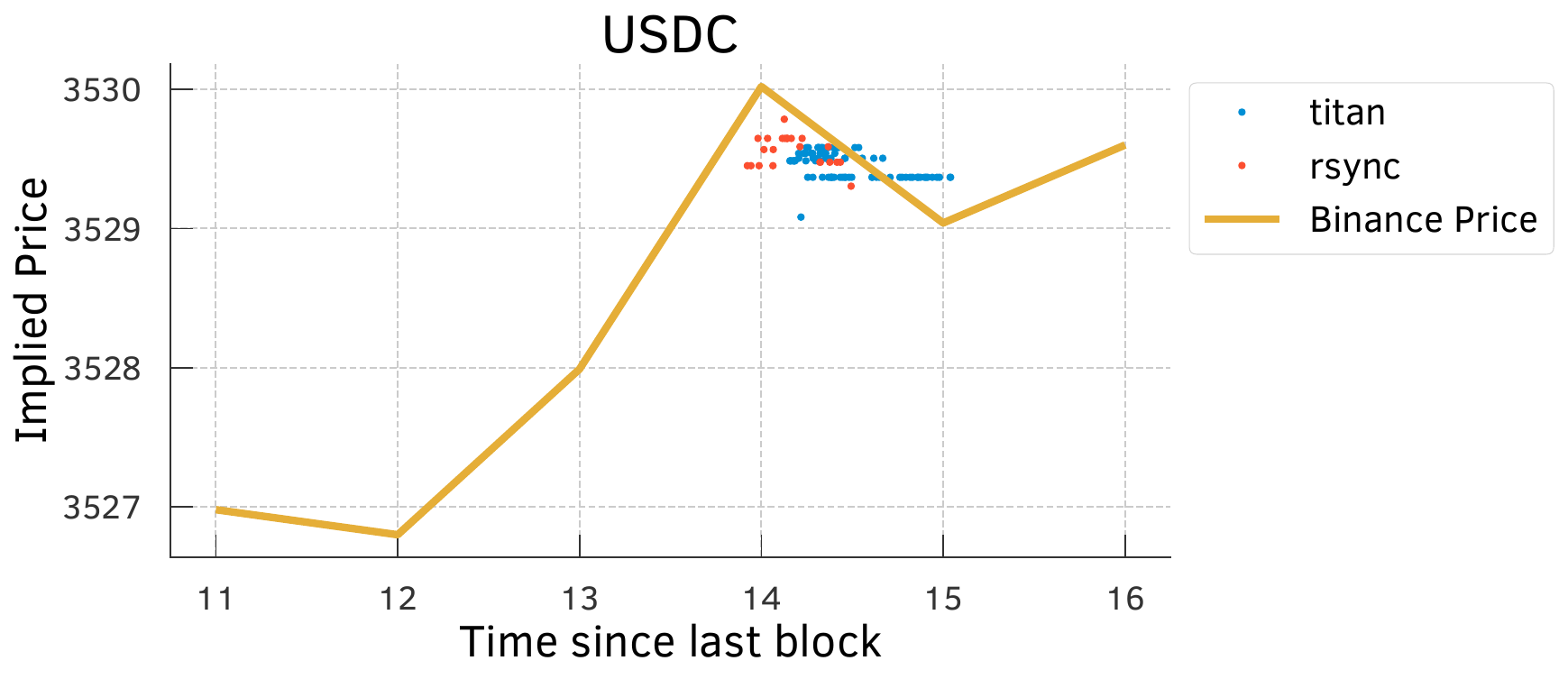}
        \caption{USDC Pool 1}
        \label{}
    \end{subfigure}%
    \begin{subfigure}[t]{0.5\textwidth}
        \centering
        \includegraphics[scale=0.25]{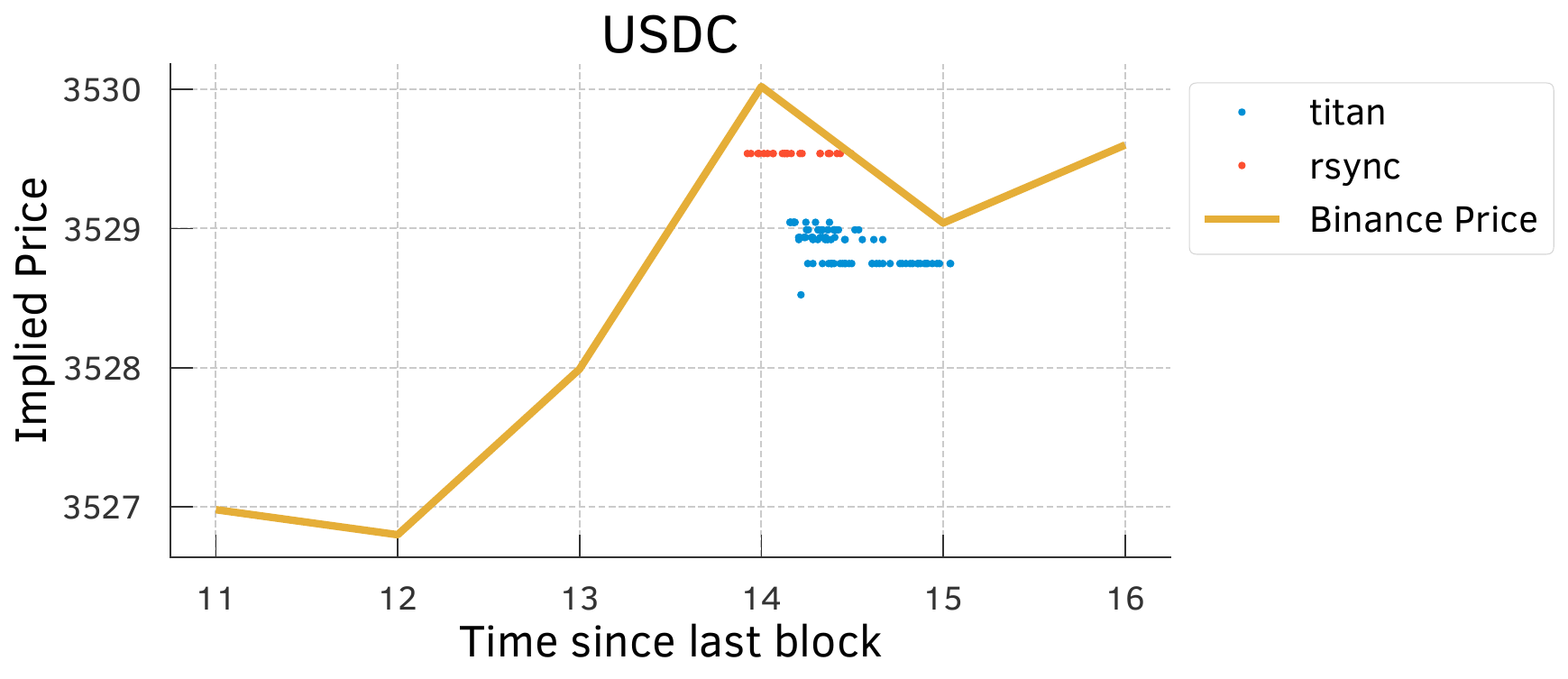}
        \caption{USDC Pool 2}
        \label{}
    \end{subfigure}
\begin{subfigure}[t]{0.45\textwidth}
        \centering
        \includegraphics[scale=0.25]{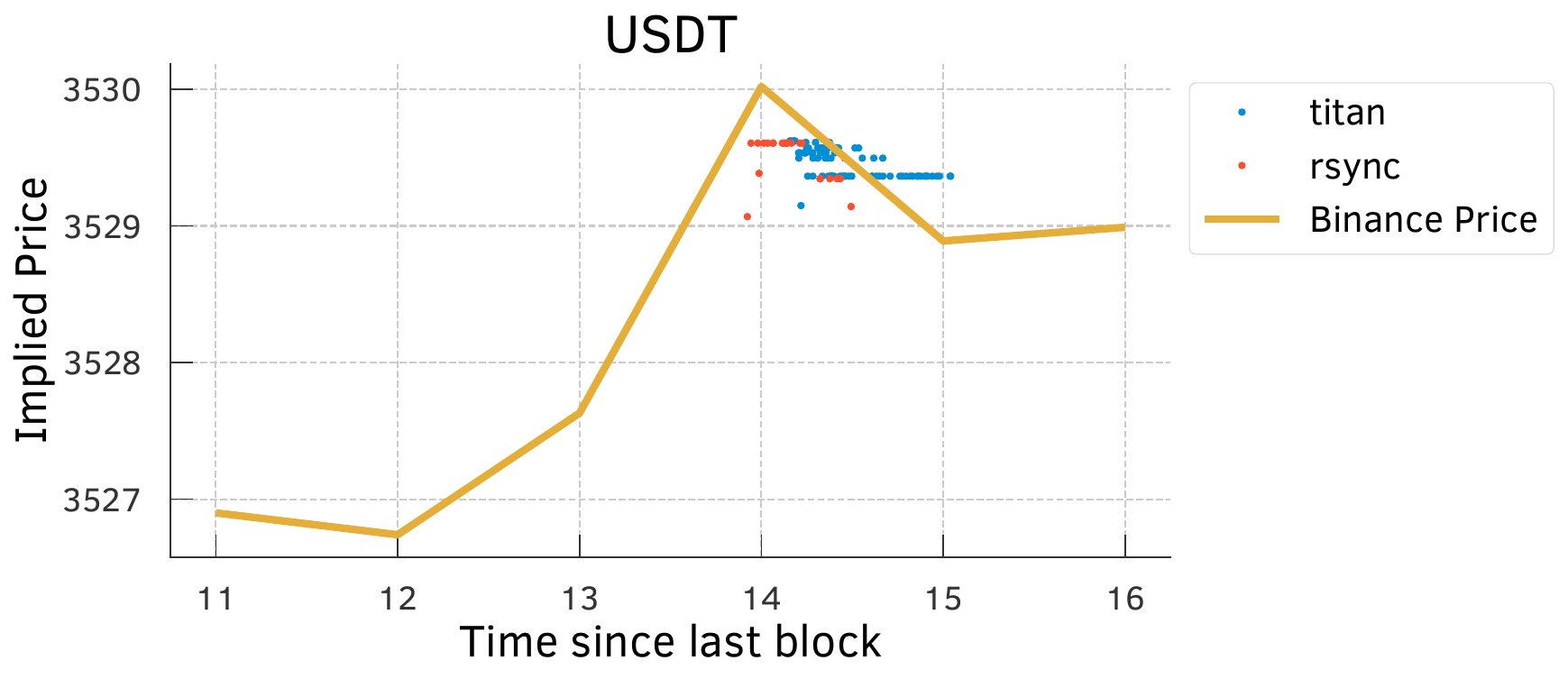}
        \caption{USDT Pool 1}
        \label{}
    \end{subfigure}%
    \begin{subfigure}[t]{0.5\textwidth}
        \centering
        \includegraphics[scale=0.25]{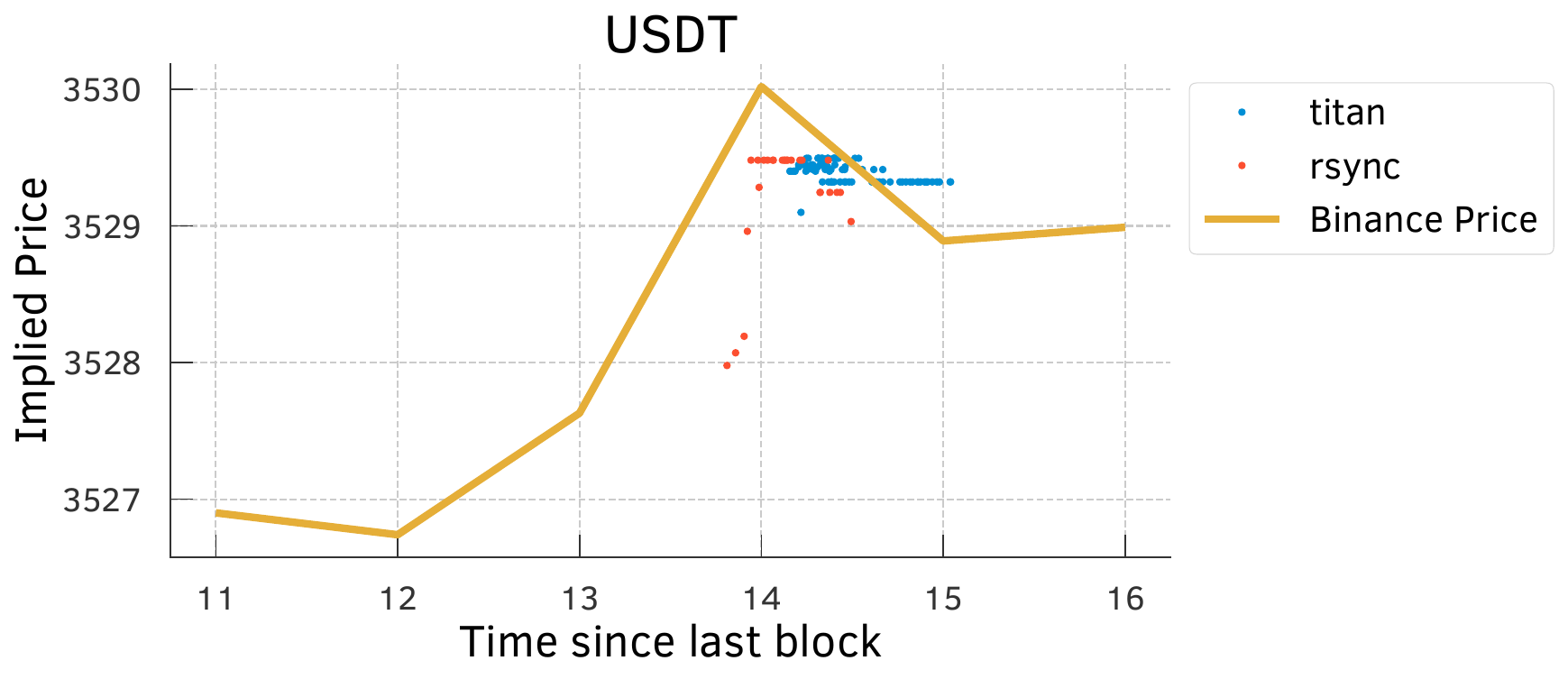}
        \caption{USDT Pool2}
        \label{}
    \end{subfigure}
    \caption{Competition between Rsyc-bot and Titan-bot during Block 21322626 (ETH as base currency)}
    \label{}
\end{figure*}

\begin{figure*}[h]
    \centering
    \begin{subfigure}[t]{0.5\textwidth}
        \centering
        \includegraphics[scale=0.25]{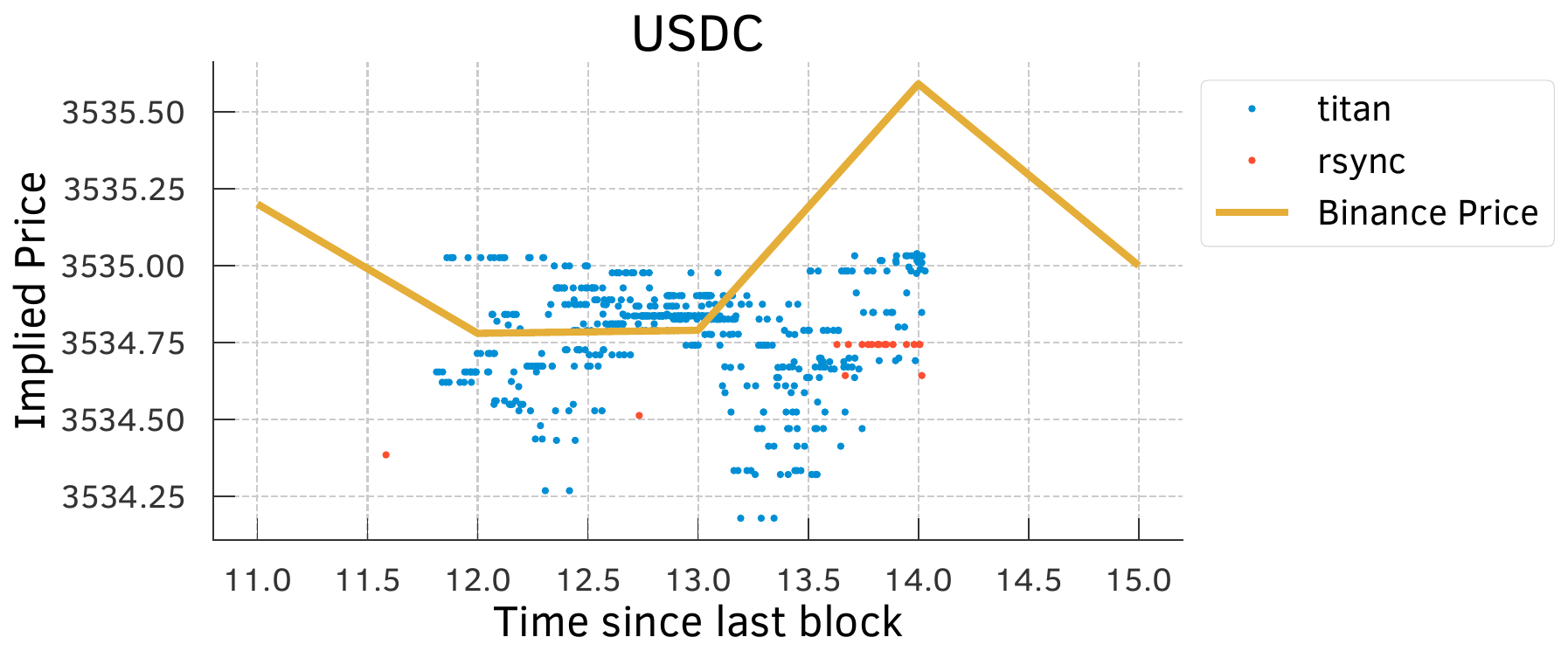}
        \caption{USDC Pool 1}
        \label{}
    \end{subfigure}%
    \begin{subfigure}[t]{0.5\textwidth}
        \centering
        \includegraphics[scale=0.25]{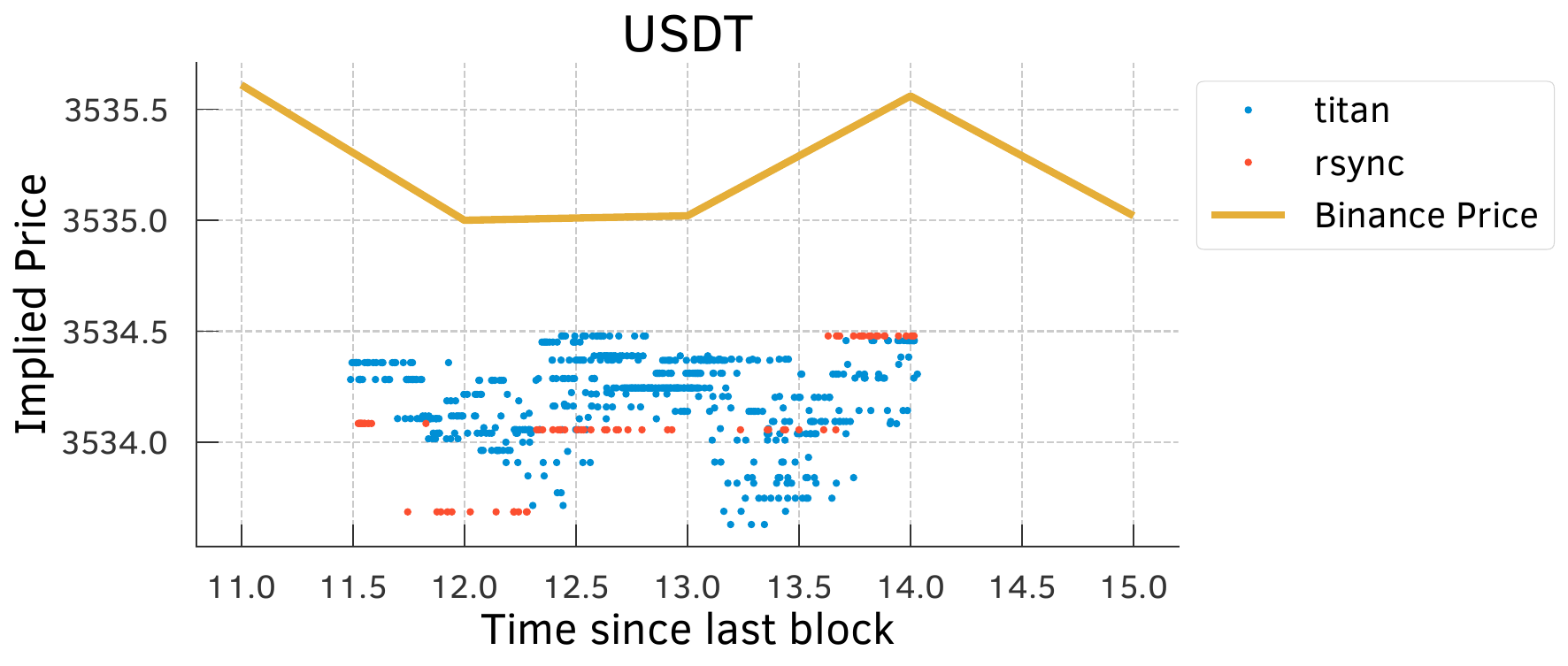}
        \caption{USDT Pool 1}
        \label{}
    \end{subfigure}
\begin{subfigure}[t]{0.45\textwidth}
        \centering
        \includegraphics[scale=0.25]{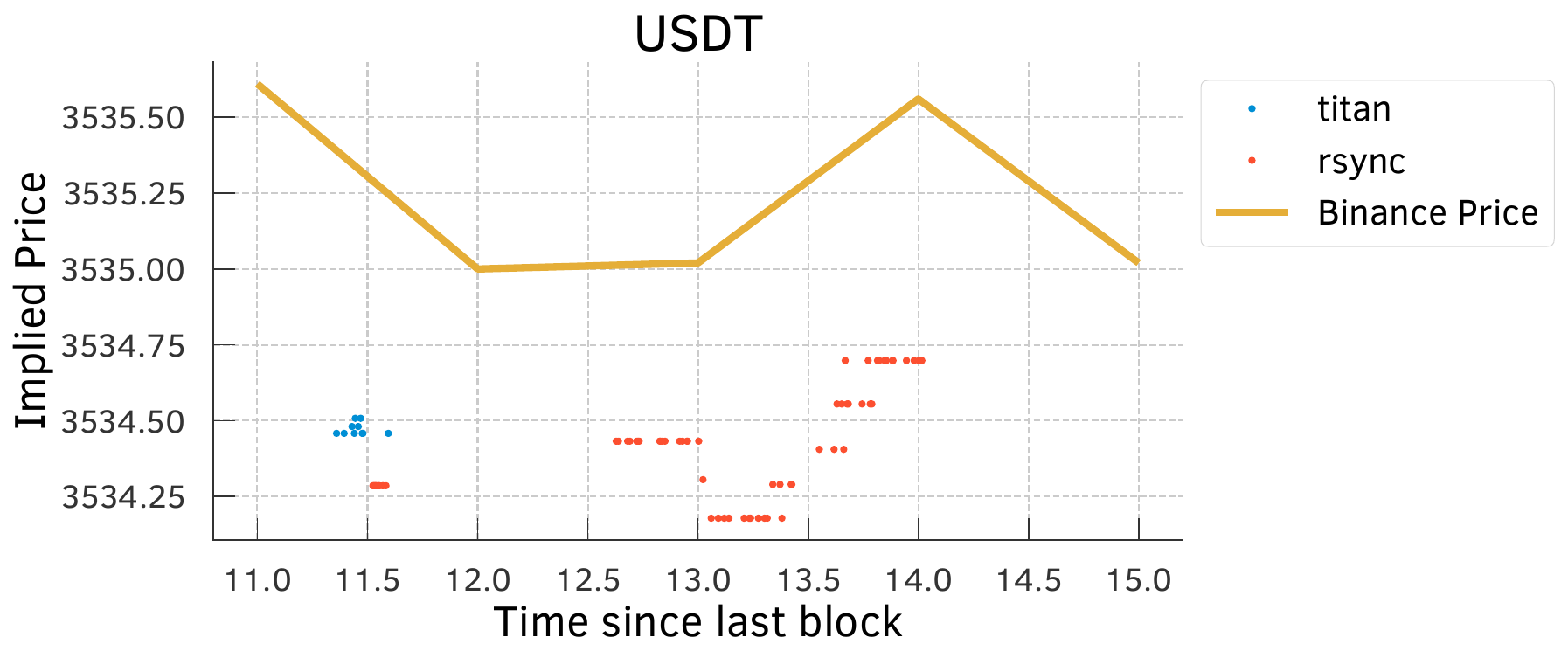}
        \caption{USDT Pool 2}
        \label{}
    \end{subfigure}%
    \caption{Competition between Rsyc-bot and Titan-bot during Block 21322630 (ETH as base currency)}
    \label{}
\end{figure*}

\begin{figure*}[h]
    \centering
    \begin{subfigure}[t]{0.5\textwidth}
        \centering
        \includegraphics[scale=0.25]{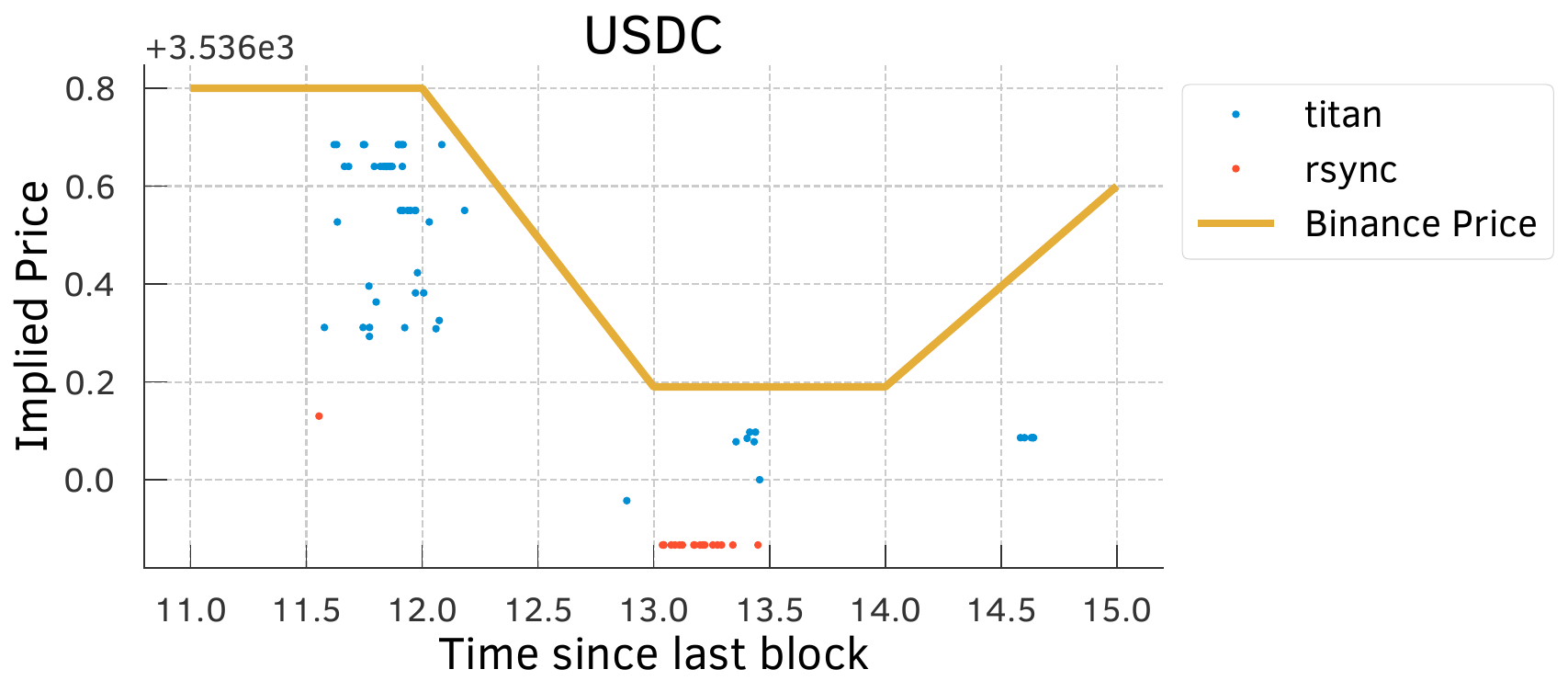}
        \caption{USDC Pool 1}
        \label{}
    \end{subfigure}%
    \begin{subfigure}[t]{0.5\textwidth}
        \centering
        \includegraphics[scale=0.25]{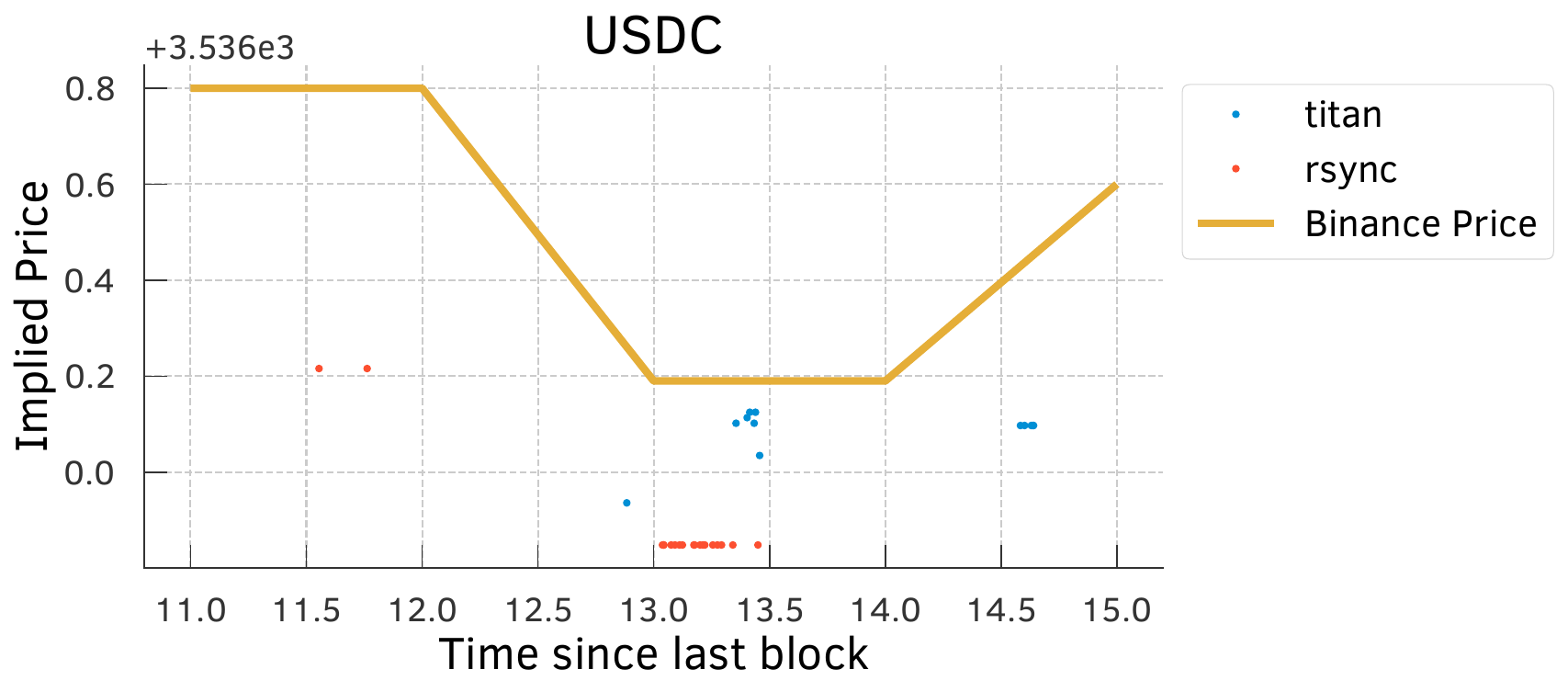}
        \caption{USDC Pool 2}
        \label{}
    \end{subfigure}
    \caption{Competition between Rsyc-bot and Titan-bot during block 21322631 (ETH as base currency)}
    \label{}
\end{figure*}

\begin{figure*}[h]
    \centering
    \begin{subfigure}[t]{0.5\textwidth}
        \centering
        \includegraphics[scale=0.25]{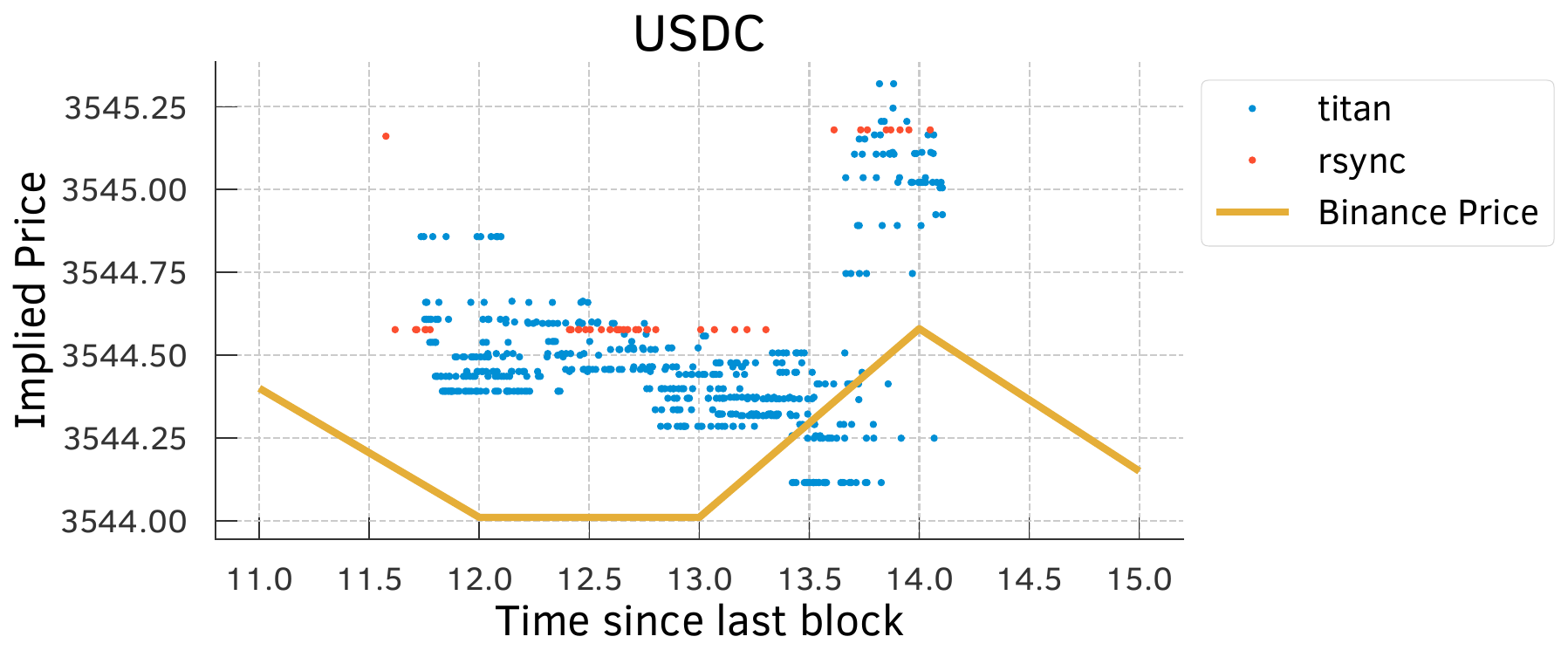}
        \caption{USDC Pool 1}
        \label{}
    \end{subfigure}%
    \begin{subfigure}[t]{0.5\textwidth}
        \centering
        \includegraphics[scale=0.25]{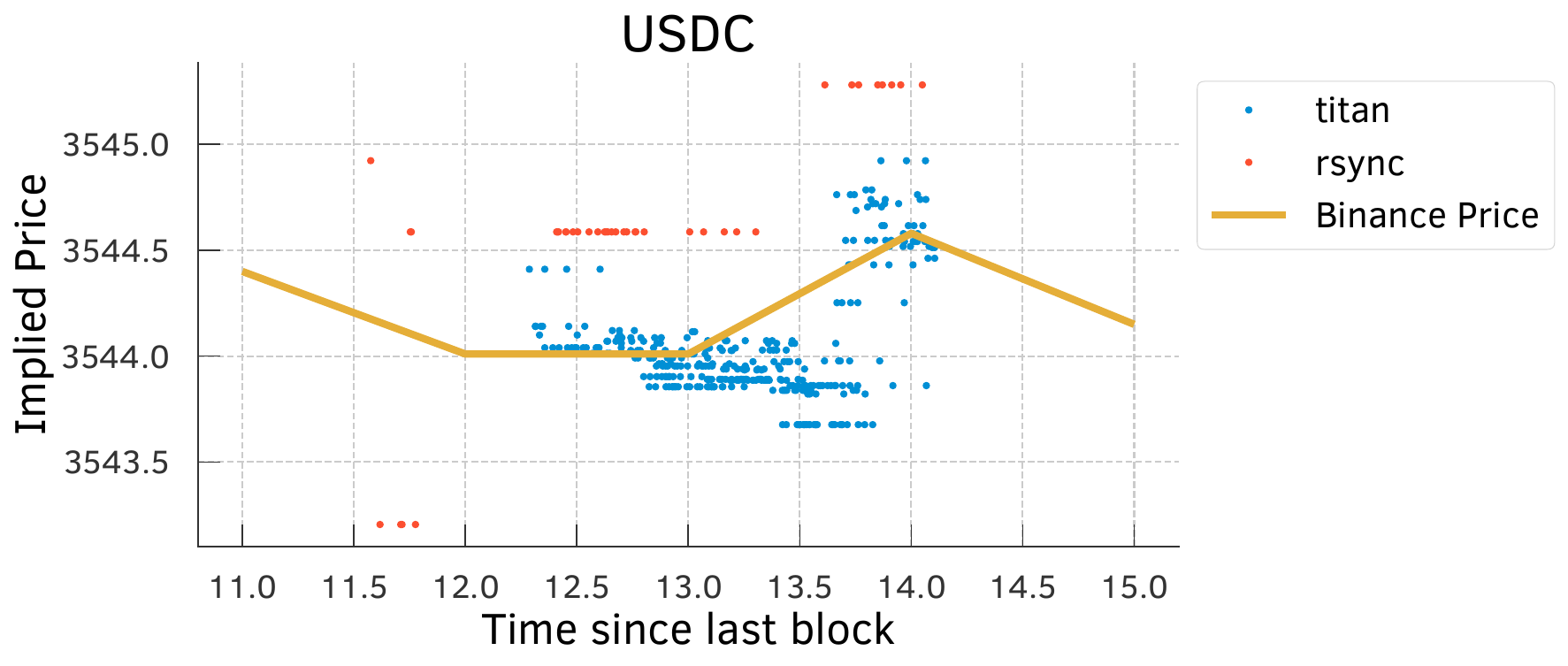}
        \caption{USDC Pool 2}
        \label{}
    \end{subfigure}
    \begin{subfigure}[t]{0.45\textwidth}
        \centering
        \includegraphics[scale=0.25]{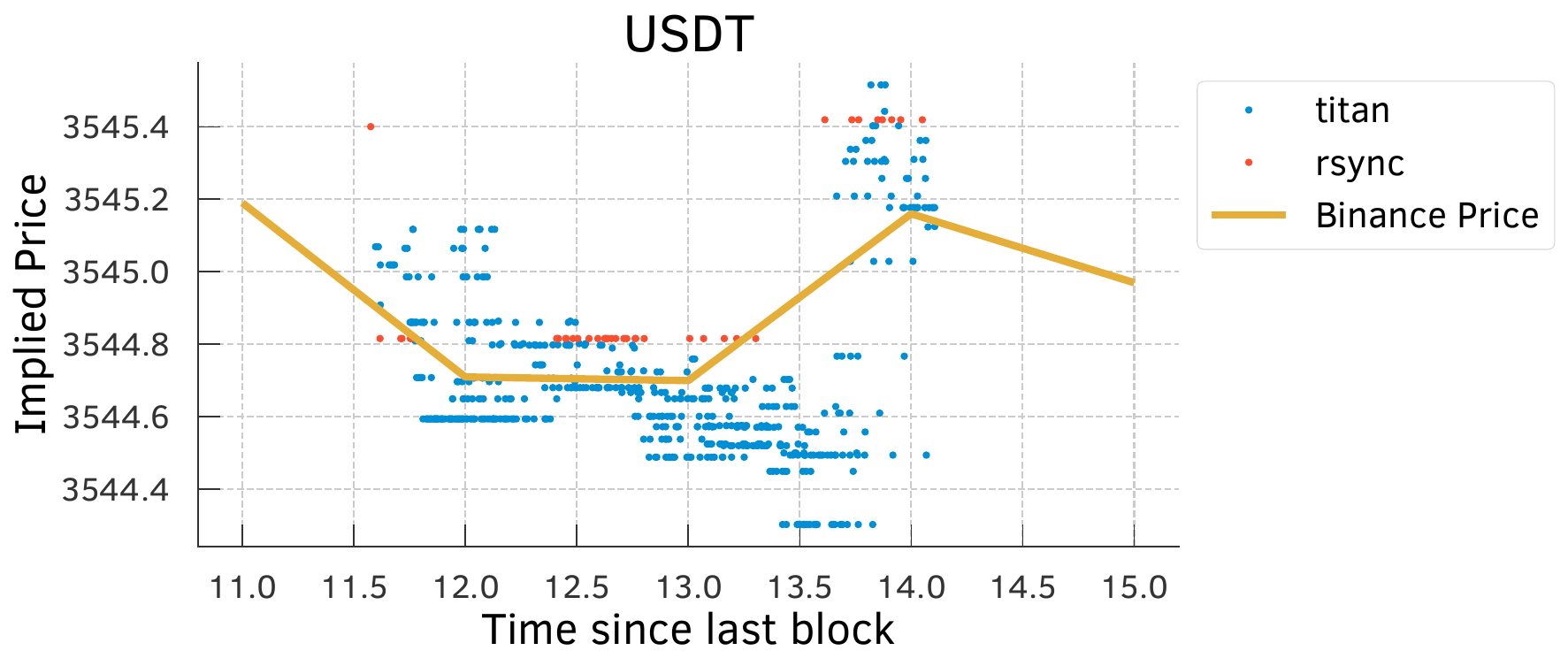}
        \caption{USDT Pool 1}
        \label{}
    \end{subfigure}%
    \caption{Competition between Rsyc-bot and Titan-bot during block 21322648 (ETH as base currency)}
    \label{}
\end{figure*}

\begin{figure*}[h]
    \centering
    \begin{subfigure}[t]{0.5\textwidth}
        \centering
        \includegraphics[scale=0.25]{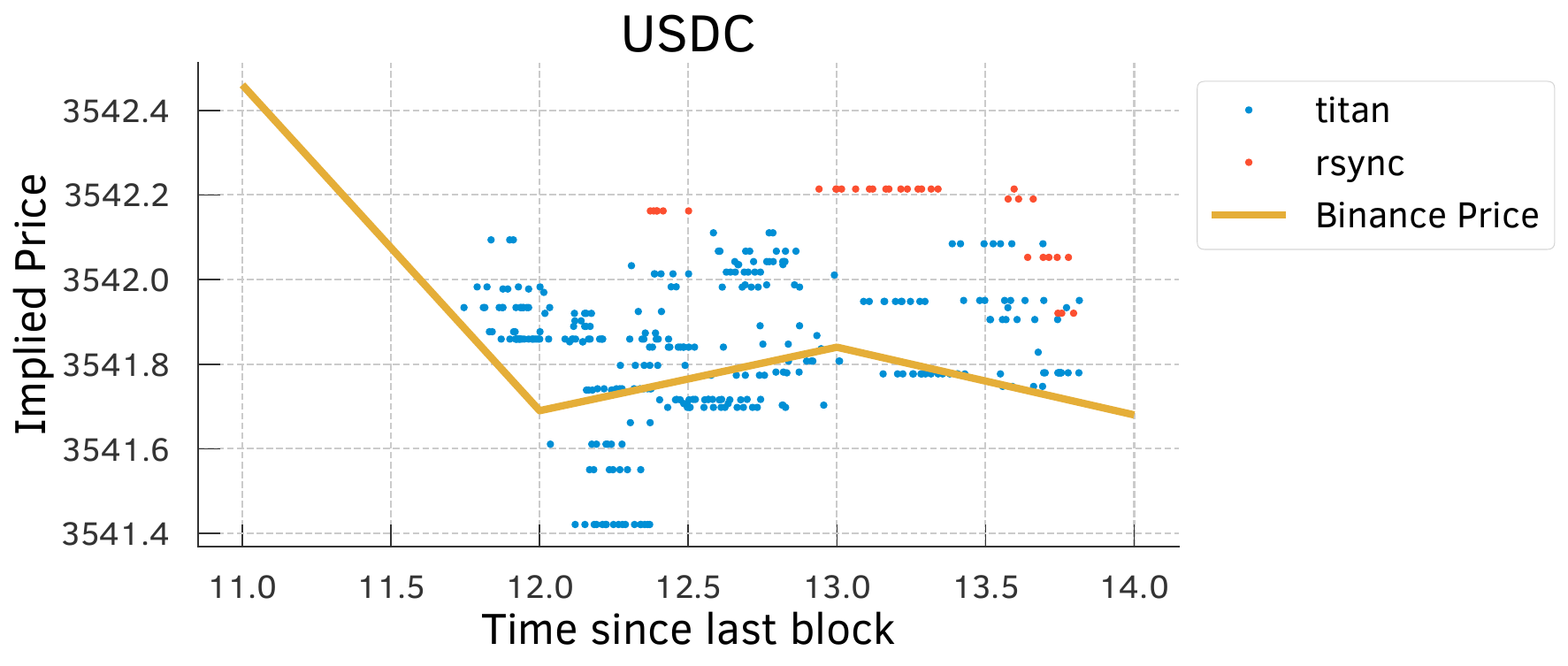}
        \caption{USDC Pool 1}
        \label{}
    \end{subfigure}%
    \begin{subfigure}[t]{0.5\textwidth}
        \centering
        \includegraphics[scale=0.25]{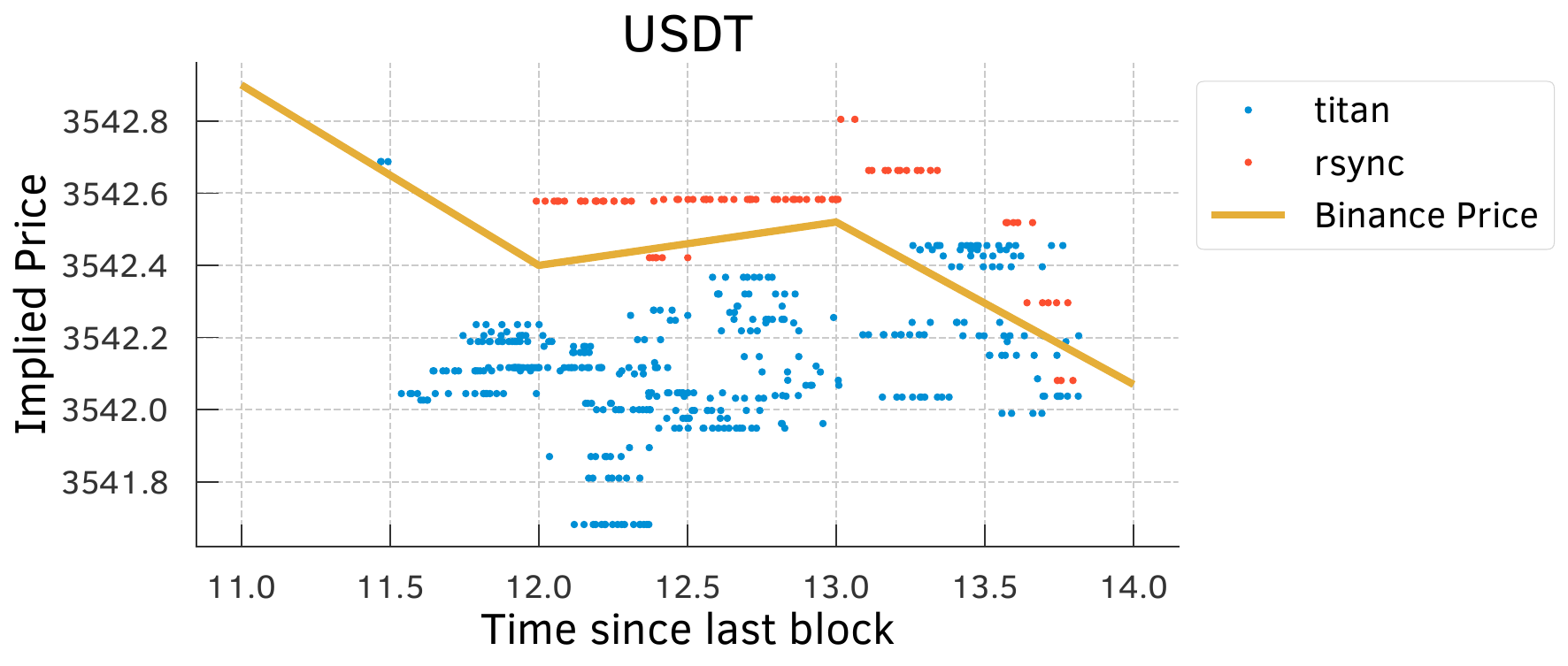}
        \caption{USDT Pool 1}
        \label{}
    \end{subfigure}
    \caption{Competition between Rsyc-bot and Titan-bot during block 21322649 (ETH as base currency)}
    \label{}
\end{figure*}

\begin{figure*}[h]
    \centering
    \begin{subfigure}[t]{0.5\textwidth}
        \centering
        \includegraphics[scale=0.25]{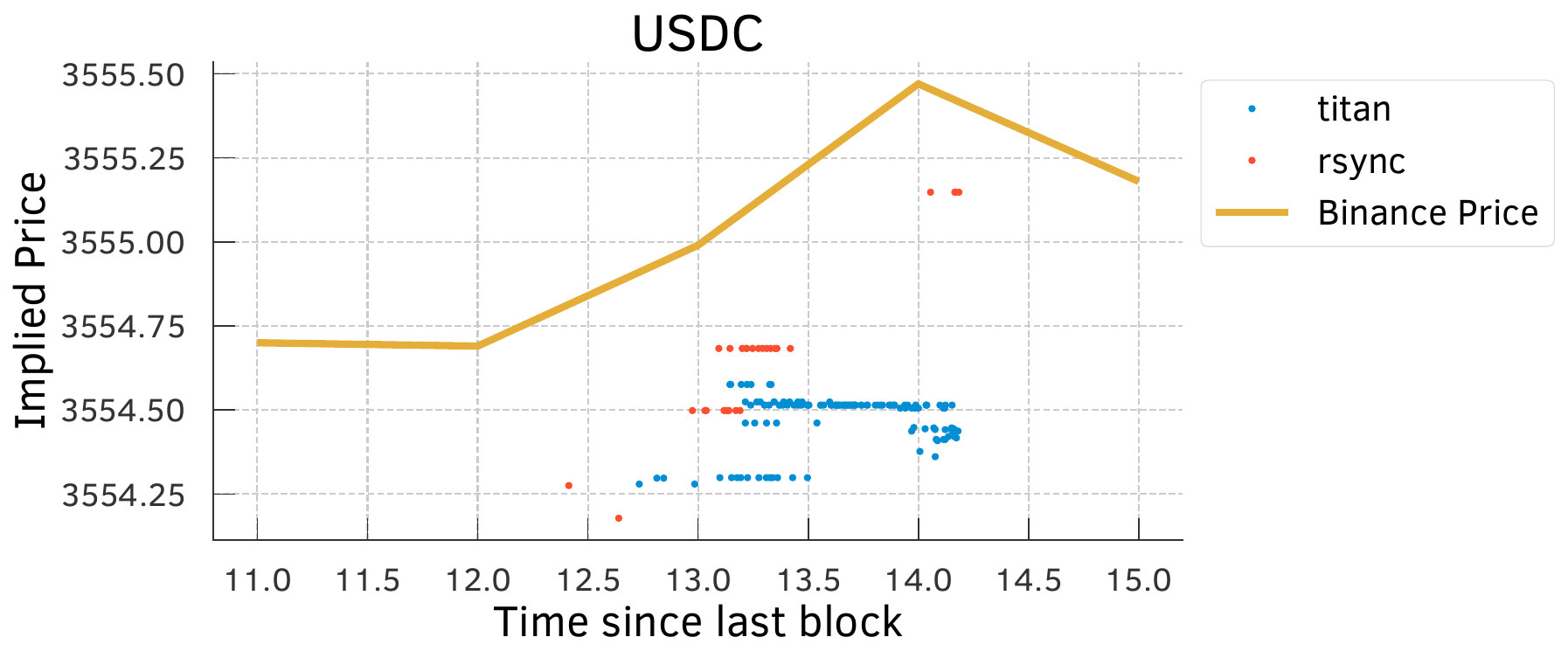}
        \caption{USDC Pool 1 Block 21322657}
        \label{}
    \end{subfigure}%
    \begin{subfigure}[t]{0.5\textwidth}
        \centering
        \includegraphics[scale=0.25]{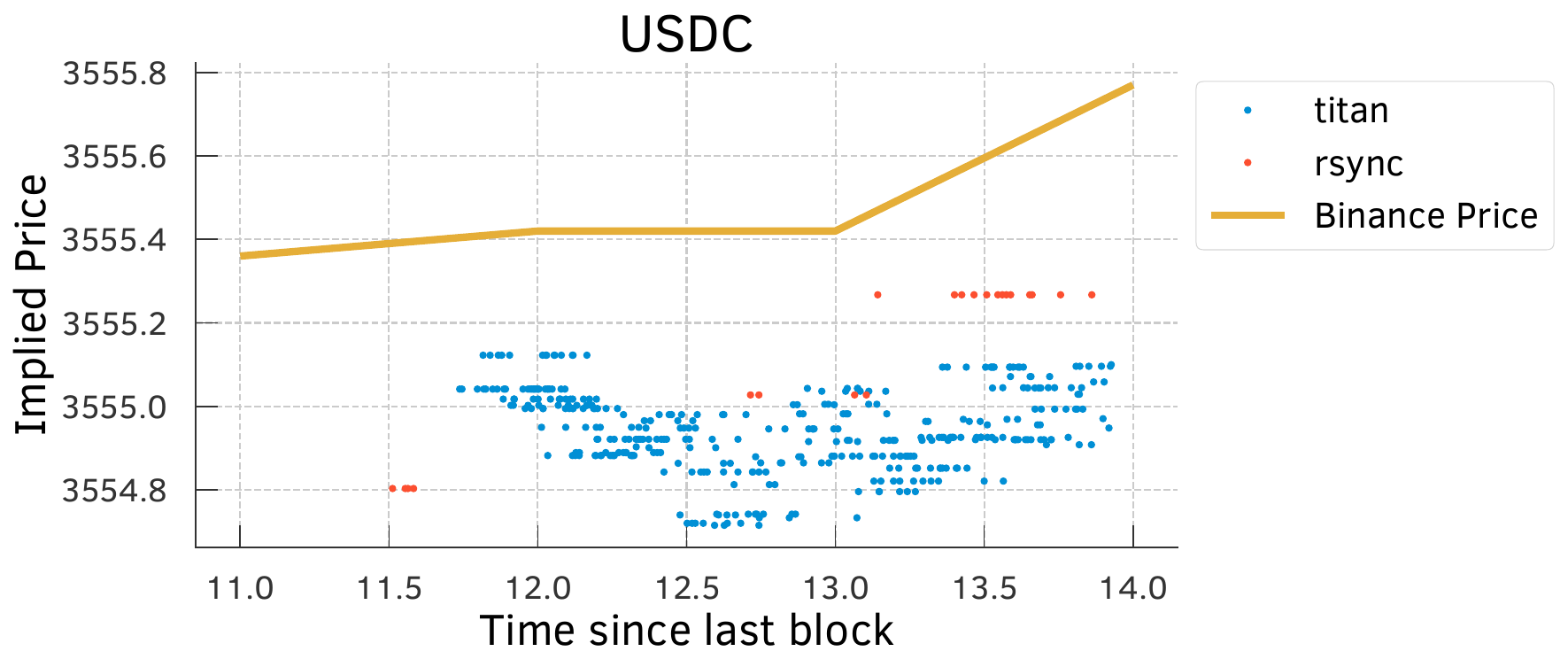}
        \caption{USDC Pool 1 Block 21322658}
        \label{}
    \end{subfigure}
\begin{subfigure}[t]{0.45\textwidth}
        \centering
        \includegraphics[scale=0.25]{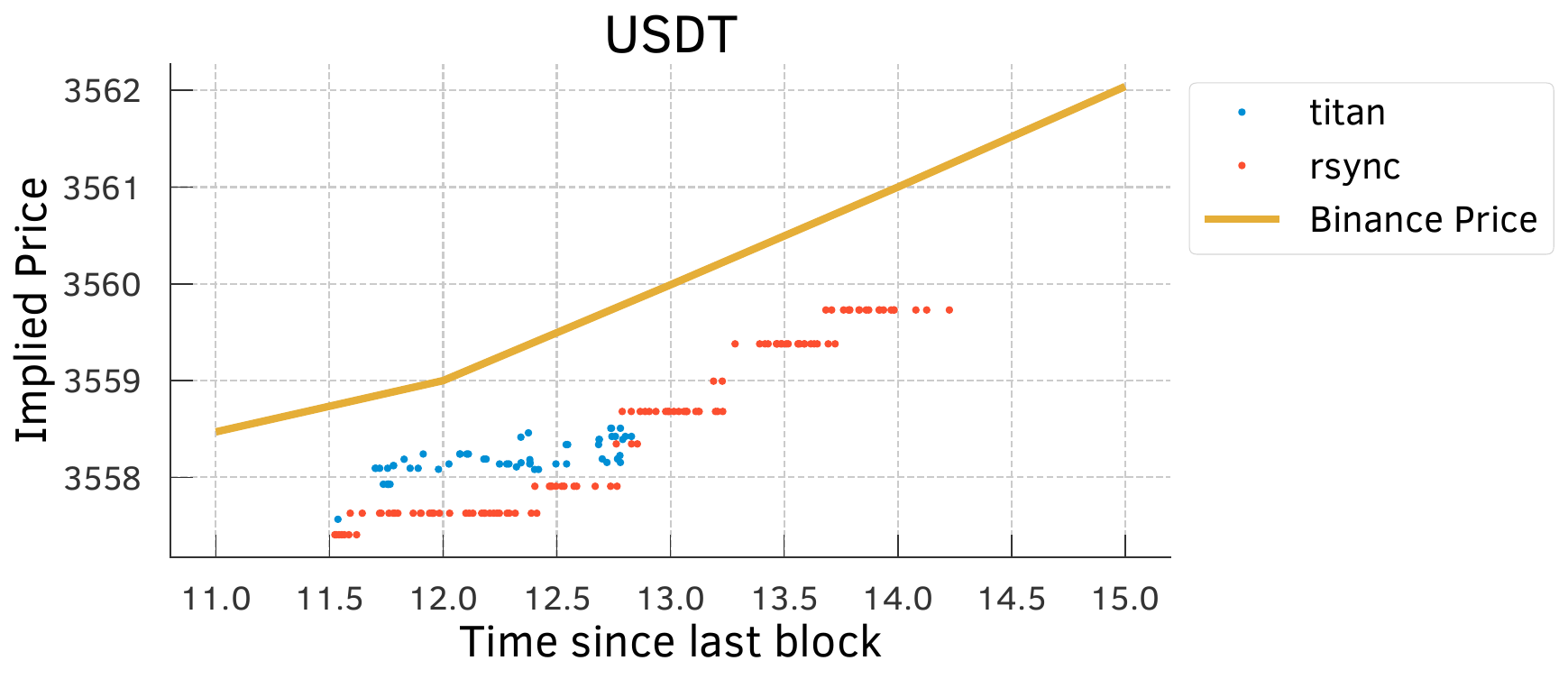}
        \caption{USDT Pool 2 Block 21322659}
        \label{}
    \end{subfigure}%
    \begin{subfigure}[t]{0.5\textwidth}
        \centering
        \includegraphics[scale=0.25]{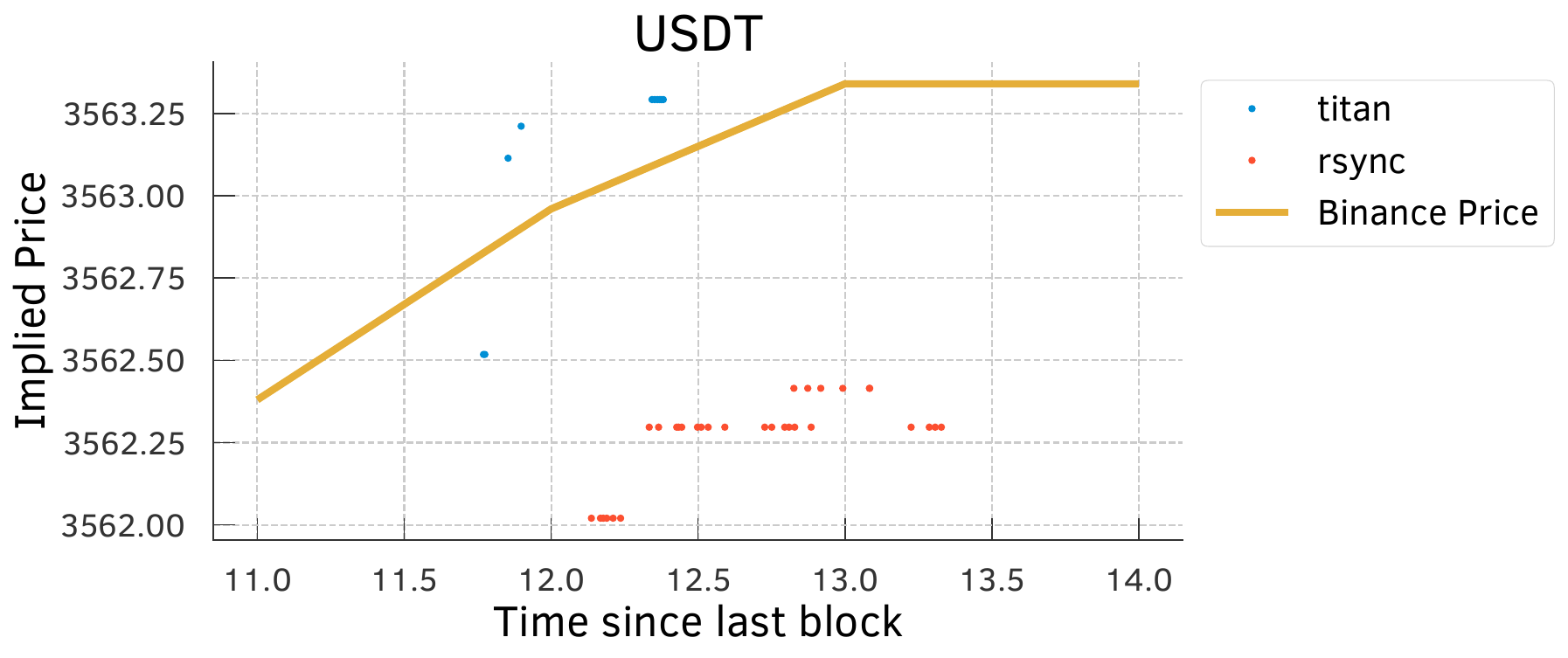}
        \caption{USDT Pool 3 Block 21322660}
        \label{}
    \end{subfigure}
    \caption{Competition between Rsyc-bot and Titan-bot during blocks 21322657 - 21322660 (ETH as base currency)}
    \label{}
\end{figure*}

\begin{figure*}[h]
    \centering
    \begin{subfigure}[t]{0.5\textwidth}
        \centering
        \includegraphics[scale=0.25]{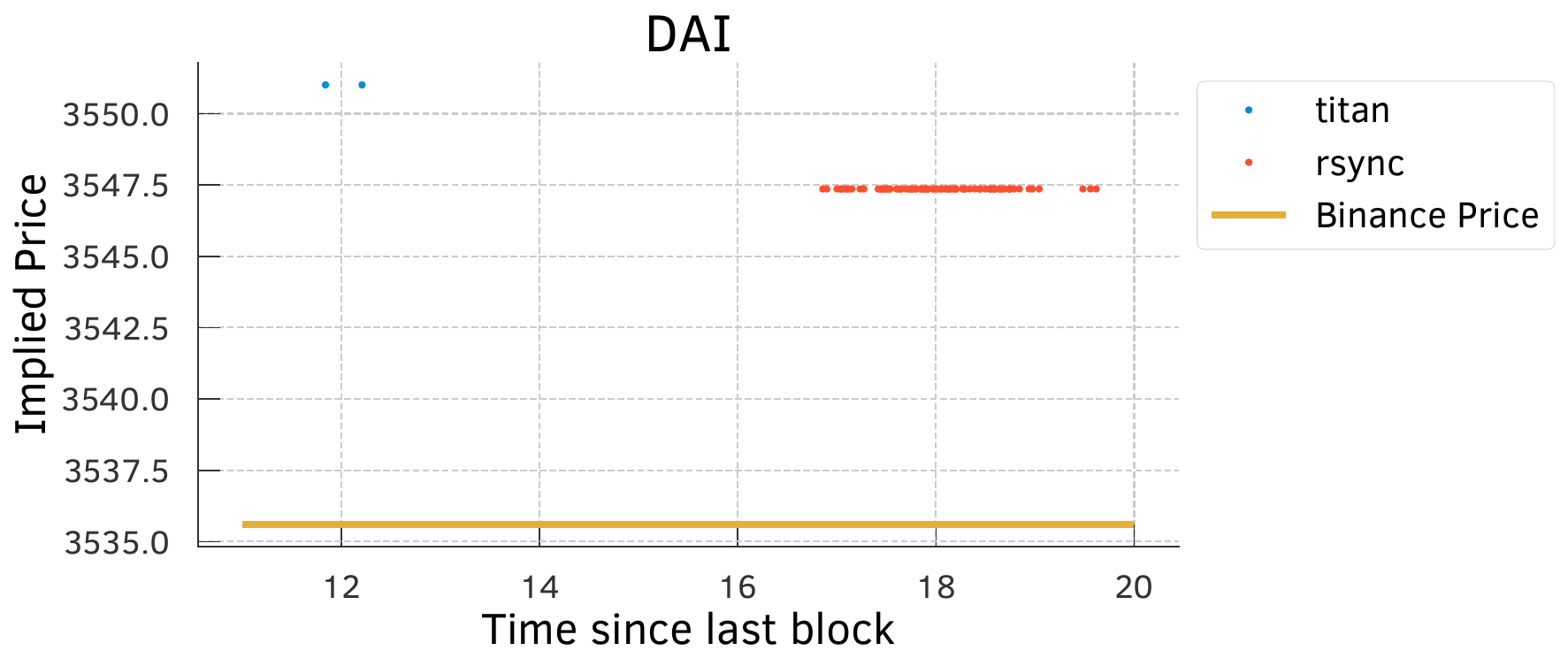}
        \caption{DAI Block 21322635}
        \label{}
    \end{subfigure}%
    \begin{subfigure}[t]{0.5\textwidth}
        \centering
        \includegraphics[scale=0.25]{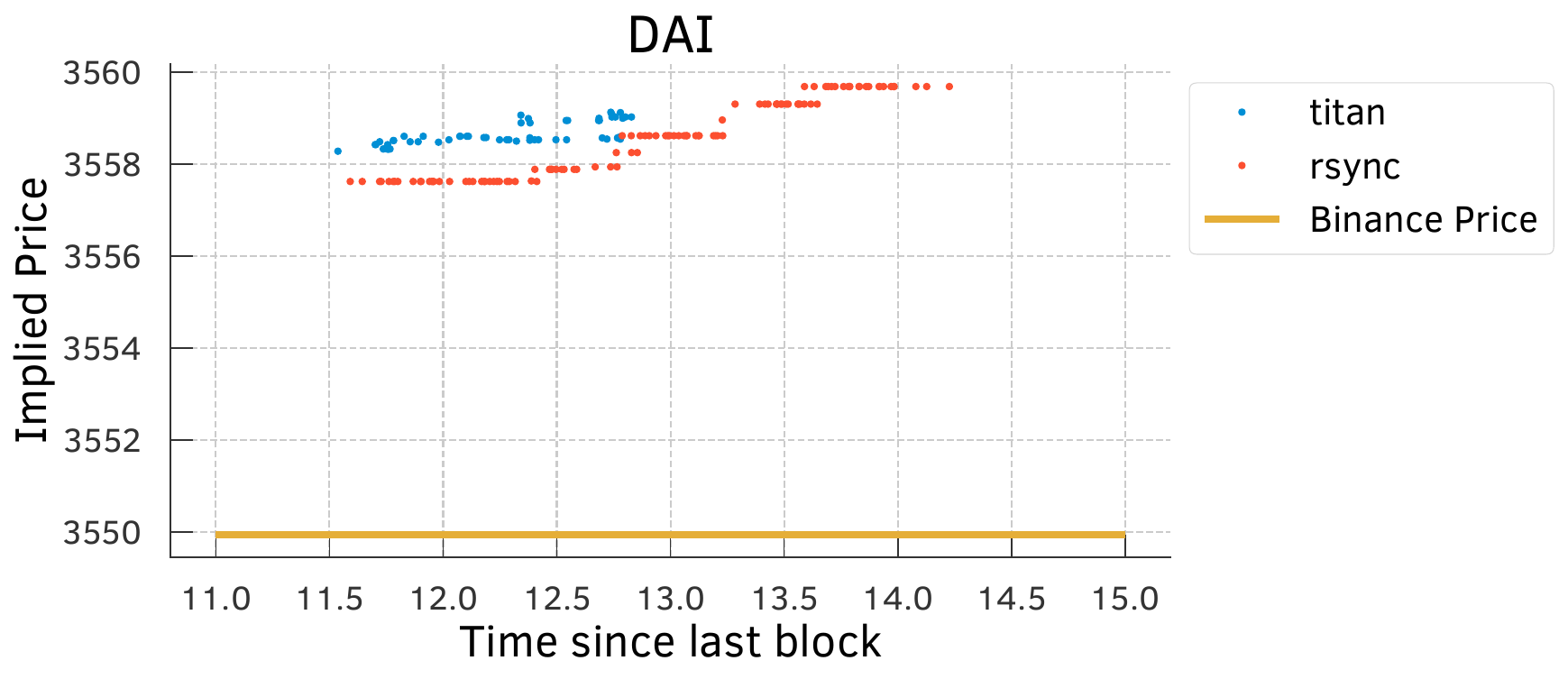}
        \caption{DAI Block 21322659}
        \label{}
    \end{subfigure}
\begin{subfigure}[t]{0.45\textwidth}
        \centering
        \includegraphics[scale=0.25]{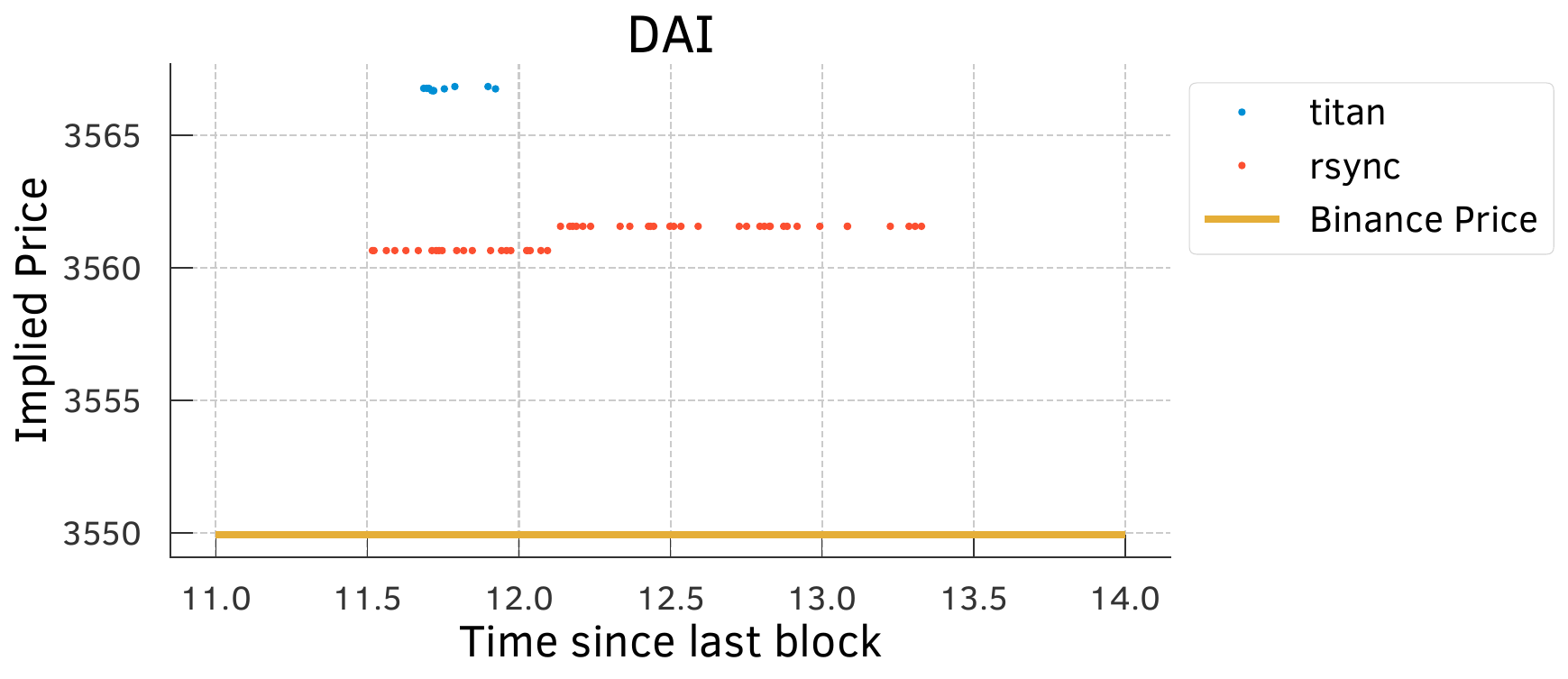}
        \caption{DAI Block 21322660}
        \label{}
    \end{subfigure}%
    \begin{subfigure}[t]{0.5\textwidth}
        \centering
        \includegraphics[scale=0.25]{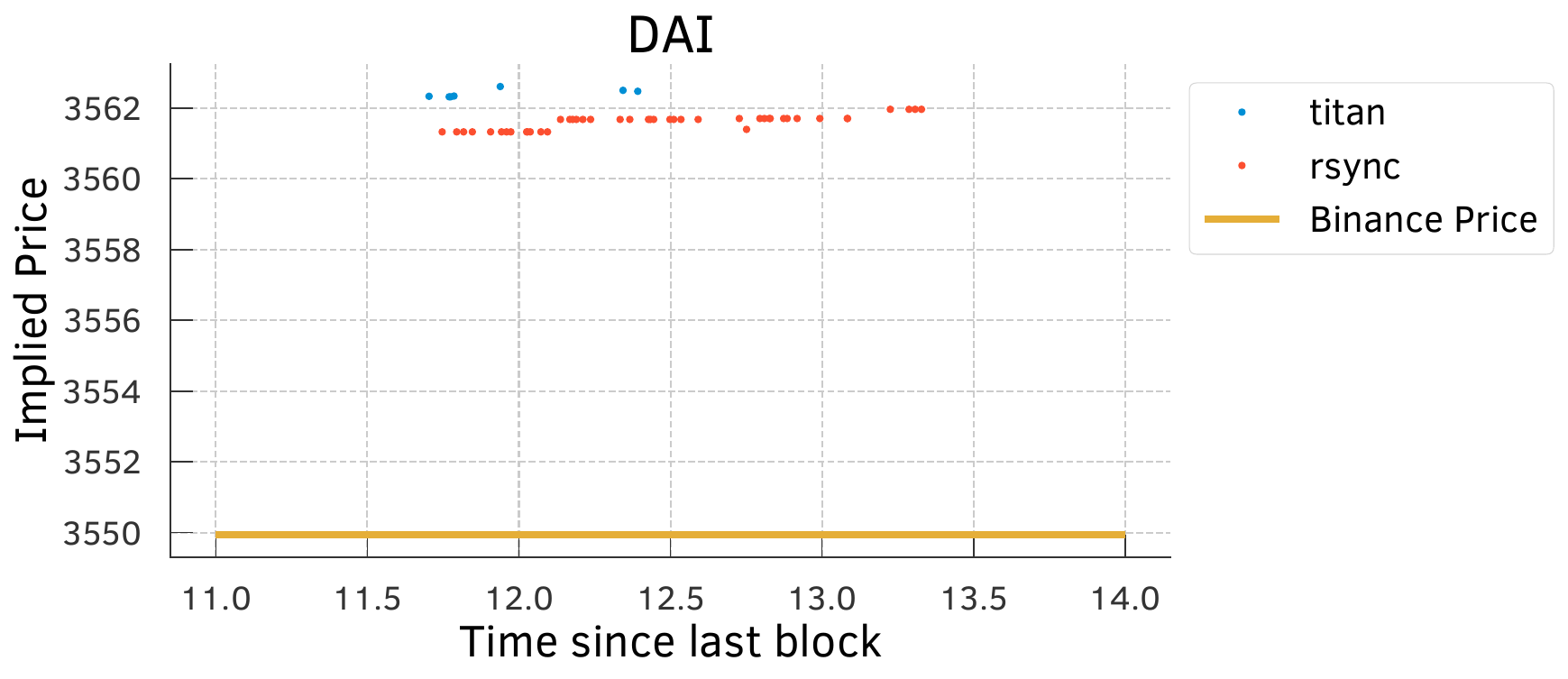}
        \caption{DAI Block 21322660}
        \label{}
    \end{subfigure}
        \begin{subfigure}[t]{0.5\textwidth}
        \centering
        \includegraphics[scale=0.25]{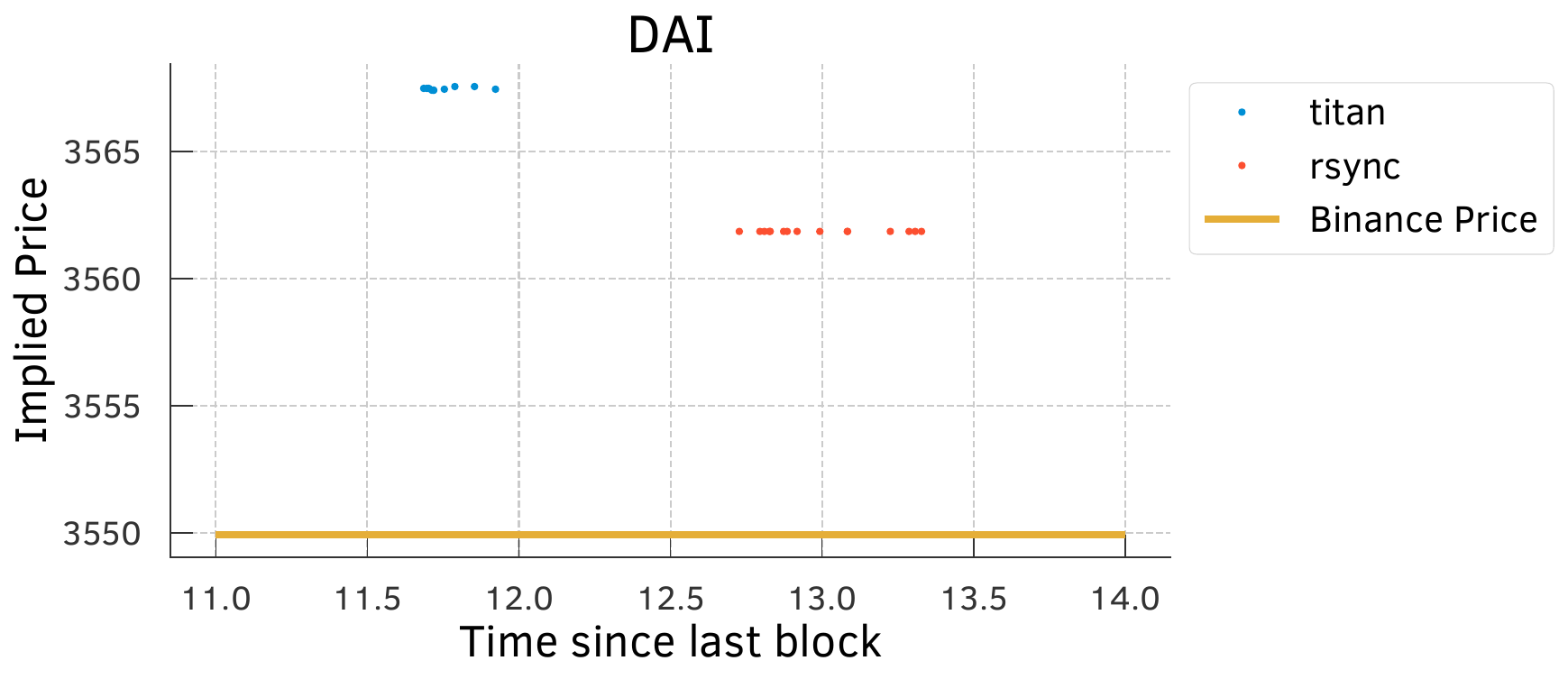}
        \caption{DAI Block 21322660}
        \label{}
    \end{subfigure}
    \caption{Competition between Rsyc-bot and Titan-bot on the DAI/ETH market (all blocks)}
    \label{}
\end{figure*}

\begin{figure*}[h]
    \centering
    \begin{subfigure}[t]{0.5\textwidth}
        \centering
        \includegraphics[scale=0.25]{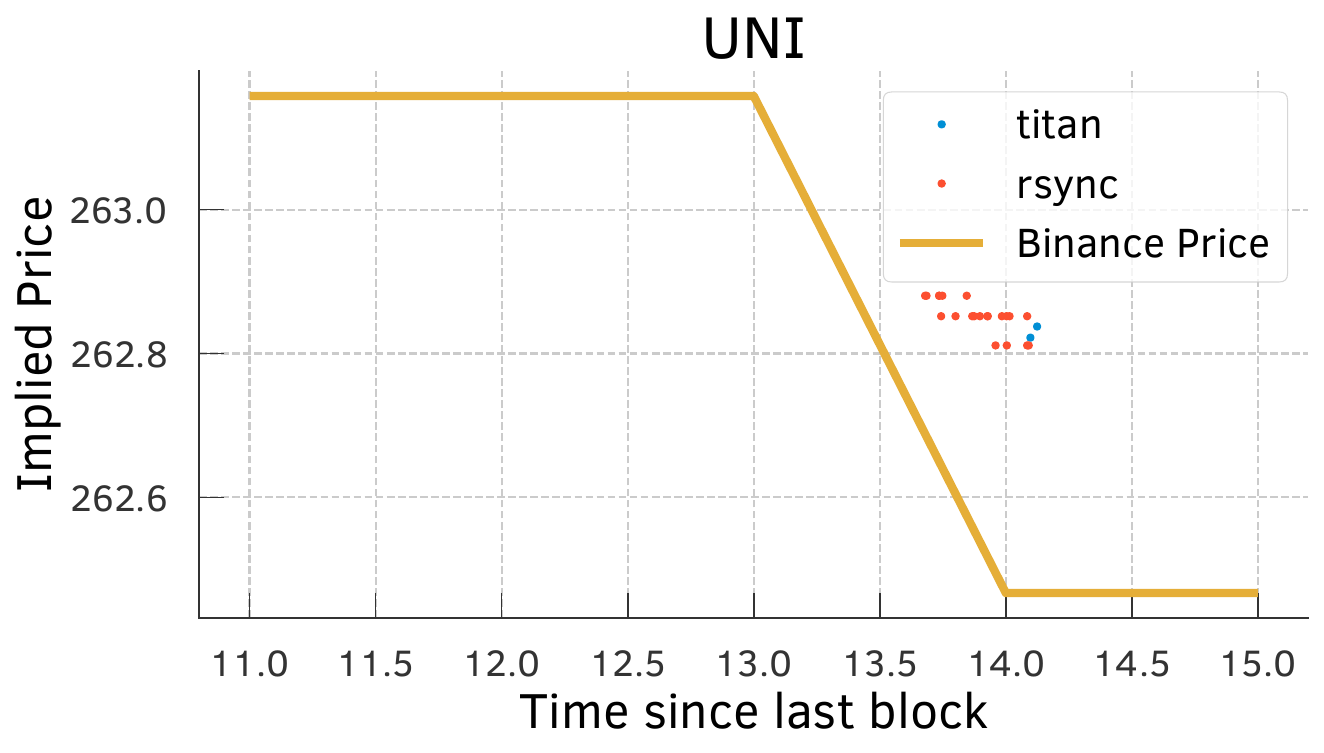}
        \caption{UNI Block 21322628}
        \label{}
    \end{subfigure}%
    \begin{subfigure}[t]{0.5\textwidth}
        \centering
        \includegraphics[scale=0.25]{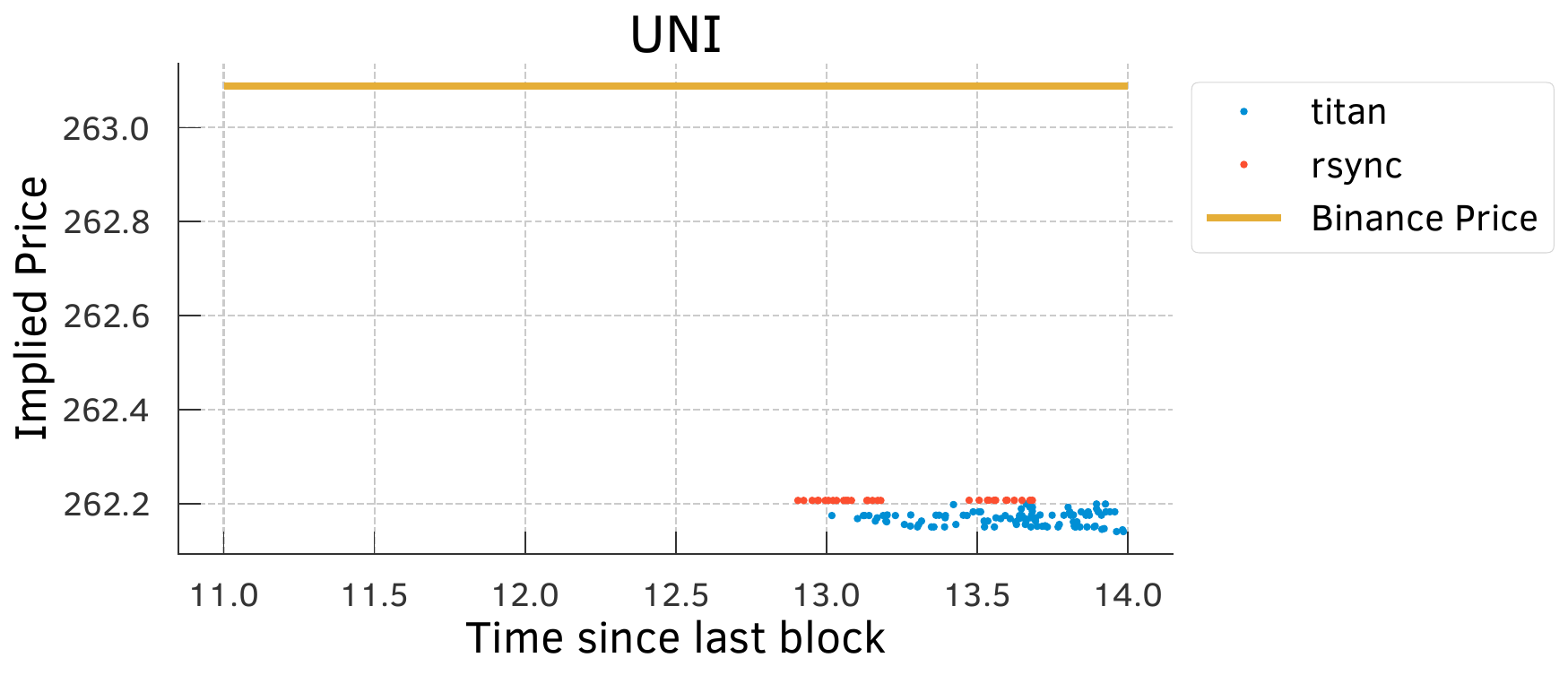}
        \caption{UNI Block 21322639}
        \label{}
    \end{subfigure}
    \begin{subfigure}[t]{0.45\textwidth}
        \centering
        \includegraphics[scale=0.25]{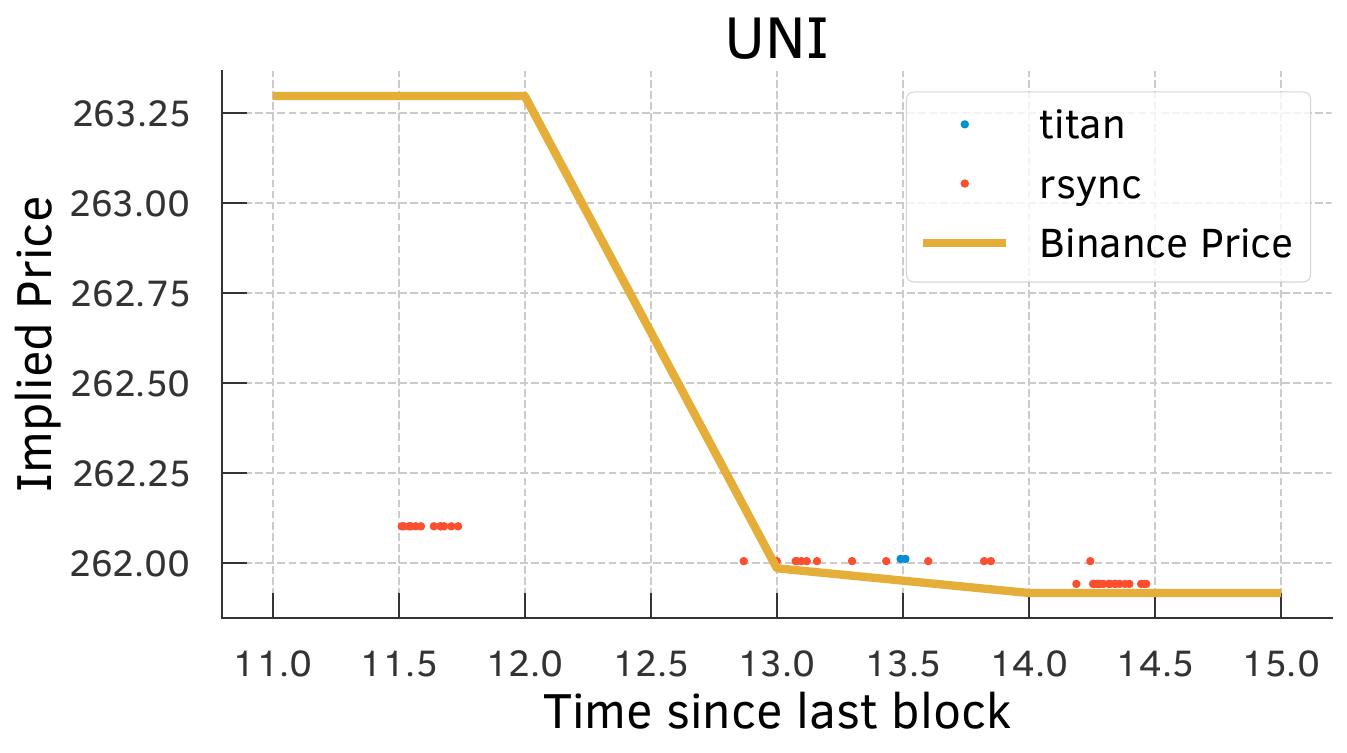}
        \caption{UNI Block 21322643}
        \label{}
    \end{subfigure}%
    \begin{subfigure}[t]{0.5\textwidth}
        \centering
        \includegraphics[scale=0.25]{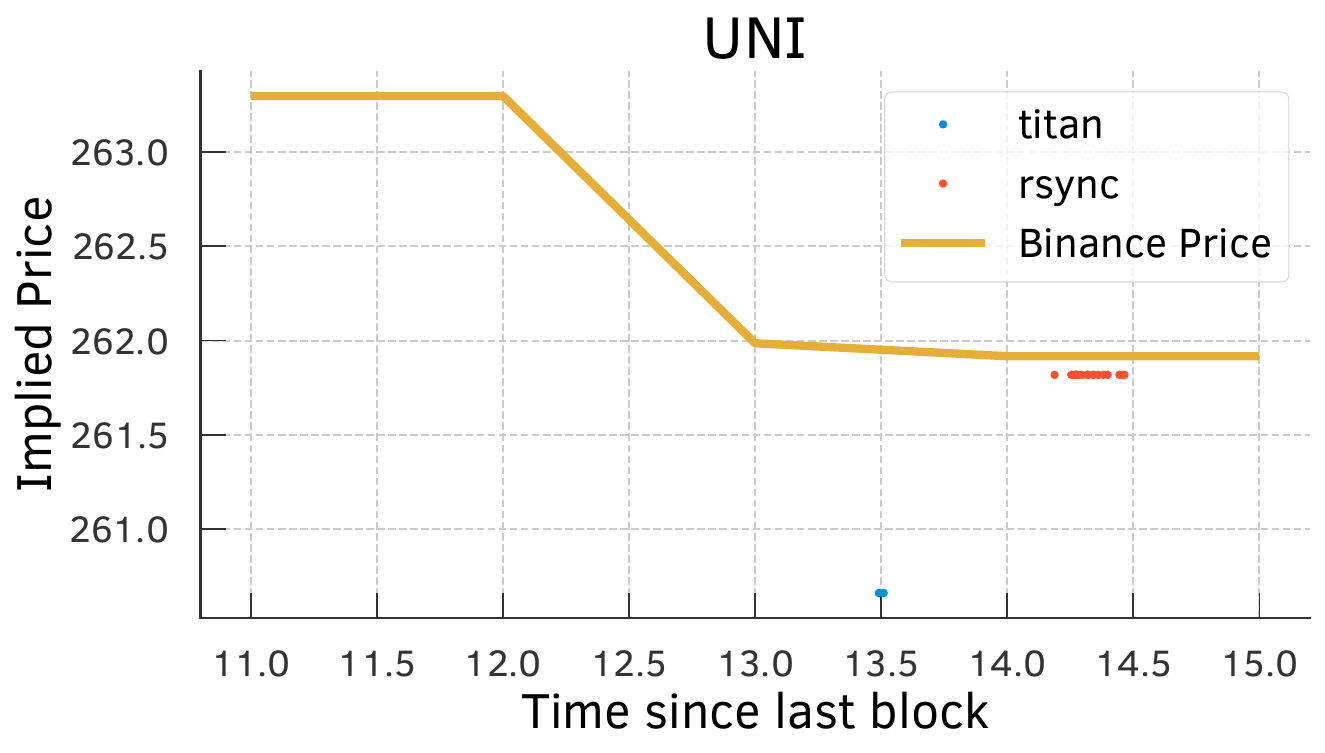}
        \caption{UNI Block 21322643}
        \label{}
    \end{subfigure}
        \begin{subfigure}[t]{0.5\textwidth}
        \centering
        \includegraphics[scale=0.25]{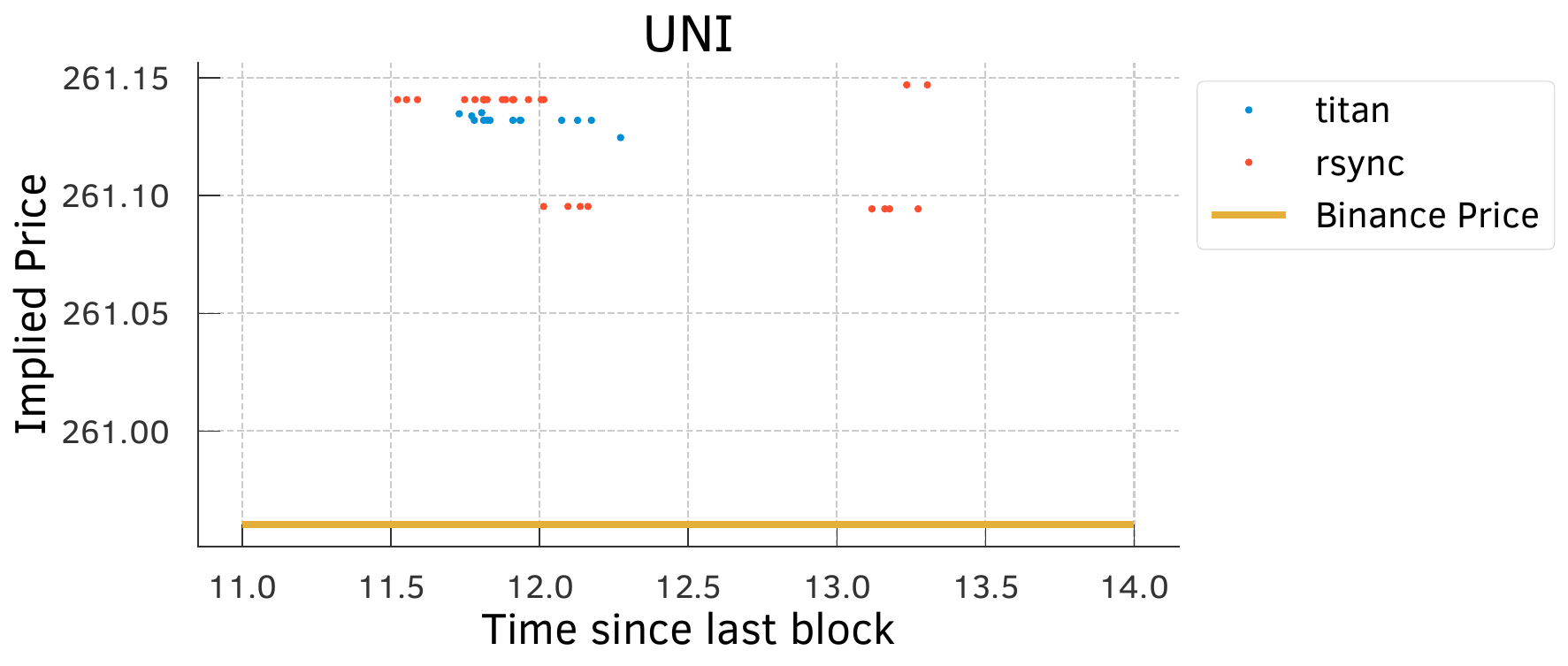}
        \caption{UNI Block 21322653}
        \label{}
    \end{subfigure}%
        \begin{subfigure}[t]{0.5\textwidth}
        \centering
        \includegraphics[scale=0.25]{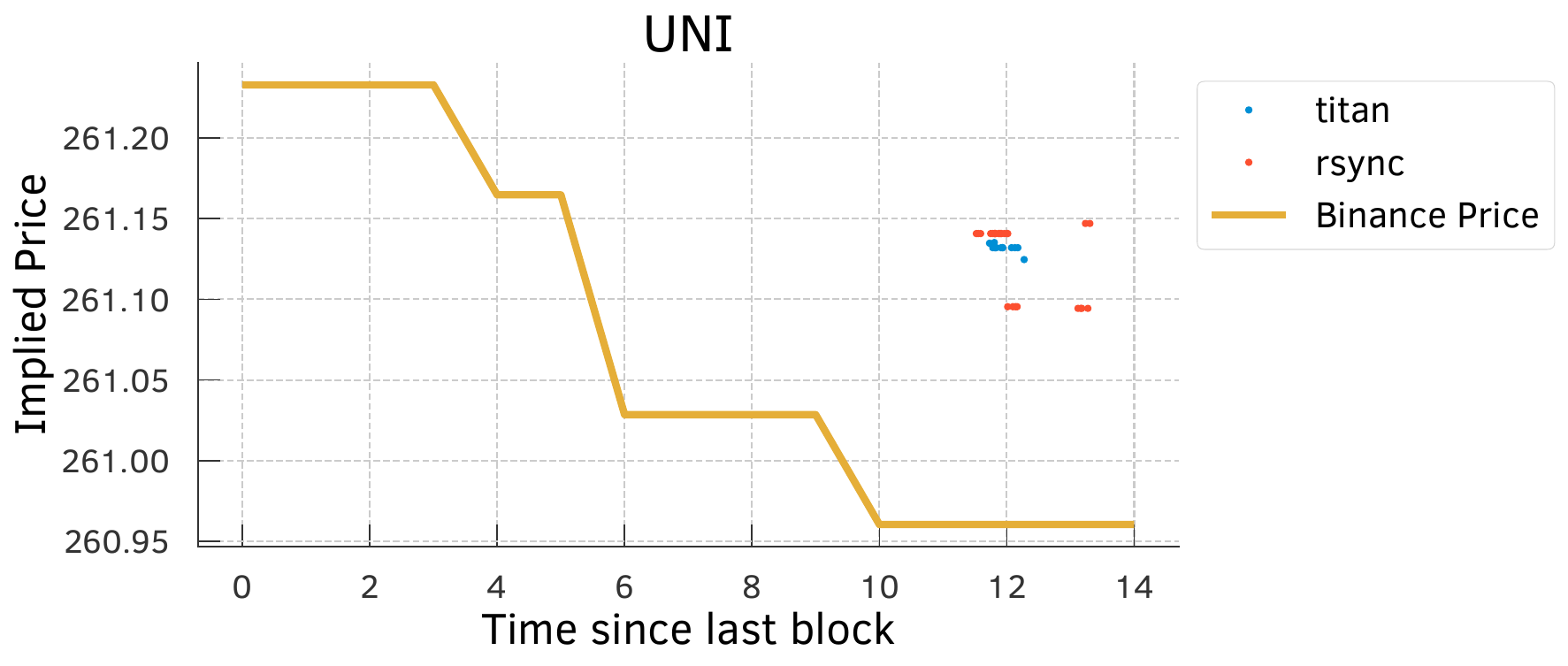}
        \caption{UNI Block 21322653}
        \label{}
    \end{subfigure}
    \caption{Competition between Rsyc-bot and Titan-bot on the UNI/ETH market (all blocks). Note that panels (e) and (f) plot the same data but at different time scales.}
    \label{}
\end{figure*}

\begin{figure*}[bth]
        \centering
        \includegraphics[scale=0.25]{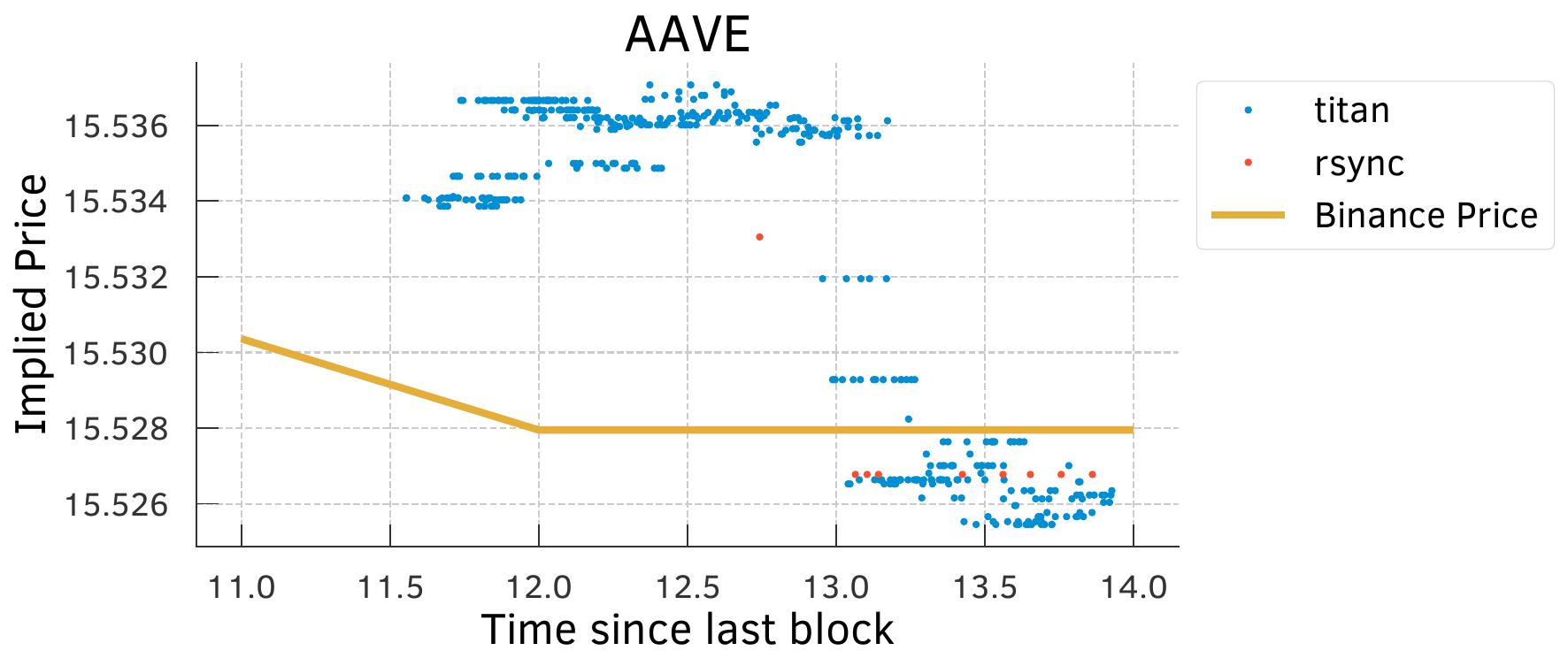}
        \caption{Competition between Rsyc-bot and Titan-bot during AAVE Block 21322658 (ETH as base currency)}
\end{figure*}

\begin{figure*}[bth]
        \centering
        \includegraphics[scale=0.25]{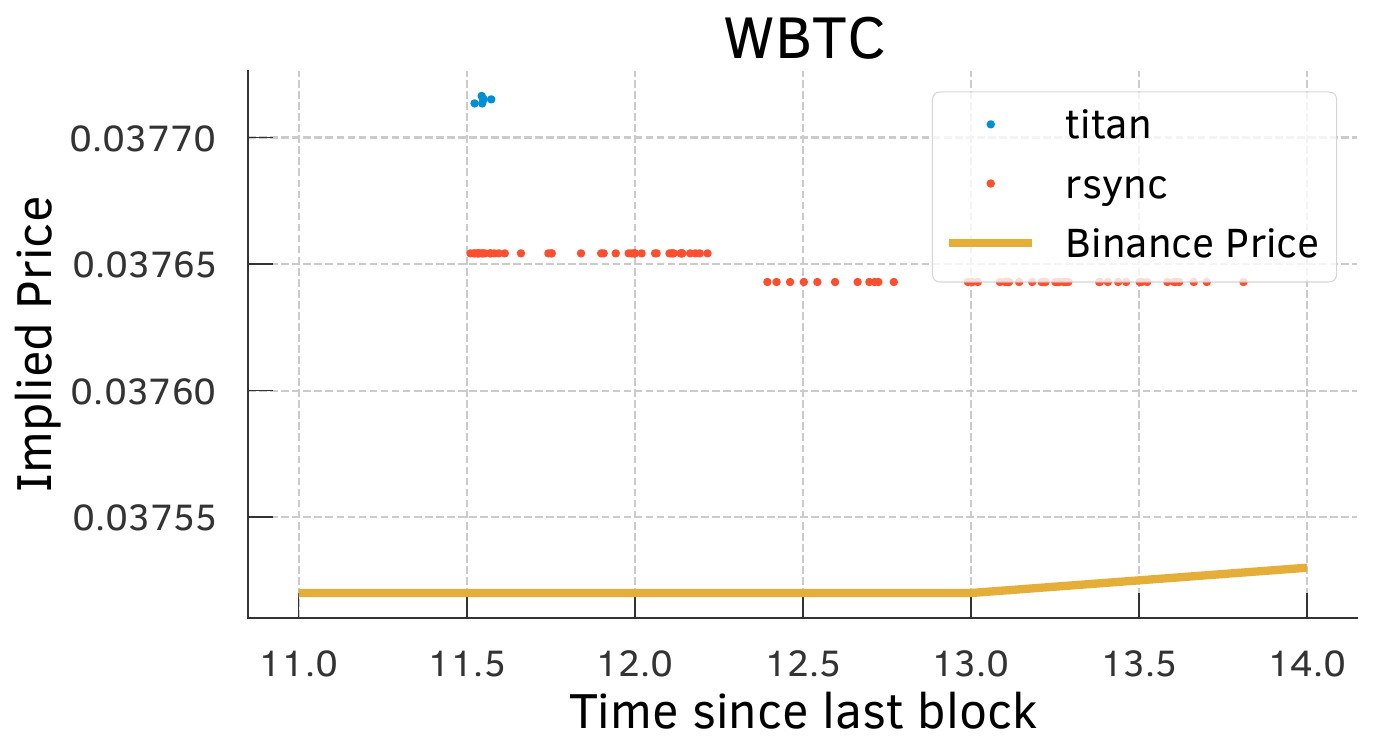}
        \caption{Competition between Rsyc-bot and Titan-bot during WBTC Block 21322641 (ETH as base currency)}
\end{figure*}

\clearpage

\end{document}